\documentclass[prd,
showpacs,preprintnumbers,amsmath,amssymb,
superscriptaddress,
a4paper,nofootinbib,longbibliography]{revtex4-2}

\usepackage{setspace}

\usepackage{nicefrac}

\usepackage{esint}

\usepackage{natbib}
\usepackage{amssymb,amsbsy,amsmath,amsfonts}
\usepackage{graphicx}
\usepackage{epsf,epsfig,float,latexsym,amsthm,fancyhdr,rotating}
\usepackage{graphics,psfrag,longtable}
\usepackage{slashed}
\usepackage[normalem,normalbf]{ulem}
\usepackage[colorlinks,citecolor=blue,linktoc=all,linkcolor=cyan,urlcolor=blue]{hyperref}
\usepackage{upgreek}
 
 \usepackage{multirow}

\def\beq{\begin{equation}}
\def\eeq{\end{equation}}
\def\bea{\begin{eqnarray}}
\def\eea{\end{eqnarray}}
\def\beqa{\begin{equation}\begin{array}{l}}
\def\eeqa{\end{array}\end{equation}}
\def\eqlab#1{\label{eq:#1}}
\def\figlab#1{\label{fig:#1}}

\def\seclab#1{\label{sec:#1}}
\def\eref#1{(\ref{eq:#1})}
\def\Eqref#1{Eq.~(\ref{eq:#1})}

\def\Figref#1{Fig.~\ref{fig:#1}}

\def\secref#1{Section \ref{sec:#1}}

\def\bv#1{\boldsymbol{#1}}

\def\barr{\left(\begin{array}{c}}
\def\earr{\end{array}\right)}
\def\bmat{\left(\begin{array}{cc}}
\def\emat{\end{array}\right)}
\def\al{\alpha}

\def\ga{\gamma}

\def\nn{\nonumber}
\def\dd{\mathrm{d}}

\DeclareMathOperator\arctanh{arctanh}
\DeclareMathOperator\arccoth{arccoth}
\DeclareMathOperator\im{Im}

\def\3d{3-D}

\def\ol#1{\overline{#1}}

\def\piEFT/{$\slashed{\pi}$EFT}

\def\2PE{2$\upgamma$}

\usepackage{xcolor}
\usepackage{bm}
\usepackage{slashed}
\usepackage{graphicx}
\allowdisplaybreaks

\makeatletter
\g@addto@macro\bfseries{\boldmath}
\makeatother

\begin{document}
\preprint{MITP-22-051}
\preprint{PSI-PR-22-19}

\author{Vadim Lensky}
\affiliation{Institut f\"ur Kernphysik,
 Johannes Gutenberg-Universit\"at  Mainz,  D-55128 Mainz, Germany}

 \author{Franziska Hagelstein}
 \affiliation{Institut f\"ur Kernphysik,
 Johannes Gutenberg-Universit\"at  Mainz,  D-55128 Mainz, Germany}
\affiliation{Paul Scherrer Institut, CH-5232 Villigen PSI, Switzerland}

\author{Vladimir Pascalutsa}
\affiliation{Institut f\"ur Kernphysik,
 Johannes Gutenberg-Universit\"at  Mainz,  D-55128 Mainz, Germany}

\title{Two-photon exchange in (muonic) deuterium  at N3LO in pionless effective field theory }

\begin{abstract}
We present a study of the two-photon-exchange (\2PE-exchange) corrections to the $S$-levels in muonic ($\mu$D) and ordinary (D) deuterium within the pionless effective field theory (\piEFT/). Our calculation proceeds up to next-to-next-to-next-to-leading order (N3LO) in the \piEFT/ expansion. The only unknown low-energy constant entering the calculation at this order corresponds to the coupling of a longitudinal photon to the nucleon-nucleon system. To minimise its correlation with the deuteron charge radius, it is extracted using the information about the hydrogen-deuterium isotope shift. We find the elastic \2PE-exchange contribution in $\mu$D larger by several standard deviations than obtained in other recent calculations. This discrepancy ameliorates the mismatch between theory and experiment on the size of \2PE-exchange effects, and is attributed to the properties of the deuteron elastic charge form factor parametrisation used to evaluate the elastic contribution. We identify a correlation between the deuteron charge and Friar radii, which can help one to judge how well a form factor parametrisation describes the low-virtuality properties of the deuteron. We also evaluate the higher-order \2PE-exchange contributions in $\mu$D, generated by the single-nucleon structure and expected to be the most important terms beyond N3LO.
The uncertainty of the theoretical result is dominated by the truncation of the \piEFT/ series and is quantified using a Bayesian approach. The resulting extractions of the deuteron charge radius from the $\mu$D Lamb shift, the $2S-1S$ transition in D, and the $2S-1S$ hydrogen-deuterium isotope shift, with the respective \2PE-exchange effects evaluated in a unified \piEFT/ approach, are in perfect agreement.

\end{abstract}

\date{\today}

\maketitle

\tableofcontents
\newpage

\section{Introduction}

Laser spectroscopy of muonic hydrogen ($\mu$H) and deuterium ($\mu$D) by the CREMA Collaboration in 2010, 2013 and 2016 enabled determinations of the proton and deuteron charge radii with unprecedented precision:
\begin{subequations}
\eqlab{CREMAresults}
\bea
r_p(\mu \text{H})&=&0.84087(26)_\text{exp}(29)_\text{theory}\ \text{fm}=0.84087(39)\ \text{fm \cite{Pohl:2010zza,Antognini:1900ns}},\label{eq:rmuh}\\
r_d(\mu \text{D})&=&2.12562(13)_\text{exp}(77)_\text{theory}\ \text{fm}=2.12562(78)\ \text{fm \cite{Pohl1:2016xoo}},
\label{eq:rmud}
\eea
\end{subequations}
while the most accurate extraction of the deuteron charge radius, 
\beq
r_d(\mu\text{H \& iso})=2.12771(22)\ \text{fm},\label{eq:Rdisotope}\\
\eeq
is an indirect achievement combining measurements from the spectroscopy of ordinary and muonic atoms~\cite{Antognini:1900ns}: the $2S$--$1S$ hydrogen-deuterium (H-D) isotope shift and the Lamb shift in $\mu$H. This result is driving the presently recommended value of the deuteron charge radius from the CODATA 2018 report~\cite{Tiesinga:2021myr}:
\beq 
r_d(\text{CODATA '18})=2.12799(74)\,\mathrm{fm}.\eqlab{CODATA18rd}
\eeq
As one can see from \Eqref{CREMAresults}, the charge radius extractions are limited by the theory uncertainty, which for muonic atoms is almost solely due to subleading nuclear-structure effects, and in particular, the $O(\alpha^5)$ two-photon exchange (\2PE exchange) discussed in this work. 

The initial tension between the $r_d(\mu\text{D})$ and $r_d(\mu\text{H \& iso})$ extractions, shown above, was resolved in 2019 by amending the $\mu$D theory~\cite{Krauth:2015nja} to include the subleading $O(\alpha^6)$ electronic vacuum polarization (VP) effects~\cite{Kalinowski:2018rmf} and the inelastic three-photon exchange ($3\gamma$ exchange)~\cite{Pachucki:2018yxe}:
\beq
r_d(\mu \text{D})=2.12710(13)_\text{exp}(81)_\text{theory}\ \text{fm}=2.12710(82)\ \text{fm \cite{Kalinowski:2018rmf}}.\eqlab{rdKalinowksi}
\eeq
The deuteron-radius extractions from deuterium spectroscopy and electron-deuteron ($ed$) scattering are less precise and lead to larger values:
\begin{subequations}
\bea
r_d(\text{D spectroscopy}) &=& 2.1415(45)\ \text{fm \cite{Pohl:2016glp}},\eqlab{rDDspec}\\
r_d(ed\text{ scattering}) &=&2.130(10)\ \text{fm \cite{Sick:1998cvq}},\\
r_d(\text{CODATA '14}) &=& 2.1413(25)\ \text{fm \cite{Mohr:2015ccw}}.
\eea
\end{subequations}
This distinct discrepancy for the deuteron radius --- the ``deuteron radius puzzle'' --- is strongly affected by the \2PE exchange.  It is thus timely to re-evaluate the \2PE-exchange effects in a model-independent manner and
try to improve their precision. While the latest developments \cite{Pachucki:2018yxe, Kalinowski:2018rmf} are certainly important, they do not provide a path to a more systematic improvement of the theory error
on the side of nuclear structure.

In this work, we consider the forward \2PE-exchange contributions to D and $\mu$D, including the accompanying electronic VP contributions,
within the pionless effective field theory (\piEFT/) of nuclear forces \cite{Kaplan:1996xu,Kaplan:1998we,Kaplan:1998tg,Chen:1999bg,Chen:1999tn,Rupak:1999rk,Phillips:1999hh}.  This framework allows one to represent the nuclear observables
in a well-defined perturbation theory, expanding in powers of the small parameter $P/m_\pi$, where $P$ is the typical momentum scale (e.g., the size of the relative momentum between two nucleons, or that of the momentum of an external probe) and $m_\pi \simeq 139$ MeV is the pion mass. The typical momentum scale in the deuteron is characterized by
the binding momentum $\gamma = \sqrt{M_N B}\simeq 45$ MeV, where $M_N$ is the nucleon mass and $B$ is the deuteron binding energy.
The momentum scale probed by the electromagnetic interaction in $\mu$D is $\sim \al m_\mu$, which is less than an MeV. This is also well below the limiting scale of the theory set by $m_\pi$. 
The atomic systems should thus be well-suited for the application of $\slashed{\pi}$EFT. In addition, it has been shown that $\slashed{\pi}$EFT provides a good description of low-energy experimental data on real deuteron Compton scattering~\cite{Griesshammer:2000mi,Chen:2004wv}, and can be used to investigate the deuteron electric polarizability and electromagnetic form factors (FFs)~\cite{Chen:1999tn}.
Finally, the effective-field-theory (EFT) expansion allows one to quantify the theoretical uncertainty using methods such as Bayesian inference~\cite{Furnstahl:2015rha}.
The basis for our \2PE-exchange calculation is provided in Ref.~\cite{Lensky:2021VVCS}, where closed analytic expressions for the unpolarized amplitudes of forward doubly-virtual Compton scattering (VVCS) off the deuteron are derived. 

The paper is organized as follows. In  \secref{Outline}, we briefly introduce the \2PE-exchange and \piEFT/ frameworks. 
 In Section \ref{sec:TPE_corrections}, we calculate the elastic finite-size, inelastic deuteron-polarizability and single-nucleon contributions to the $\mu$D Lamb shift, and compare 
 our results to other recent predictions.
 In Section \ref{sec:isotope_shift}, we repeat the same calculation for D and use the H-D isotope shift to fix the unknown low-energy constant (LEC) $l_1^{C0_S}$ that enters the VVCS amplitude. In Section \ref{sec:Total_DeltaE}, we utilize the unique possibility to cross-check the theoretical predictions for the \2PE exchange in $\mu$D with an empirical determination. The latter is extracted from the measured $\mu$D Lamb shift by fixing the deuteron charge radius to the independent value $r_d(\mu\text{H \& iso})$. We also compile an update for the theory prediction of the $\mu$D Lamb shift that will be used to extract $r_d(\mu\text{D})$ from the measurement of the CREMA Collaboration. In Section \ref{sec:radii}, we discuss deuteron and proton charge radii extractions from $\mu$D, D and the H-D isotope shift. A discussion of the neutron charge radius is postponed to Appendix \ref{app:neutroncr}. In \secref{Discussion}, we finish with conclusion and outlook. The appendices cover: \ref{sec:error}) the Bayesian error analysis, \ref{app:ff}) the inclusion of nucleon FFs beyond the \piEFT/ framework, \ref{sec:TPE_eVP}) electronic VP corrections, and updated theory compilations for the: \ref{app:iso}) $2S-1S$ H-D isotope shift, \ref{RydbergSection}) $2S-1S$ in H, and \ref{1S2SD}) $2S-1S$ in D. Appendix \ref{RydbergSection} also has a determination of the Rydberg constant $R_\infty$ from $2S-1S$ in H and the Lamb shift in $\mu$H. A concise summary of our main results and their implications is published in Ref.~\cite{Lensky:2022fif}.

\section{Theoretical Framework}
\seclab{Outline}

\subsection{\2PE Exchange in (Muonic) Deuterium} \seclab{TPEintro}

\begin{figure}[t]
\includegraphics[width=0.25\textwidth]{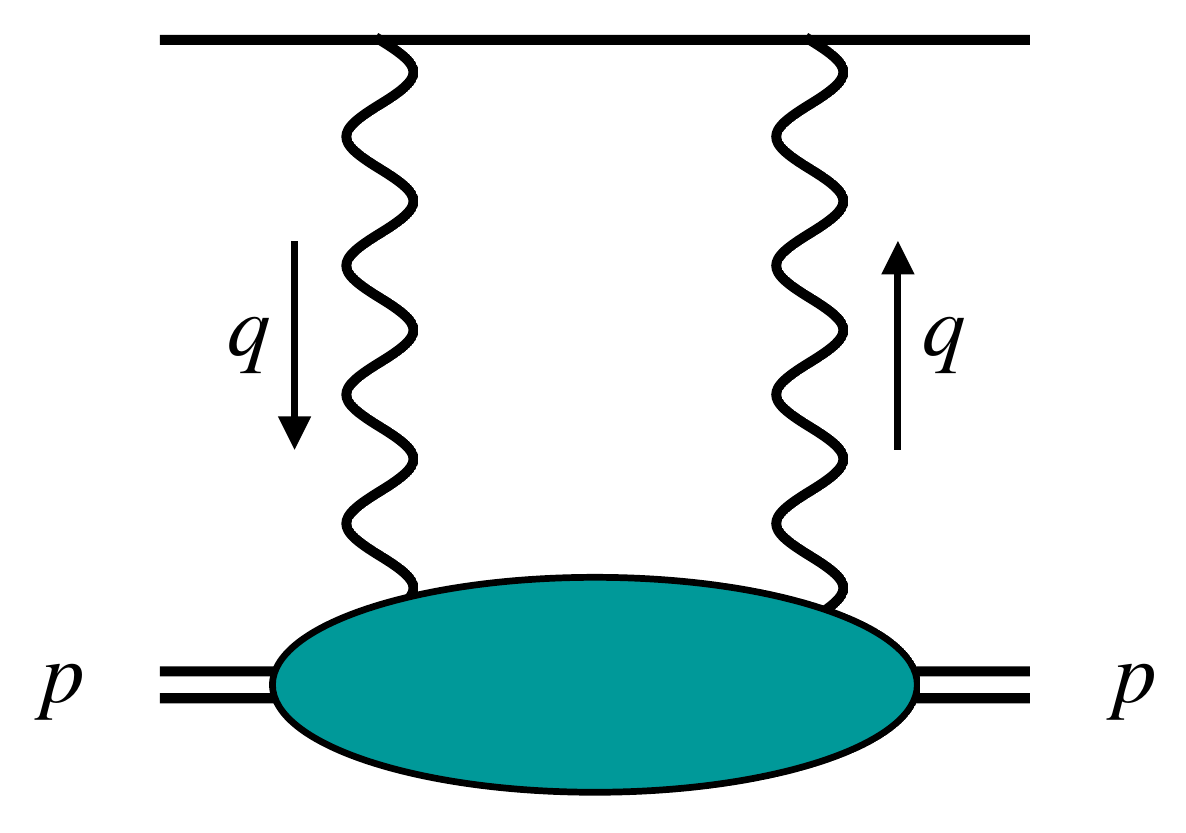}
\caption{The leading order in $\alpha$ \2PE-exchange potential. \figlab{TPE}}
\end{figure}

The leading order (LO) in $\alpha$ \2PE-exchange correction corresponds to the forward kinematics, shown in \Figref{TPE}. It gives a $\delta(\boldsymbol{r})$-function correction to the Coulomb potential, thus, only shifts the $S$-levels, which have a non-vanishing atomic wave function at the origin. The spin-independent forward \2PE exchange is related to the VVCS amplitude off an unpolarized deuteron: 
\begin{equation}
    T_{fi} = \varepsilon_0\,\varepsilon_0^{\,\prime*}\,f_{L}(\nu,Q^2) +(\bv{\varepsilon} \cdot \bv{\varepsilon}^{\,\, \prime *}) \,f_{T}(\nu,Q^2),
\label{eq:VVCS_amplitude_LandT}
\end{equation}
where $f_L(\nu,Q^2)$ and $f_T(\nu,Q^2)$ are the longitudinal and transverse scalar amplitudes with $Q^2=-q^2$ and $\nu = p\cdot q/M_d$ the photon virtuality and lab frame energy, and $M_d$ the deuteron mass. The modified photon polarization vector components are defined as
\begin{align}
\varepsilon_0&=\left[\epsilon_0-\frac{\nu}{\left|\bv{q}\right|}\, (\bv{\epsilon}\cdot\bv{\hat q})\right]\frac{\left|\bv{q}\right|}{Q},
&\bv{\varepsilon} = \bv{\epsilon}-\bv{\hat q}\,(\bv{\epsilon}\cdot\bv{\hat q}),\eqlab{modified_polarization_vectors}
\end{align}
with $\bv{q}$ and $\bv{\hat{q}}=\bv{q}/|\bv{q}|$ being the photon three-momentum in the lab frame and its unit vector, and $(\epsilon_0,\bv{\epsilon})$ the time and space components of the photon polarization vector. This description in terms of $f_L(\nu,Q^2)$ and $f_T(\nu,Q^2)$ is natural for the $\slashed{\pi}$EFT framework, but not unique. The explicitly covariant tensor decomposition with two other scalar amplitudes $T_i(\nu,Q^2)$ related via
\begin{equation}
    f_L(\nu,Q^2) = -T_1(\nu,Q^2)+\left(1+\frac{\nu^2}{Q^2}\right)T_2(\nu,Q^2), \qquad f_T(\nu,Q^2) = T_1(\nu,Q^2),
\end{equation}
is widely used in, e.g., the dispersive \2PE-exchange evaluations~\cite{Carlson:2011zd,Carlson:2013xea}. We start from the covariant expression for the forward $O(\alpha^5)$ \2PE-exchange correction to the energy of a $nS$ state in (muonic) deuterium, given in these references, and rewrite them in terms of the longitudinal and transverse amplitudes:
\begin{subequations}
\bea
   E_{nS}^\mathrm{fwd}&=& -8i\pi \al m \,\left[\phi_{n}(0)\right]^2\,
\int \!\!\frac{\dd^4 q}{(2\pi)^4}   \frac{\left(Q^2-2\nu^2\right)T_1(\nu,Q^2)-(Q^2+\nu^2)\,T_2(\nu,Q^2)}{Q^4(Q^4-4m^2\nu^2)}
\label{eq:TPE_LS}\\
&=& \hphantom{-}8i\pi \al m \,\left[\phi_{n}(0)\right]^2\,
\int \!\!\frac{\dd^4 q}{(2\pi)^4}   \frac{f_L(\nu,Q^2)+2(\nu^2/Q^2)f_T(\nu,Q^2)}{Q^2(Q^4-4m^2\nu^2)},\label{eq:TPE_LT}
\eea
\end{subequations}
where  $m$ is the electron or muon mass, $[\phi_{n}(0)]^2=1/(\pi n^3 a^3)$ is the (Coulomb) wave function of the $nS$ atomic state at the origin, $a=1/(\mathcal{Z} \al m_r)$ is the Bohr radius, $\mathcal{Z} $ is the nuclear charge ($\mathcal{Z}=1$ for the deuteron), and $m_r=m M_d/(m + M_d)$ is the atomic reduced mass. Separating the scalar amplitudes into the deuteron-pole and non-pole parts, one splits the \2PE-exchange correction into the elastic and inelastic part~\cite{Carlson:2013xea}. The inelastic part, after doing the Wick rotation $\nu = iq_0$ and introducing the hyperspherical coordinates, takes the form:
\begin{align}
E_{nS}^\mathrm{inel}
&=-\frac{\alpha}{2\pi^2 m}\left[\phi_{n}(0)\right]^2
\int\limits_0^\infty\frac{\mathrm{d}Q}{Q}\int\limits_{-1}^{1} \mathrm{d}x\,\sqrt{1-x^2}\,
\frac{f_L(-iQx,Q^2)-2x^2 f_T(-iQx,Q^2)}{
\tau_l+x^2},
\label{eq:LS_from_LT}
\end{align}
with $\tau_l=Q^2/(4m^2)$. Here we assume that the pole-part is subtracted from the scalar VVCS amplitudes.
The elastic part of the \2PE exchange is readily obtained via the deuteron electromagnetic FFs --- charge, magnetic, and quadrupole --- $G_C(Q^2)$, $G_M(Q^2)$, and $G_Q(Q^2)$,
resulting in~\cite{Carlson:2013xea}:
\begin{align}
E_{nS}^\mathrm{elastic} & = \frac{m \alpha^2}{M_d(M_d^2-m^2)}[\phi_{n}(0)]^2
\int\limits_0^\infty2\frac{\mathrm{d}Q}{Q} 
 \times\left\{
\frac{2}{3}G_M^2(Q^2)
(1+\tau_d)\hat{\gamma}_1(\tau_d,\tau_l)
 \right. \nonumber\\
& \qquad \left. 
-\left[\frac{G_C^2(Q^2)-1}{\tau_d}+\frac{2}{3}G_M^2(Q^2)+\frac{8}{9}\tau_d G_Q^2(Q^2)\right]
\hat{\gamma}_2(\tau_d,\tau_l)
\right. +16M_d^2\frac{M_d-m}{Q}G_C'(0)
\bigg\},\label{eq:contrib_elastic}
\end{align}
where $\tau_d=Q^2/(4M_d^2)$, and the weighting functions are defined by:
\begin{subequations}
\begin{align}
\hat{\gamma}_{1,2}(x,y) & = \frac{\gamma_{1,2}(x)}{\sqrt{x}}-\frac{\gamma_{1,2}(y)}{\sqrt{y}},\\
\gamma_1(x) & = (1-2x)\sqrt{1+x}+2x^{3/2},\\
\gamma_2(x) & = (1+x)^{3/2}-x^{3/2} -\frac{3}{2}\sqrt{x}.
\end{align}
\end{subequations}
Note that the contributions of point-like charge and charge radius of the deuteron are removed from the elastic part to avoid double counting~\cite{Carlson:2013xea}. This is done by subtracting the unity and the term proportional to $G_C'(0)$ in Eq.~\eqref{eq:contrib_elastic}.

\subsection{Unpolarized Deuteron VVCS in Pionless EFT}\seclab{pEFT}

In our analysis, we use results from the \piEFT/ calculation of the unpolarized deuteron VVCS amplitudes $f_L(\nu, Q^2)$ and $f_T(\nu,Q^2)$ presented in Ref.~\cite{Lensky:2021VVCS}. This section gives a brief recap of the \piEFT/ framework applied to the deuteron VVCS, as well as a description of the technicalities relevant to the \2PE-exchange calculation.

\piEFT/ is an EFT for nucleon interactions at low energies, where the high-energy scale is set by the pion mass $m_\pi$. If the momentum transfer between two nucleons is $P\ll m_\pi$, one can treat a pion-exchange interaction as a contact one. In \piEFT/ nucleons are thus interacting through contact interactions~\cite{Chen:1999tn,Kaplan:1996xu,Kaplan:1998we,Kaplan:1998tg}. The Lagrangian is constructed performing a non-relativistic expansion in the one-nucleon sector and writing out the relevant two-nucleon interactions \cite{Chen:1999tn,Kaplan:1996xu,Kaplan:1998we,Kaplan:1998tg,Chen:1999bg,Rupak:1999rk,Phillips:1999hh}.
To assign a particular order to a Feynman graph, one counts powers of momenta [$Q=O(P)$] and energies [$\nu=O(P^2)$] coming from the interaction vertices, nucleon propagators [$O(P^{-2})$], and loops [$O(P^5)$]. The small expansion parameter is the ratio $P/m_\pi$. For the deuteron, where the typical momentum scale is the binding momentum $\gamma$, this corresponds to $P/m_\pi\simeq 1/3$. Note that different momentum scales can count as different powers of the typical momentum $P$, depending on the problem setting. For instance, the counting we use has the photon three-momentum $|\bv{q}|=O(P)$, whereas its energy is $\nu=O(P^2)$, and hence also its virtuality $Q=O(P)$.
This reflects our expectation that the virtual photons in the $2\gamma$-exchange, as viewed in the lab frame, mostly transfer three-momentum, and very little energy, to the intermediate deuteron state, and is in contrast to, e.g., a typical real Compton scattering setting where $\nu=|\bv{q}|$, implying they have to be of the same size in the counting. 

Regarding the description of the deuteron state, one can use different prescriptions to perform the expansion around the deuteron pole of the nucleon-nucleon ($NN$) scattering amplitude. The $z$-parametrisation~\cite{Phillips:1999hh}, chosen in Ref.~\cite[Sec.\ II B]{Lensky:2021VVCS}, is particularly well-suited for quantities such as the deuteron electric dipole polarizability $\alpha_{E1}$ that receive mostly long-range contributions and are thus sensitive to the correct description of the long-range tail of the deuteron wave function.
This parametrisation reproduces the residue $Z$ of the $NN$ scattering amplitude at the deuteron pole at next-to-leading order (NLO). The residue
is related to the effective range $\rho_d$ in the $NN$ triplet channel via $Z=(1-\gamma\rho_d)^{-1}$, and is also connected to the asymptotic normalisation of the deuteron $S$-wave via
 \begin{align}
 \psi(r)\xrightarrow{r\rightarrow \infty}\sqrt{\frac{\gamma Z}{2\pi}}\frac{e^{-\gamma r}}{r}.
 \end{align}
It is therefore straightforward to see that this procedure also reproduces the correct large-distance asymptotics at NLO. Note also that it introduces a new formal expansion parameter $(Z-1)=O(P)$. 

Analysing the counting for the VVCS  shows~\cite[Sec.\ II B]{Lensky:2021VVCS} that the longitudinal amplitude, driven by the deuteron electric polarizability $\alpha_{E1}$, is dominant, starting at $O(P^{-2})$, whereas the transverse amplitude starts two orders higher at $O(P^0)$. In the context of the \2PE-exchange correction, Eq.~\eqref{eq:TPE_LT} shows that the $f_T$ contribution is additionally suppressed, compared to the contribution of $f_L$, by the factor $\nu^2/Q^2 = O(P^2)$. The transverse contribution to the \2PE-exchange correction therefore starts only at $O(P^2)$, or N4LO compared to the leading longitudinal contribution. It is also at N4LO that, as explained in Ref.~\cite{Lensky:2021VVCS}, higher powers of momenta entering the \piEFT/ expansion render the \2PE-exchange correction naively divergent. This divergence ought to be absorbed by a four-nucleon two-lepton contact term entering at this order, and, since there is no data that would allow one to pinpoint the corresponding coupling other than the \2PE-exchange correction itself, this is where the predictive power of \piEFT/ is exhausted. This motivated us to calculate the longitudinal amplitude up to N3LO in Ref.~\cite{Lensky:2021VVCS}. We also calculated the transverse amplitude up to $O(P)$, or its respective NLO; this allows us to quantify here the corresponding \2PE-exchange contribution.

Further details of the \piEFT/ framework used to calculate $f_L$ and $f_T$ at their respective N3LO and NLO can be found in Ref.~\cite[Sec.\ II]{Lensky:2021VVCS}. The results for the VVCS amplitudes are given in a closed analytic form in Ref.~\cite[Sec.\ III]{Lensky:2021VVCS}, in terms of the longitudinal and transverse four-point functions $\mathcal{M}_{L,T}(\nu,Q^2)$ and the inverse of the derivative of the deuteron self-energy (SE) at the deuteron pole $\left[\Sigma'(E_d)\right]^{-1}$. We use them here to calculate the \2PE-exchange correction. At N3LO in the \piEFT/ expansion of $f_L$, one encounters a previously undetermined coupling $l_1^{C0_S}$ of a longitudinal photon to the two-nucleon system, which contributes, in particular, to $G_C(Q^2)$ and $r_d$. The latter quantity was used in Ref.~\cite{Lensky:2021VVCS} to extract $l_1^{C0_S}$ from a fit to $r_d(\mu\text{D})$ in \Eqref{rmud}. This procedure is potentially problematic due to the fact that $l_1^{C0_S}$ also enters the \2PE-exchange correction, both in $\mu$D and the isotope shift. We investigate the resulting correlations below and demonstrate explicitly that they are negligible at the current level of theoretical and experimental precision.

\section{\2PE Exchange in Muonic Deuterium}
\label{sec:TPE_corrections}

In the following section, we will present in detail our calculation of the elastic and inelastic \2PE-exchange contributions to the Lamb shift in $\mu$D. A summary of our  results can be found in \secref{SummaryTPEmuD}.

\subsection{Elastic Contribution}
\label{sec:TPE_elastic}

We start by considering the elastic contribution to the \2PE-exchange correction based on the \piEFT/ deuteron FFs. Taking the N3LO result for $G_C(Q^2)$ in Ref.~\cite[Eq.~(75)]{Lensky:2021VVCS} and expanding in  Eq.~\eqref{eq:contrib_elastic} also to N3LO results in
 \begin{align}
 E_{2S}^\mathrm{elastic}=\left[-0.4482-0.9938\, l_1^{C0_S} \right]\text{ meV}.
 \label{eq:TPE_elastic_ct_separated}
 \end{align}
 This neglects the magnetic and quadrupole FFs, whose contributions are subleading in the \piEFT/ counting and are indeed numerically very small, see below in  Table~\ref{tab:FF_numbers_TPE}.
 The electric contact term coupling $l_1^{C0_S}$ can be fixed through the deuteron charge radius:
  \beq
  r_d^2\equiv  - 6\, G_C'(0)  =
    \frac{1}{8 \gamma ^2}
    +\frac{Z-1}{8 \gamma ^2}
    +2r_0^2
    +\frac{3(Z-1)^3}{\gamma ^2}\,l_1^{C0_S}, \eqlab{rdl1}
    \eeq
with $r_0^2 = \nicefrac{1}{2}\left[r_p^2 + \nicefrac{3}{4}\,M_p^{-2} + r_n^2 \right]$ being the isoscalar nucleon charge radius, with the proton Darwin-Foldy term $\nicefrac{3}{8}\,M_p^{-2}$ added to it.
Previously, $l_1^{C0_S}$ was chosen to reproduce the deuteron charge radius from $\mu$D spectroscopy, resulting in
    \begin{align}
    l_1^{C0_S}=-2.32(41)\times 10^{-3}~\text{\cite{Lensky:2021VVCS},}
\label{eq:contact_term_value}
\end{align}
where the uncertainty stems from the error of the deuteron radius, Eq.~\eqref{eq:rmud},  and the uncertainty of $Z$. However, the extraction of $r_d^2$ from $\mu$D spectroscopy depends on the theory result for the \2PE-exchange correction (even though the contribution of $l_1^{C0_S}$ to the \2PE-exchange correction is small).
This correlation can be practically eliminated if the deuteron radius extracted from the combination of the proton radius and the H-D $2S-1S$ isotope shift, as given in Eq.~\eqref{eq:Rdisotope}, is used as the reference data point. One has to note that the isotope shift also has a \2PE-exchange contribution, but its relative importance as well as its correlation with $r_d^2$ is much smaller. To investigate this issue quantitatively, we perform a re-analysis of the isotope shift using the \piEFT/ formalism to predict the \2PE-exchange correction in ordinary D, see Section~\ref{sec:isotope_shift} and Appendix~\ref{app:iso}. Our calculation confirms that the contribution of $l_1^{C0_S}$ to the isotope shift can indeed be safely neglected. The corresponding result for the electric contact term coupling, which will be used throughout this work, is
\begin{align}
    l_1^{C0_S}=-1.80(38)\times 10^{-3}.
\label{eq:contact_term_value2}
\end{align}
This agrees with the result that we deduced from \Eqref{Rdisotope} \cite{Lensky:2022tue}, but differs from  \Eqref{contact_term_value} by about $1\, \sigma$, since  the extraction via the isotope shift gives a value of $r_d(\mu\text{H \& iso})$ slightly different from $r_d(\mu\text{D})$ in Eq.~\eqref{eq:rmud}. The related effect on $E_{2S}^\mathrm{elastic}$ is small. 
The final numerical result for the elastic contribution is:
\begin{align}
    E_{2S}^\mathrm{elastic}=[-0.2043-0.1582-0.0626-0.0213]~\mathrm{meV} = -0.4463(77) \text{ meV},
\label{eq:elastic_piEFT}
\end{align}
where the numbers here stand for the order-by-order contributions.
The uncertainty of $E_{2S}^\mathrm{elastic}$ is due to higher orders in the \piEFT/ expansion;
we quantify it as explained  in Appendix~\ref{sec:error}.

{\renewcommand{\arraystretch}{1.5}
\begin{table}[tbh]
    \centering
    \begin{tabular}{c||r|r|r}
         & $E_{2S}^\mathrm{elastic}$&$E_{2S}^{\mathrm{inel},L}$&$E_{2S}^\mathrm{sum}$\\
 \hline
 \hline
&\multicolumn{3}{c}{LO}\\
\hline
$E^{(-3)}$ &$-0.2043$&$-0.9433$&$-1.1476$\\
         \hline
         \hline
&\multicolumn{3}{c}{NLO}\\
\hline
$(Z-1)E^{(-3)}$ &$-0.1408$&$-0.6502$&$-0.7910$\\
$E^{(-2)}$ &$-0.0174$&$0.0153$&$-0.0021$\\
         \hline
         \hline
&\multicolumn{3}{c}{N2LO}\\
\hline
$(Z-1)E^{(-2)}$ &$-0.0120$&$0.0106$&$-0.0014$\\
$E^{(-1)}$ &$0$&$-0.0006$&$-0.0006$\\
$E^{(-1)}_{r_N^2}$ &$-0.0506$&$0.0389$&$-0.0117$\\
         \hline
         \hline
&\multicolumn{3}{c}{N3LO}\\
\hline
$(Z-1)E^{(-1)}$ &$0$&$-0.0004$&$-0.0004$\\
$(Z-1)E^{(-1)}_{r_N^2}$ &$-0.0349$&$0.0268$&$-0.0081$\\
$E^{(0)}$ &$0$&$-0.0009$&$-0.0009$\\
$E^{(0)}_{w_2}$ &&$0.0002$&$0.0002$\\
$E^{(0)}_{P}$ &&$0.0068$&$0.0068$\\
$ E^{(0)}_{l_1^{C0_S}}$ &$0.0018$&$-0.0012$&$0.0006$\\
$E^{(0)}_{r_N^2}$ &$0.0118$&$-0.0063$&$0.0055$\\
   \end{tabular}
    \caption{$E_{2S}^\mathrm{elastic}$, $E_{2S}^{\mathrm{inel},L}$ and their sum $E_{2S}^\mathrm{sum}$ in detail: contributions appearing at each order in the expansion. Values are in meV.
    Upper indices indicate the order of $\mathcal{M}_L$ that generates the corresponding contribution, see~Ref.~\cite{Lensky:2021VVCS}. Quantities without labels are the contributions at the respective order excluding the labelled terms listed separately. Labels indicate specific terms within $\mathcal{M}_L$: $r_N^2$, $w_2$, $P$, and $l_1^{C0_S}$ stand, in order, for the nucleon charge radii correction, the contribution proportional to the $NN$ triplet $S$-wave shape parameter $w_2$, the $NN$ $P$-wave contribution, and the contribution proportional to $l_1^{C0_S}$.
    }
    \label{tab:TPE_contributionsElInel}
\end{table}}

To study the elastic (and inelastic) contribution in detail, we split them as shown in Table~\ref{tab:TPE_contributionsElInel}, keeping track of different terms appearing both due to the $(Z-1)$ factors coming from the NLO piece of $\left[\Sigma^\prime(E_d)\right]^{-1}$ and due to new sources at each order in the longitudinal four-point function $\mathcal{M}_L$.
This representation will also be useful below in the investigation of the theoretical uncertainty.
In order to split the elastic term this way, it is convenient to rewrite the last term in Eq.~\eqref{eq:contrib_elastic} replacing $G_C'(0)$ by $G_C(0)\, G_C'(0)=\nicefrac{1}{2}\left[G_C^2(Q^2)\right]'|_{Q^2=0}$, where the normalization of $G_C(0)=1$ is used. This reflects the fact that the elastic part of the VVCS amplitude is proportional to the deuteron FFs squared, and allows one to separate the contributions in the integrand without generating spurious singularities at $Q=0$.
One can see from the table that the most important contributions to $E_{2S}^\mathrm{elastic}$ come, as expected, from the LO part of $\mathcal{M}_L$, with the nucleon charge radius contributions providing the most important correction at N2LO. One can also see that the only new contributions beyond NLO come either from the nucleon structure or from the N3LO contact term proportional to $l_1^{C0_S}$. The nucleon charge radius contributions may seem somewhat larger than expected at N2LO and N3LO; to judge whether this is an indication of potentially sizeable corrections to $E_{2S}^\mathrm{elastic}$ beyond N3LO, it is instructive to look at the details of the deuteron charge FF at small photon virtualities. Indeed, it is evident from Eq.~\eqref{eq:contrib_elastic} that the \2PE-exchange integrand is strongly weighted towards low $Q^2$. Therefore, it is the slopes and the curvatures of the deuteron FFs at $Q^2=0$ that will have significant influence on the elastic contribution. The slope of the charge FF, proportional to $r_d^2$, is reproduced at N3LO; based on that alone, a sizeable modification of the shape of $G_C(Q^2)$ at small $Q^2$ could come from higher-order coefficients in its expansion in powers of $Q^2$. To look into this issue, we review the calculation of $E_{2S}^\mathrm{elastic}$ using several different deuteron FFs along with the N3LO \piEFT/ result, and investigate how the features of those FFs affect the result.

\begin{figure}[hbt]
    \centering
    \begin{tabular}{cc}
    \includegraphics[height=0.35\textwidth]{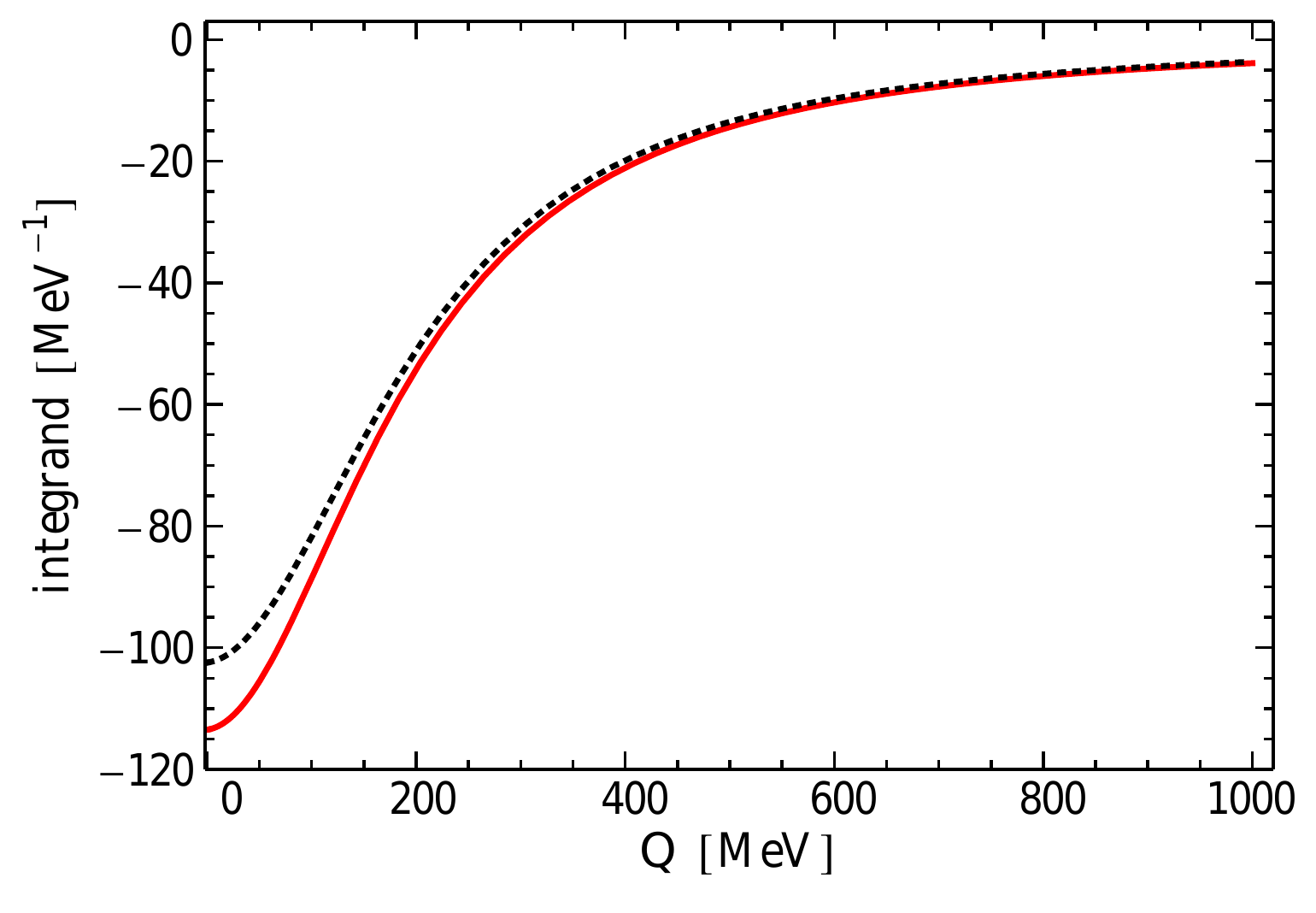}
    &  
    \includegraphics[height=0.35\textwidth]{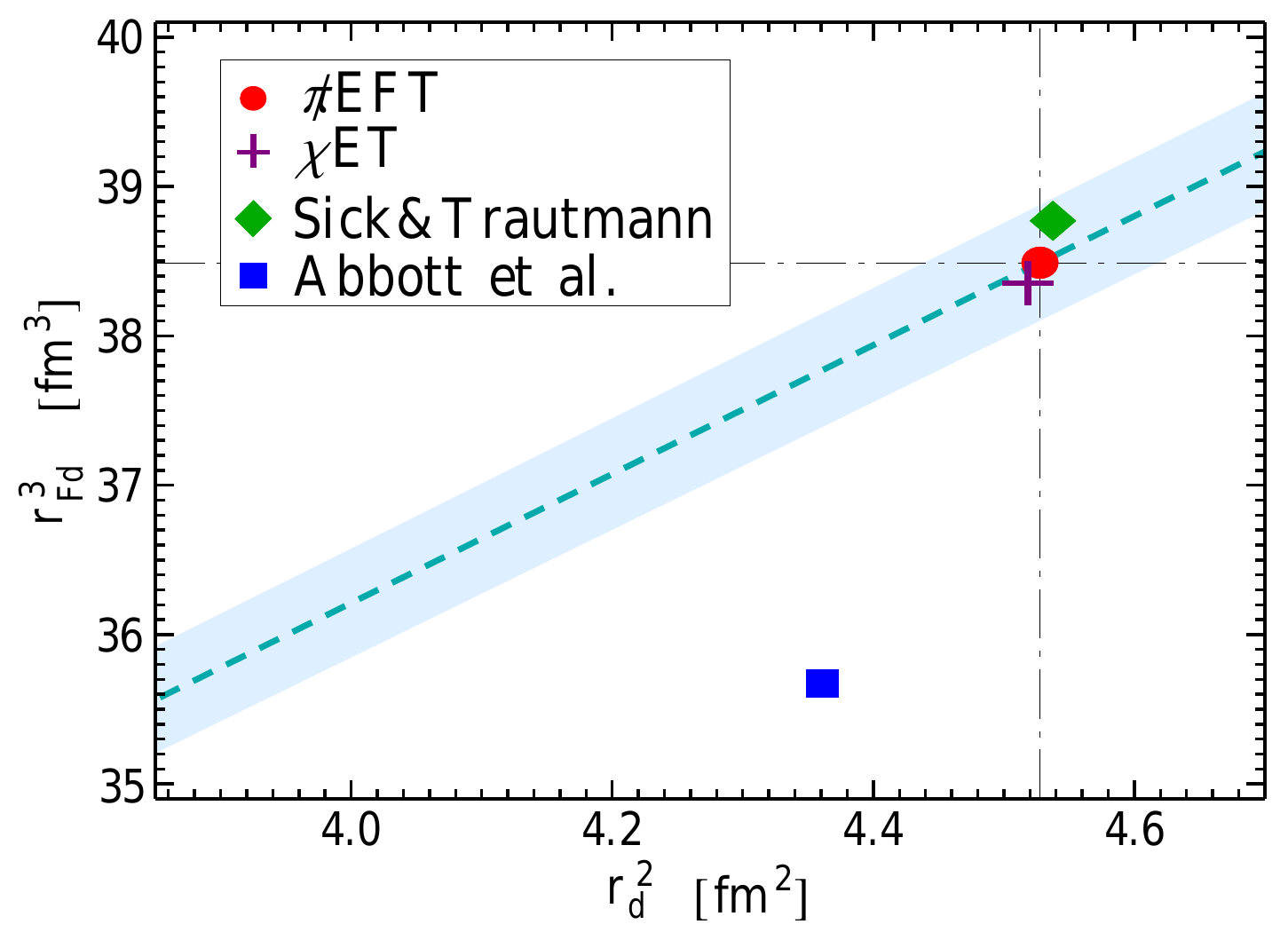}
    \end{tabular}
    \caption{Left panel: Integrand of Eq.~\eqref{eq:contrib_elastic} as function of $Q$. Black dotted: deuteron form factor parametrisations from Ref.~\cite{Abbott:2000ak}; red solid: result of the \piEFT/ calculation. 
    Right panel: Correlation of $r_{\mathrm{F}d}^3$ and $r_d^2$. The dashed line shows the correlation obtained from the \piEFT/ results at N3LO, with the band showing the estimated $~1\%$ N3LO uncertainty; the red disc, purple cross, green diamond, and blue square show the values obtained, respectively, from \piEFT/ at N3LO, the N4LO $\chi$ET form factor~\cite{Filin:2020tcs}, the parametrisation of Ref.~\cite{Sick:1998cvq}, and the parametrisation of Ref.~\cite{Abbott:2000ak}.
    }
    \label{fig:RF_correlation}
\end{figure}

Starting with the recent higher-order, N4LO in the respective counting, chiral effective theory ($\chi$ET) calculation of Refs.~\cite{Filin:2019eoe,Filin:2020tcs}, a good agreement between the N3LO \piEFT/ and the N4LO $\chi$ET results for $G_C(Q^2)$ at low $Q$ was pointed out in Ref.~\cite{Lensky:2021VVCS}. As expected, our result for $E_{2S}^\mathrm{elastic}$ perfectly agrees with what one obtains using the $\chi$ET charge FF from Ref.~\cite{Filin:2020tcs}:
\begin{align}
    E_{2S}^\mathrm{elastic}(\chi\text{ET}) & = -0.4456(18)\text{ meV},
\label{eq:elastic_chiral}
\end{align}
where we neglected the  magnetic and quadrupole contributions, and evaluated the  uncertainty using the uncertainty of the $\chi$ET result for $G_C(Q^2)$.
On the other hand, using the recent empirical deuteron FFs from Ref.~\cite{Abbott:2000ak}, Carlson et al.\ obtained a considerably smaller value~\cite{Carlson:2013xea}:
\begin{align}
    E_{2S}^\mathrm{elastic}(\text{emp.\ FF~\cite{Abbott:2000ak}}) & = -0.417(2) \text{ meV},\eqlab{elasticAbbott}
\end{align}
with the uncertainty estimated using the different FF parametrisations derived in~\cite{Abbott:2000ak}. The same result has been adopted in Ref.~\cite{Acharya:2020bxf}. We repeat this calculation, separating the contributions from the charge, magnetic and quadrupole FFs. 
The results are presented in Table \ref{tab:FF_numbers_TPE} (using parametrisation II of Abbott et al.), along with values obtained by us based on the Sick and Trautmann parametrisation~\cite{Sick:1998cvq}, as well as the $\chi$ET and \piEFT/ FFs.
One can see that the contributions of both the magnetic and quadrupole FFs to the elastic part of the \2PE-exchange correction can be safely neglected at the current level of precision. While the values of $E_{2S}^\mathrm{elastic}$ obtained in \piEFT/, $\chi$ET, and with the Sick and Trautmann parametrisation of the deuteron FF agree, the parametrisation of Abbott et al.\ gives a significantly smaller value for the elastic contribution. 
The left panel of Fig.~\ref{fig:RF_correlation} shows that this discrepancy is due to the behaviour of the parametrisation of Ref.~\cite{Abbott:2000ak} being very different from the other three calculations (which would all overlap) at low $Q$. 

{\renewcommand{\arraystretch}{1.5}
\begin{table}[t]
    \centering
    \begin{tabular}{c||l|l|l|l}
   \multicolumn{1}{c||}{Deuteron form factor } & \multicolumn{1}{c|}{ $G_C$}&\multicolumn{1}{|c|}{ $G_M$}&\multicolumn{1}{|c|}{ $G_Q$}&\multicolumn{1}{|c}{$E_{2S}^\mathrm{elastic}$}\\
    \hline
    \hline
Abbott et al.~\cite{Abbott:2000ak}& $-0.4153$ & $<10^{-4}$ &$-0.0007$ & $-0.417(2)$ \cite{Carlson:2013xea}\\
Sick and Trautmann~\cite{Sick:1998cvq}& $-0.4503 $&$<10^{-4}$&$-0.0006$ & $-0.4509$\\
$\chi$ET N4LO \cite{Filin:2020tcs}&$-0.4456(18)$&\multicolumn{1}{|c|}{ /}&\multicolumn{1}{|c|}{ /}&$-0.4456(18)$\\
\piEFT/ N3LO &$-0.4463(77)$&\multicolumn{1}{|c|}{$0$} &\multicolumn{1}{|c|}{$0$}&$-0.4463(77)$\\
    \end{tabular}
    \caption{Deuteron form factor contributions to the elastic \2PE exchange. Values are in meV. The magnetic and quadrupole contributions are omitted in the $\chi$ET calculation. In \piEFT/, those contributions first start at N4LO.}
    \label{tab:FF_numbers_TPE}
\end{table}}%

Expanding the integrand in Eq.~\eqref{eq:contrib_elastic} at small $Q$ [neglecting the $G_M(Q^2)$ and $G_Q(Q^2)$ contributions], one obtains
\begin{align}
    \frac{2}{Q}&\left[-\frac{G_C^2(Q^2)-1}{\tau_d}\hat{\gamma}_2(\tau_d,\tau_l)
+16M_d^2\frac{M_d-m}{Q}G_C'(0)\right] \nonumber \\
&=
4 M_d (M_d-m) \left[4 M_d\, G_C''(0)-4 M_d\, G_C'(0)^2+3 G_C'(0)/m\right] +O(Q).
\end{align}
Therefore, the bulk of the difference can be further traced down to the deuteron charge radius and the $4$\textsuperscript{th} moment of the deuteron charge density: $G_C'(0)=-r_d^2/6$ and $G_C''(0)=\big<r^4_d\big>/60$.  Also interesting are two further quantities related to the elastic \2PE-exchange contribution, namely, the cubic and the Friar radii, defined respectively as~\cite{Hagelstein:2015egb}:
\begin{subequations}
\begin{align}
  \big<r^3_d\big>  & = \frac{48}{\pi}\int\limits_0^\infty \frac{\dd Q}{Q^4}\left[G_C(Q^2)-1-G'_C(0)\,Q^2\right], \\
   r_{\mathrm{F}d}^3  & = \frac{48}{\pi}\int\limits_0^\infty \frac{\dd Q}{Q^4}\left[G_C^2(Q^2)-1-2G'_C(0)\,Q^2\right].
\end{align}
\end{subequations}
In \piEFT/ at N3LO, the considered moments have the following analytic expressions, obtained using $G_C(Q^2)$ from Ref.~\cite[Eq.~(75)]{Lensky:2021VVCS} (note again that the integrand in $r_{\mathrm{F}d}^3$ has to be expanded up to N3LO):
\begin{subequations}
\begin{align}
    r_{\mathrm{F}d}^3 & = \frac{3}{80\gamma^3}\left\{
    Z \left[5 - 2 Z (1-2\ln 2)\right]-\frac{320}{9} r_0^2\gamma ^2 \left[Z(1-4\ln 2)-2+2\ln 2\right]
    +
    80 (Z-1)^3\, l_1^{C0_S}\right\}, \\
    \big<r^3_d\big> & = \frac{Z}{32\gamma^3}\left(3+32\,r_0^2\gamma^2\right),\\
     \big<r^4_d\big> & =\frac{Z}{96\gamma^4 }\left(9+80\, r_0^2 \gamma^2\right).
\end{align}
\end{subequations}
Table~\ref{tab:FF_numbers} shows the values of these quantities for the considered FFs.
It is evident that
 parametrisation II of Abbott et al.\ \cite{Abbott:2000ak} gives smaller values for all radii. 
 Smaller $r_d$ and $\left\langle r_d^4\right\rangle$ lead to a significantly smaller value of the integrand at low $Q$, as seen in 
 the left panel of Fig.~\ref{fig:RF_correlation}, and consequently a smaller $E_{2S}^\mathrm{elastic}$ as well as smaller Friar and cubic radii.
Note that, neglecting recoil corrections, the elastic contribution can be approximated through the Friar radius as~\cite{Hagelstein:2015egb}
\begin{align}
    E^\mathrm{elastic,\ F}_{2S}&=-\frac{m_r^4\alpha^5}{24} r_{\mathrm{F}d}^3.
\label{eq:Friar_radius_to_energy}
\end{align}
This approximation, however, results in a noticeable underestimation of $E_{2S}^\mathrm{elastic}$. The \piEFT/ value, for instance, turns out to be $E_{2S}^\mathrm{elastic,\ F} = -0.4323 \text{ meV}$, which has to be compared to \Eqref{elastic_piEFT}. We therefore conclude that at the present level of theoretical precision it is  important to retain the full weighting function $\hat{\gamma}_2(\tau_d,\tau_l)$ in Eq.~\eqref{eq:contrib_elastic} instead of only taking the leading Friar radius term.

 {\renewcommand{\arraystretch}{1.5}
\begin{table}[t]
    \centering
    \begin{tabular}{c||c|c|c|c}
Radii [fm]      & \piEFT/ N3LO & $\chi$ET N4LO~\cite{Filin:2020tcs}  & Sick and Trautmann~\cite{Sick:1998cvq}  & Abbott et al.~\cite{Abbott:2000ak}\\
    \hline
    \hline
     $r_d$                                 & $2.128$ & $2.126$ & $2.130(10)$& $2.094(9)$ \\
     $r_{\mathrm{F}d}$                     & $3.376$ & $3.372$ & $3.385$ & $3.292$\\
     $\big<r^3_d\big>^{\nicefrac{1}{3}}$   & $2.468$ & $2.468$ & $2.480$ & $2.401$ \\
     $\big<r^4_d\big>^{\nicefrac{1}{4}}$   & $2.820$ & $2.837$ & $2.844$ & $2.726$
    \end{tabular}
    \caption{Various radii corresponding to the different deuteron charge form factors.}
    \label{tab:FF_numbers}
\end{table}}%

The dependence of both $r_d^2$ and $r_{\mathrm{F}d}^3$ on $l_1^{C0_S}$ can be represented as a linear correlation between these quantities.
 We show the correlation line in the right panel of Fig.~\ref{fig:RF_correlation}, where we also plot a $\pm 1\%\sim(\gamma/m_\pi)^4$ band as a simple estimate of terms beyond N3LO in the \piEFT/ expansion. One can see that the N4LO $\chi$ET result lies almost on the correlation line, very close to the \piEFT/ results fixed by the H-D $2S-1S$ isotope shift, see Section \ref{sec:isotope_shift} and Appendix \ref{app:iso}. The parametrisation of Ref.~\cite{Sick:1998cvq} lies some distance from the line, albeit reasonably close to it, whereas that of Ref.~\cite{Abbott:2000ak} is much further away. It would be interesting to see if this correlation line can be reproduced in a $\chi$ET calculation.

The above considerations indicate that the FF parametrisation of Ref.~\cite{Abbott:2000ak}, used in Refs.~\cite{Carlson:2013xea,Acharya:2020bxf}, might not adequately describe the behaviour of the deuteron charge FF at low virtualities. The agreement between the N3LO \piEFT/ and N4LO $\chi$ET calculations, see Ref.~\cite[Sec.~IV]{Lensky:2021VVCS} for a detailed comparison of the FFs, is not entirely surprising as both these EFTs are expected to well reproduce low-momenta/long-range properties of the deuteron, and both calculations are of sufficiently high orders in the respective expansions. This vindicates our choice of the \piEFT/ as the calculational framework. One can also conclude that the correlation shown in Fig.~\ref{fig:RF_correlation} can serve as a diagnostic criterion for a realistic parametrisation of the deuteron charge FF. Furthermore, one can note that the \piEFT/ expression for the deuteron charge FF at N3LO, as given in Ref.~\cite[Sec.~IV]{Lensky:2021VVCS}, can serve as an analytic one-parameter fit to the electron-deuteron scattering data in the low-$Q^2$ range that is to be covered in the planned DRad experiment~\cite{Zhou:2020cdt}.

\subsection{Inelastic Contribution}
\label{sec:TPE_inelastic}
The calculation of the longitudinal contribution to  inelastic part of the  \2PE-exchange correction with the known $f_L(\nu,Q^2)$ is straightforward. The only technical complication is that the longitudinal term in the integral of \Eqref{LS_from_LT} goes as $f_L(0,Q^2)/Q^3\propto 1/Q$ for $Q\to 0$ when one sets $x=0$. This singularity, however, is spurious, and can be avoided by subtracting from $f_L(\nu,Q^2)$ its static part:
\begin{align}
    f_L(\nu,Q^2) & = f_L(0,Q^2) + \left[f_L(\nu,Q^2)-f_L(0,Q^2)\right].
\end{align}
The integration over $x$ in the integral of $f_L(0,Q^2)$ can be done analytically, resulting in a $f_L(0,Q^2)/Q^2\propto Q^0$ behaviour for $Q\to 0$. At the same time, the difference in the square brackets is $O(x^2)$ at small $x$ and therefore cancels the singularity in the weighting function.
The longitudinal contribution then results in:
\begin{equation}
E_{2S}^{\mathrm{inel},L} = \left[-1.5032 + 0.6350\, l_1^{C0_S}\right]\ \text{meV}.
\label{eq:TPE_inelastic_ct_separated}
\end{equation}
One can see that the coefficients in front of $l_1^{C0_S}$ here and in Eq.~(\ref{eq:TPE_elastic_ct_separated}) partially cancel each other. The resulting contribution of the N3LO contact term to the \2PE exchange in $\mu$D is rather small. The numerical order-by-order result for $E_{2S}^{\mathrm{inel},L}$, using $l_1^{C0_S}$ as obtained from the H-D isotope shift,
is:
\begin{equation}
E_{2S}^{\mathrm{inel},L}  = [-0.943 - 0.635 + 0.049 + 0.025]~\mathrm{meV} = - 1.504(16) \ \text{meV}.
\label{eq:TPE_inelastic_orders}
\end{equation}
The uncertainty here is due to higher-order terms in the \piEFT/ expansion, calculated as explained in Appendix~\ref{sec:error}.
The individual terms of the inelastic contribution are shown in Table~\ref{tab:TPE_contributionsElInel}, in an analogy to what is shown for the elastic part.
While the bulk of $E_{2S}^{\mathrm{inel},L}$ is given by the LO part of $\mathcal{M}_L$, the most important correction comes from the nucleon charge radii, with the second-biggest correction driven by the NLO term of $\mathcal{M}_L$. The remaining mechanisms all give much smaller contributions.

The above results, Eqs.~\eqref{eq:TPE_inelastic_ct_separated}, \eqref{eq:TPE_inelastic_orders}, and Table~\ref{tab:TPE_contributionsElInel}, are obtained with the substitution $|\bv{q}|\to Q$ in the expressions for $\mathcal{M}_L$; using $|\bv{q}|=\sqrt{Q^2+\nu^2}$ brings the total value to $-1.507$~meV. This gives an estimate of the relativistic corrections at N4LO. The smallness of the effect corroborates the choice of  counting scheme in our calculation, namely, that the energy transfer is suppressed and $\nu/Q=O(P)$. This statement can be made more quantitative by observing that shrinking the $x$ integration interval in Eq.~\eqref{eq:LS_from_LT} to $x\in[-\gamma/m_\pi,\gamma/m_\pi]\simeq[-\nicefrac{1}{3},\nicefrac{1}{3}]$ retains $\sim 96\%$ of the LO$+$NLO contribution. Furthermore, the transverse contribution to $E_{2S}^\mathrm{inel}$, calculated at NLO for $f_T(\nu,Q^2)$, is small in accordance with the prediction of the \piEFT/ counting:
\begin{align}
E_{2S}^{\mathrm{inel},T}=-0.005\text{ meV}.
\end{align}
It is also in a very good agreement with the existing dispersive $\chi$ET-based evaluations~\cite{Hernandez:2019zcm,Acharya:2020bxf}. Despite the smallness of the transverse contribution, we add it to the total inelastic contribution, since it is included in most of the alternative calculations, thus having
\begin{equation}
    E_{2S}^\mathrm{inel}=E_{2S}^{\mathrm{inel},L}+E_{2S}^{\mathrm{inel},T}=-1.509(16)\,\mathrm{meV}.
\end{equation}
The uncertainty of the transverse contribution is neglected.

Based on the observations above, we conclude that the \piEFT/ counting used by us works well for the present calculation. We also do not expect any higher-order corrections that would change the pattern that one sees at N3LO; a quantification of this statement follows through the Bayesian procedure in Appendix~\ref{sec:error}. 

In Table~\ref{tab:results_comparison_parts}, we compare our $E_{2S}^\mathrm{inel}$ result with other recent evaluations. Our result agrees with the recent covariant dispersive calculation~\cite{Acharya:2020bxf} as well as with the value quoted in Ref.~\cite{Hernandez:2019zcm} within the uncertainties. The latter has a slightly larger in magnitude central value. These two results obtain the deuteron response functions at N3LO in the $\chi$ET expansion to calculate $E_{2S}^\mathrm{inel}$ from a dispersive integral. The data-driven evaluation of Carlson et al.~\cite{Carlson:2013xea} also uses a dispersive approach, but extracts information on the deuteron response functions from experimental data. It calculates an even larger $E_{2S}^\mathrm{inel}$ with a large uncertainty that makes it compatible with all other results. 
In addition, we compare the results in the point-nucleon limit, where the contributions from the nucleon charge radii are removed (in which case we also omit the contribution of $l_1^{C0_S}$). Our result here is compatible with the earlier N3LO $\chi$ET result~\cite{Hernandez:2019zcm}, as well as with that obtained from the N2LO \piEFT/  deuteron longitudinal response function in the point-nucleon limit~\cite{Emmons:2020aov}. 

{\renewcommand{\arraystretch}{1.5}
\begin{table}[t]
    \centering
    \begin{tabular}{l||c|c|c|c|c}
    &\ \piEFT/ N3LO \
    & Acharya et al.~\cite{Acharya:2020bxf}
    & Hernandez et al.~\cite{Hernandez:2019zcm}
    & Emmons et al.~\cite{Emmons:2020aov}
    & Carlson et al.~\cite{Carlson:2013xea}
    \\
    \hline\hline
    $E_{2S}^\mathrm{inel}$
    & $-1.509(16)$
    & $-1.511(12)$
    & $-1.531(12)$
    &
    & $-1.566(740)$
    \\
    \hline
    $E_{2S}^\mathrm{inel,\ p.N.}$
    &$-1.567$
    & 
    & $-1.571$
    & $-1.574(80)$
    &
    \end{tabular}
    \caption{Comparison of our results with other recent calculations for the inelastic contribution $E_{2S}^\mathrm{inel}$ and the inelastic contribution in the point-nucleon limit $E_{2S}^\mathrm{inel,\ p.N.}$. The latter is the inelastic contribution with point-like nucleons (calculated up to N3LO, with the contribution of $l_1^{C0_S}$ omitted).  Values are in meV. To compare with Ref.~\cite{Hernandez:2019zcm}, we subtract the subleading $O(\alpha^6\log\alpha)$ Coulomb correction from their ``$\eta$-less'' result. The uncertainty given here for their prediction is obtained using the relative uncertainties of individual error sources from Ref.~\cite[Table 8]{Ji:2018ozm} (nuclear model, isospin symmetry breaking,
relativistic, higher $\mathcal{Z}\al$) summed in quadrature.
    The value quoted for Ref.~\cite{Emmons:2020aov} is their ``$\mathcal{Z}_d$-improved'' result.
      }
    \label{tab:results_comparison_parts}
\end{table}}

\subsection{Single-Nucleon Effects Beyond N3LO}
\label{sec:nucleon_effects}

The results of Secs.~\ref{sec:TPE_elastic} and \ref{sec:TPE_inelastic} show that single-nucleon contributions generated by the hadron structure, such as the nucleon FFs, are the most important corrections beyond the LO and NLO nuclear-structure effects. They are also potentially the most problematic, since they tend to be enhanced by factors of $\bv{q}^2$ compared with the corresponding amplitude with point-like nucleons. For instance, an N4LO correction with two insertions of the nucleon charge radius operator in the LO $\mathcal{M}_L$ diagrams, shown in Ref.~\cite[Fig.~7]{Lensky:2021VVCS}, would be enhanced by a factor of $\bv{q}^4$ and would lead to a contribution to $E_{2S}$ that is divergent at large $Q$. Another potentially sizeable single-nucleon effect, first appearing also at N4LO, is that of the nucleon polarizabilities. Their inclusion into the deuteron VVCS amplitude also leads to a similar divergent contribution. A \piEFT/ consideration would therefore introduce four-nucleon and two-lepton contact terms at N4LO to regularise the divergence generated by the single-nucleon terms. Those contact terms, as pointed out in Sec.~\ref{sec:pEFT}, limit the predictive powers of \piEFT/ in the study of the \2PE-exchange corrections to N3LO. In this section, we quantify these hadron structure effects, expected to be the most important ones beyond N3LO, using alternative methods that go beyond the \piEFT/ expansion.

Starting from the nucleon FF, one alternative that can improve the bad behaviour of the nucleon FF correction would be to insert the full nucleon FFs in the nucleon charge operator vertex, replacing its LO term according to:
\begin{align}
\frac{1}{2}(1+\tau_3)\to \frac{\hat{G}^N_E(Q^2)}{\sqrt{1+\frac{Q^2}{4M_p^2}}},
\label{eq:nucleonFFs}
\end{align}
where $\hat{G}^N_E(Q^2)=G_E^{0}(Q^2)+G_E^1(Q^2)\,\tau_3$ with $G_E^{0,1}$ being the isoscalar and isovector nucleon electric FFs, $G_E^{0,1}(Q^2)=\left[G_E^p(Q^2)\pm G_E^n(Q^2)\right]/2$.  This procedure obviously represents a departure from the strict \piEFT/ treatment. It provides, however, a viable workaround and allows one to estimate the effects generated by the higher-order terms in the expansion of the nucleon FFs.
It also is routinely used in $\chi$ET calculations of electromagnetic processes in nuclei, since the nucleon FFs do not converge well in a chiral expansion, either, see Refs.~\cite{Acharya:2020bxf,Filin:2020tcs} for recent examples. The specific substitution of Eq.~\eqref{eq:nucleonFFs}, strictly speaking, breaks the electromagnetic gauge invariance. The violating terms are, however, of higher orders than we consider. The modified VVCS amplitudes can be found in Appendix \ref{app:ff}.

The N3LO \piEFT/ prediction for the deuteron charge radius, given in \Eqref{rdl1}, does not change with \Eqref{nucleonFFs}, as long as we make sure that the parametrisation of the isoscalar nucleon FF agrees with our choice of $r_0^2$. We chose the nucleon FF parametrisations from Borah et al.~\cite{Borah:2020gte}, since their slopes are constrained by the nucleon radii used by us: the proton charge radius from $\mu$H spectroscopy given in \Eqref{rmuh}, and the neutron charge radius \cite{Kopecky:1995zz,Kopecky:1997rw}:
\beq
r_n^2=-0.1161(22)\text{ fm}^2.\eqlab{rnOLD}
\eeq
The elastic \2PE-exchange correction resulting from inserting the full nucleon FFs can be calculated using Eq.~\eqref{eq:contrib_elastic} with the re-summed deuteron FFs given in Ref.~\cite[Eqs.~(77) and (78)]{Lensky:2021VVCS}, and it differs only marginally from the result in Eq.~\eqref{eq:elastic_piEFT} (specifically, by about $-0.001$~meV); we neglect this difference.

The inelastic part changes more significantly. Introducing the nucleon FFs results in the following modifications to the LO and NLO contributions to $E_{2S}^{\mathrm{inel},L}$ in Table~\ref{tab:TPE_contributionsElInel}, using the nucleon FF parametrisation of Ref.~\cite{Borah:2020gte}:
\begin{align}
    E^{(-3)}\to  E^{(-3)}_\mathrm{FF} =-0.9156~\text{meV}, \qquad 
    E ^{(-2)}\to E^{(-2)}_\mathrm{FF} = 0.0125~\text{meV}, \qquad 
\end{align}
which at the same time absorbs both $E ^{(-1)}_{r_N^2}$ and $E ^{(0)}_{r_N^2}$.  Using a different nucleon FF parametrisation~\cite{Bradford:2006yz} results in:
\begin{align}
    E ^{(-3)}\to E ^{(-3)}_\mathrm{FF} =-0.9151~\text{meV}, \qquad 
    E ^{(-2)}\to  E^{(-2)}_\mathrm{FF} = 0.0125~\text{meV}. \qquad 
\end{align}
This amounts to a correction of $E _{2S}^\mathrm{hadr,\ FF}=-0.0129$~meV
with the nucleon FFs from Ref.~\cite{Borah:2020gte}; the parametrisation of Ref.~\cite{Bradford:2006yz} gives
$E _{2S}^\mathrm{hadr,\ FF}=-0.0121$~meV.
In the following, we will adopt 
\begin{align}
E _{2S}^\mathrm{hadr,\ FF} = -0.013(1)~\mathrm{meV}. \eqlab{HadrFF}
\end{align}
This effect is within our N3LO uncertainty estimate; one can also notice that it is significantly larger than a similar difference obtained in a $\chi$ET calculation replacing linearised (expanded in $Q^2$) nucleon FFs by a realistic parametrisation~\cite{Acharya:2020bxf}. At the same time, the difference due to the different nucleon FF parametrisations is negligibly small. The replacement of the charge operator by the nucleon FFs in the contributions to $E_{2S}^{\mathrm{inel},L}$ beyond NLO would also give a negligible effect on the total result.

{\renewcommand{\arraystretch}{1.5}
\begin{table}[t]
    \centering
      \begin{tabular}{l||c|c|c}
   
   &$\mu\text{H}$&$\mu n$&$\mu \text{D}$\\
   \hline\hline
 $\chi$PT \cite{Alarcon:2013cba,Lensky:2017bwi}&$0.0035(26)$&$0.0043(25)$&$0.0091(60)$\\
 data-driven  &$0.0023(13)$ \cite{Tomalak:2018uhr}&$0.0043(20)$ \cite{Tomalak:2018uhr}&$0.0078(37)$
    \end{tabular}
    \caption{Single-nucleon subtraction-function contributions from \2PE exchange between muon and proton ($\mu$H) or neutron ($\mu n$), respectively. The last column gives the $E _{2S}^\mathrm{hadr,\ subt}$ contribution to $\mu$D, obtained by rescaling the muon-nucleon values. Values are in meV.
    }
    \label{tab:nucleon_contr}
\end{table}}

Coming to the other effect we consider here, that of the nucleon polarizabilities, it consists of two parts, the inelastic and the subtraction hadronic corrections. The first one of the two can be calculated from a dispersive relation, using the empirical deuteron structure functions, at energies starting from the pion production threshold, as done in Ref.~\cite{Carlson:2013xea}:
\begin{align}
E _{2S}^\mathrm{hadr,\ inel}=-0.028(2)~\mathrm{meV}.\eqlab{HadrInel}
\end{align}
See also Ref.~\cite{Eskin:2015brf} for a similar evaluation.
Another method to calculate it, similar to inserting the nucleon FFs in the consideration above, is to apply a rescaling procedure to the inelastic \2PE-exchange effect in $\mu$H and the analogous inelastic \2PE-exchange effect between a muon and a neutron ($\mu$n), adding them together and correcting for the different atomic wave functions, rescaling the sum by the factor $[\phi_{2S}^{\mu\mathrm{D}}(0)/\phi_{2S}^{\mu\mathrm{H}}(0)]^2$, as done in Ref.~\cite{Krauth:2015nja} and references therein. Using the single-nucleon values from Ref.~\cite{Tomalak:2018uhr}, we obtain a value of
$-0.030(2)$~meV, where we added uncertainties linearly to be conservative. This perfectly agrees with the dispersive evaluation given by Eq.~\eqref{eq:HadrInel}, which is an indication that the rescaling procedure works well in this setting.

The second part of the single-nucleon polarizability effect, the subtraction contribution, cannot be directly related to empirical data. It has to be either modelled or predicted from baryon chiral perturbation theory ($\chi$PT). With the rescaling procedure described above, and the 
covariant $\chi$PT results for the proton VVCS subtraction function \cite{Lensky:2017bwi} and its neutron counterpart, we obtain for the subtraction-function contribution to $\mu$D:\footnote{Note that the leading pion-nucleon loop contribution to the subtraction-function correction was previously calculated with an approximate formula \cite[Eq.~(17a)]{Alarcon:2013cba} and has been updated here.}
\begin{align}
E _{2S}^\mathrm{hadr,\ subt}=0.009(6)~\mathrm{meV},\eqlab{HadrSubt}
\end{align}
which agrees well with the value adopted in Ref.~\cite{Krauth:2015nja}: $0.0098(98)$~meV.
As one can see from Table~\ref{tab:nucleon_contr}, our predictions agree with the dispersive estimates from Ref.~\cite{Tomalak:2018uhr}. It is also instructive to compare our result for the proton subtraction contribution, $0.0035(26)$~meV, to predictions in the framework of heavy-baryon $\chi$PT: $0.0042(10)$~meV \cite{Birse:2012eb} and $0.0029(12)$~meV \cite{Peset:2014jxa}.

The above considerations take into account the most significant higher-order nucleon structure corrections that start to appear at N4LO in the \piEFT/ expansion. One can notice that each one of the corrections,
$E _{2S}^\mathrm{hadr,\ FF}=-0.013(1)$~meV  from \Eqref{HadrFF}, and the nucleon polarizability corrections, $E _{2S}^\mathrm{hadr,\ subt}+E _{2S}^\mathrm{hadr,\ inel}=-0.019(6)$~meV from Eqs.~\eref{HadrInel} and \eref{HadrSubt}, is separately smaller or of the size of the estimated N3LO uncertainty of the inelastic contribution, $0.016$~meV, Eq.~\eqref{eq:TPE_inelastic_ct_separated}. Their total, however,
\begin{equation}
    E _{2S}^\mathrm{hadr} = E _{2S}^\mathrm{hadr,\ FF} + E _{2S}^\mathrm{hadr,\ subt} + E _{2S}^\mathrm{hadr,\ inel} =- 0.032(6)~\mathrm{meV},
    \label{eq:higher_order_hadron_correction}
\end{equation}
is about twice as large as that uncertainty. Nevertheless, we expect the higher-order nuclear effects, as well as the relativistic corrections, to be much smaller, and we expect the remaining higher-order effects to be within our N3LO uncertainty estimate. Erring on the side of caution, we refrain from going as far as performing an N4LO adjustment of the uncertainty.

\subsection{Summary of Results} \seclab{SummaryTPEmuD}

\begin{figure}[t]
    \centering
    \includegraphics[width=0.95\textwidth]{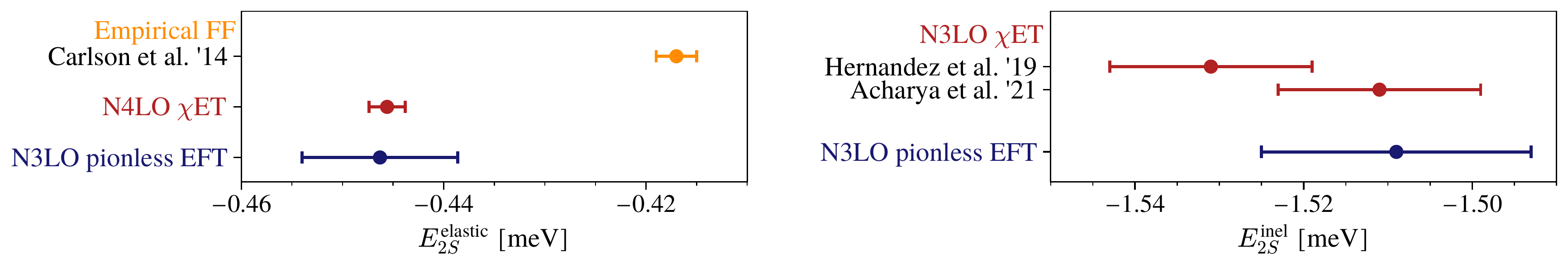}
    \caption{Comparison of predictions for the elastic and inelastic contributions to the \2PE exchange in $\mu$D. Values are the same as in Tables \ref{tab:FF_numbers_TPE} and \ref{tab:results_comparison_parts}.}
    \label{fig:TPELitComp}
\end{figure}

We conclude this section by summarizing our \piEFT/ predictions of the nuclear-structure effects on the $2S$ level in $\mu$D from the forward \2PE exchange, and including the accompanying electronic VP contributions. At N3LO, we derived the dominant \2PE-exchange effects coming from the elastic deuteron charge FF $G_C$ and the non-pole part of the  deuteron VVCS amplitude:
\begin{subequations}
\bea
    E _{2S}^\mathrm{elastic} &=& -0.446(8) \text{ meV},\\
E _{2S}^{\mathrm{inel}}  &=&  - 1.509(16) \ \text{meV},
\eea
\end{subequations}
see Sections \ref{sec:TPE_elastic} and \ref{sec:TPE_inelastic} for details.
The uncertainties have been quantified through the Bayesian error estimate described in Appendix \ref{sec:error}.
As mentioned above, the value of $E_{2S}^\mathrm{inel}$ contains the transverse contribution.

In Fig.~\ref{fig:TPELitComp}, our \piEFT/ predictions are compared to data-driven and $\chi$ET results. The disagreement with Carlson et al.~\cite{Carlson:2013xea} for $E _{2S}^\mathrm{elastic}$ is due to the deuteron charge FF parametrisation from Ref.~\cite{Abbott:2000ak}. As one can see from  Table \ref{tab:FF_numbers_TPE}, our prediction is in good agreement with the data-driven approach if the Sick \& Trautmann parametrisation \cite{Sick:1998cvq} is used instead.

Beyond N3LO, we also take into account the single-nucleon effects discussed in \secref{nucleon_effects}. They can be split into the nucleon-polarizability contribution, the single-nucleon subtraction-function contribution, and the insertion of the nucleon FFs in the nucleon charge operator vertex of \piEFT/. In total, they amount to:
\begin{equation}
    E _{2S}^\mathrm{hadr}  =- 0.032(6)~\mathrm{meV}.
\end{equation}

On top of the above forward \2PE-exchange effects,
\begin{equation}
    E _{2S}^\mathrm{fwd}=E_{2S}^\mathrm{elastic}+E_{2S}^\mathrm{inel}+E_{2S}^\mathrm{hadr}  =-1.987(20)~\mathrm{meV},\eqlab{TPEforVP}
\end{equation}
there are the electronic VP corrections to the \2PE-exchange, described in Appendix \ref{sec:TPE_eVP}:
\beq
    E _{2S}^\mathrm{eVP} =-0.027\text{ meV}
\eeq
(their uncertainty also being negligibly small).
In total this adds up to:
\begin{align}
    E _{2S}^{\text{fwd+eVP}} =E _{2S}^\mathrm{fwd} + E _{2S}^\mathrm{eVP}= -2.014(20)\text{ meV}\, .\eqlab{TPE+}
\end{align}
 In~\secref{Total_DeltaE}, we will discuss all the relevant deuteron-structure effects, including also the Coulomb distortion from the off-forward \2PE exchange \cite{Krauth:2015nja} and the $3\gamma$-exchange effect \cite{Pachucki:2018yxe}.

\section{Hydrogen-Deuterium  Isotope Shift}\label{sec:isotope_shift}

In this section, we will use the isotope shift between $1S$ and $2S$ states in H and D:
\beq
E _\text{iso}=h \, f_\text{iso}=E_{2S-1S}^\text{D}-E_{2S-1S}^\text{H}\,,
\eeq
where $h$ is the Planck constant,
to get a prediction for the deuteron charge radius, cf.\ \Eqref{Rdisotope}, and, in turn, determine the LEC $l_1^{C0_S}$ as given by~\Eqref{contact_term_value2}.
The empirically measured value of the isotope shift is very precise~\cite{Jentschura:2011is}, 
\beq
f_\mathrm{iso}=670\,994\,334.605(15)\,\mathrm{kHz}\,.\eqlab{fexpt}
\eeq
To extract from it $r_d$ and $l_1^{C0_S}$, we will  update the theoretical prediction for the isotope shift. Our notation generally follows the work of Jentschura et al.~\cite{Jentschura:2011is}. It is, along with most of the features of the consideration in this section, such as a list of all contributions relevant to the isotope shift, presented in Appendix~\ref{app:iso}. Here, we focus on our \piEFT/ result for the \2PE-exchange correction to the $S$-levels in D. The pertinent calculation proceeds analogously to Section~\ref{sec:TPE_corrections}, where the \2PE exchange in $\mu$D is evaluated, hence its details are largely omitted.

\subsection{\2PE Exchange in Deuterium}

The longitudinal part of the inelastic contribution to the $2S-1S$ shift in D is:
\begin{subequations}
\bea
    \nu_{9,L}^\mathrm{D} & =&\left[16.612 -0.769\, l_1^{C0_S}\right]\mathrm{kHz}\\
    &=& \left[9.929+6.825-0.062-0.078\right]\ \mathrm{kHz} =16.613(191)\ \mathrm{kHz}.
\eea
\end{subequations}
In the second line, we show our numerical order-by-order result with the LEC $l_1^{C0_S}$ determined in the following Section~\ref{sec:l1det}.
Note that all forward \2PE-exchange contributions scale through the atomic wave function at the origin as $\nicefrac{1}{n^3}$. Thus, to deduce the shift of the $n^\mathrm{th}$ $S$-level in D, one simply has to multiply the isotope shift value by $-\nicefrac{8}{7n^3}$. The uncertainty of our result is obtained in a simplified way by multiplying the total by $(\gamma/m_\pi)^4$. This is justified by the smallness of the N2LO and N3LO contributions (with the NLO contribution given by $(Z-1)$ times the LO result plus a small correction, cf.\ Table~\ref{tab:TPE_contributionsElInel} for the case of $\mu$D).

The transverse \2PE-exchange contribution appears to be relatively more important in D than in $\mu$D:
\begin{align}
    \Delta\nu_{9,T}^\mathrm{D} & = 1.859(65)\ \mathrm{kHz}.
\end{align}
The uncertainty is obtained here by multiplying the total with $(\gamma/m_\pi)^3$, where the usual NLO factor of $(\gamma/m_\pi)^2$ is multiplied with another $\gamma/m_\pi$ to take into account that the transverse amplitude is well reproduced already at NLO~\cite[Sec.~V]{Lensky:2021VVCS}. The full N3LO \piEFT/ prediction for the inelastic contribution to the forward \2PE exchange is then given by
\begin{align}
    \Delta \nu_{9,L+T}^\mathrm{D} = 18.472(202)\, \mathrm{kHz}.
\end{align}

The hadronic contributions to the shift of levels in D are as follows. Inserting the nucleon FFs as in Eq.~\eqref{eq:nucleonFFs} leads to a negligible shift:
\begin{align}
    \Delta\nu_{9, \mathrm{\ hadr.\ FF}}^\mathrm{D}=0.014(1)\ \mathrm{kHz}.
\end{align}
The inelastic part, calculated in the same way as done for $\mu$D~\cite{Carlson:2013xea}, gives~\cite{Gorchtein:2020prv}
\begin{align}
    \Delta\nu_{9\mathrm{,\ hadr.\ inel}}^\mathrm{D}=0.148(11)\ \mathrm{kHz}.
\end{align}
This is in perfect agreement with the result from rescaling the single-nucleon values obtained in Ref.~\cite{Tomalak:2018uhr}:
$\Delta\nu_{9\mathrm{,\ hadr.\ inel}}^\mathrm{D}=0.145(12)$~kHz. 
The subtraction part is calculated by us in the same way as done for $\mu$D by rescaling the single-nucleon values from $\chi$PT:
\begin{align}
    \Delta\nu_{9\mathrm{,\ hadr.\ subt}}^\mathrm{D}= -0.069(29)\ \mathrm{kHz}. \eqlab{hadrsubtD}
\end{align}
The subtraction function contributions found in Ref.~\cite{Tomalak:2018uhr} tend to be smaller, cf.\ Table \ref{tab:nucleon_contr_eD}.

{\renewcommand{\arraystretch}{1.5}
\begin{table}[t]
    \centering
      \begin{tabular}{l||l|l|l}
      &$\text{H}$&$e n$&$\text{D}$\\
   \hline
      \hline
 $\chi$PT \cite{Alarcon:2013cba,Lensky:2017bwi}&$-0.032(15)$&$-0.037(14)$&$-0.069(29)$\\
 data-driven  &$-0.016(4)$ \cite{Tomalak:2018uhr}&$-0.025(9)$ \cite{Tomalak:2018uhr}&$-0.041(13)$
    \end{tabular}
    \caption{Single-nucleon subtraction-function contributions from \2PE exchange between electron and proton (H) or neutron ($e n$), respectively. The last column gives the $\Delta\nu_{9, \mathrm{\ hadr.\ subt}}$ contribution to D, obtained by rescaling the electron-nucleon values. Values are in kHz.
    }
    \label{tab:nucleon_contr_eD}
\end{table}}

The off-forward \2PE-exchange correction, known as the Coulomb distortion, can be estimated by rescaling the results for $\mu$D presented in Ref.~\cite{Pachucki:2011xr}. For the $2S-1S$ shift in D, we obtain a very small result of $\simeq -0.034$~kHz. Adding all contributions together, we find
\begin{align}
    \Delta \nu_{9}^\mathrm{D} 
    = \left[18.530 - 0.769\,l_1^{C0_S} \right]~\mathrm{kHz} 
    = 18.531(204)~\mathrm{kHz}  . \eqlab{nu9D}
\end{align}
This has to be compared to $\Delta \nu_{9}^\mathrm{D} = 18.70(7)~\mathrm{kHz}$  used in Ref.~\cite{Mohr:2008fa} and based on Ref.~\cite{Friar:1997tr}.

The N3LO \piEFT/ prediction for the elastic contribution to the $2S-1S$ shift in D is
\begin{subequations}
\eqlab{nubD}
\bea
    \Delta\nu_{(b)}^\mathrm{D}&=&\left[0.625+1.607\,l_1^{C0_S}\right]\mathrm{kHz}\\
  &=&[0.292+0.221+0.087+0.023]~\mathrm{kHz}=0.622(7)\ \mathrm{kHz},
\eea
\end{subequations}
where the uncertainty is estimated as above for $\Delta\nu_{9,L}^\mathrm{D}$. This is slightly bigger than the pure Friar-radius contribution appearing in Ref.~\cite{Mohr:2008fa}, which gives $\Delta \nu_{(b)}^\mathrm{D}=0.507$~kHz. 

Adding all \2PE-exchange corrections to the $2S-1S$ transition in D together, we find 
\begin{align}
    \Delta \nu_{9+(b)}^\mathrm{D} = \left[19.155+0.838\,l_1^{C0_S}\right]~\mathrm{kHz}= 19.153(204)\mathrm{kHz}  .
\end{align}
One can see that the elastic and inelastic contributions proportional to $l_1^{C0_S}$ partially cancel each other, 
making the total slightly less sensitive to the value of the N3LO contact term, similarly to what happens in $\mu$D. In any case, the effect of it (assuming the maximal magnitude of $l_1^{C0_S}\simeq 10^{-2}$) is at most $0.01\,\mathrm{kHz}$, which is far smaller than the total uncertainty of the isotope shift. Therefore, the contribution of $l_1^{C0_S}$ to the \2PE exchange in the isotope shift can be safely neglected (at the current level of precision), and the deuteron charge radius extracted from the isotope shift is a good quantity to determine $l_1^{C0_S}$.

\subsection{\2PE Exchange in Isotope Shift and Determination of Low-Energy Constant $l_1^{C0_S}$}\label{sec:l1det}

For the isotope shift, we also need the \2PE-exchange correction to the $2S-1S$ transition in H. For the elastic contribution, we use the results from Ref.~\cite{Tomalak:2018uhr}:
\beq
\Delta \nu_{(b)}^\mathrm{H} = 0.035(6)~\mathrm{kHz},\eqlab{nubH}
\eeq
which is in perfect agreement with the Friar-radius contribution $\Delta \nu_{(b)}^\mathrm{H} =0.035 ~\mathrm{kHz}$ appearing in Ref.~\cite{Mohr:2008fa}. For the inelastic contribution, it is important that $\Delta \nu_{9}^\mathrm{H}$ is consistent with the single-proton contributions entering D through $\Delta\nu_{9\mathrm{,\ hadr.\ inel}}$ and $\Delta\nu_{9\mathrm{,\ hadr.\ subt}}$. Therefore, we will use the subtraction-function contribution  predicted by $\chi$PT, see Table \ref{tab:nucleon_contr_eD}, and the inelastic contributions from Ref.~\cite{Tomalak:2018uhr}:
\beq
    \Delta \nu_{9}^\mathrm{H} = \left[-0.032(15)+0.073(5)\right]~\mathrm{kHz} =  0.041(16)\,\mathrm{kHz}\eqlab{nu9H}.
\eeq
 This compares to $\Delta \nu_{9}^\mathrm{H} = 0.061(11)~\mathrm{kHz}$ used in Ref.~\cite{Mohr:2008fa} and based on Ref.~\cite{Khriplovich:1999ak}. Using instead the subtraction-function contribution from Ref.~\cite{Tomalak:2018uhr}, we would find:
\beq
\Delta \nu_{9}^\mathrm{H}=0.057(6)\,\mathrm{kHz}.\eqlab{nu9Tomalak}
\eeq
In total, the \2PE-exchange correction to the $2S-1S$ transition in H amounts to
\beq
    \Delta \nu_{9+(b)}^\mathrm{H} =   0.076(17)\,\mathrm{kHz}.
\eeq
The combined results for the isotope shift are given in Eqs.~\eref{isonu9} and \eref{isonub} of Appendix \ref{app:iso}. 

In the appendix, we give a full updated list of all contributions entering the isotope shift, together with a comparison to the values used in Ref.~\cite{Jentschura:2011is}. Besides theoretical updates, e.g., of the VP and recoil contributions, we discuss the impact of refined values for the electron, proton and deuteron masses, and the role of the Rydberg constant. Our final result for the theoretical prediction of the $2S-1S$ deuterium-hydrogen isotope shift reads:
\beq
f_\mathrm{iso}=\left[671\,000\,534.811(924)+ 0.838 \,l_1^{C0_S}-1369.346\, \left(\frac{r_d}{\mathrm{fm}}\right)^2\right]~ \mathrm{kHz}.\eqlab{theoryPredictionISO}
\eeq
Note that, in the calculation of the \2PE-exchange corrections, we used the value of the proton charge radius $r_p(\mu\text{H})$ published by the CREMA Collaboration, \Eqref{rmuh}. This value is consistent with the nucleon FF parametrisations from Ref.~\cite{Borah:2020gte}, used in \secref{nucleon_effects} to estimate the single-nucleon effects beyond N3LO in \piEFT/. The proton finite-size corrections to the isotope shift use instead a refined value \cite{Antognini:2022xoo}, extracted from the Lamb shift measurement of the CREMA Collaboration \cite{Pohl:2010zza,Antognini:1900ns} accounting for the recent updates of the $\mu$H theory \cite{Korzinin:2013uia,Karshenboim:2015,Karshenboim:2018iyl,Karshenboim:2021jsc}:
\beq
r_p(\mu\text{H})=0.84099(36)\,\mathrm{fm}.\eqlab{rpreview}
\eeq
The effect of the updated $r_p$ value on the \2PE-exchange corrections would be negligibly small compared to the estimated theoretical uncertainties of the latter.

The LEC $l_1^{C0_S}$ is small (again, a reasonable estimate of its maximal magnitude being $\simeq 10^{-2}$). It is therefore justified to use the N3LO \piEFT/ prediction for the deuteron radius, given in \Eqref{rdl1}, as an exact relation to express $l_1^{C0_S}$ in \Eqref{theoryPredictionISO} through $r_d$. 
We can then extract $r_d$ by comparing our theory prediction and the experimental value for the isotope shift \eref{fexpt}:
\beq
r_d(\mu\text{H \& iso})=2.12788(16)\ \text{fm},\label{eq:OURRdisotope}\\
\eeq
where the error is completely dominated by the theory. 
Our result for $r_d$ is in perfect agreement with the previous extraction in \Eqref{Rdisotope}.
Setting $l_1^{C0_S}=0$ in \Eqref{theoryPredictionISO} leads to the same result, which proves that the error generated by applying \Eqref{rdl1} as an exact relation can indeed be safely neglected. A comparison and consistency check of state-of-the-art deuteron charge radius extractions from $\mu$D, D and the H-D isotope shift can be found in Section \ref{sec:radiid}. From \Eqref{rdl1}, we then find:
\begin{align}
    l_1^{C0_S}=-1.80(4)(36)(12)\times 10^{-3},\eqlab{l1iso}
\end{align}
where the uncertainties in the brackets stem from our extracted value of the deuteron radius, the uncertainty of $Z=1.6893(30)$~\cite{Epelbaum:2019kcf}, and the isoscalar nucleon charge radius $r_0=0.5586(10)$ fm, respectively.

\section{Muonic Deuterium Lamb Shift}
\label{sec:Total_DeltaE}

In this section, we will extract an empirical value for the \2PE-exchange effects in the $\mu$D Lamb shift from the high-precision Lamb shift measurement by the CREMA Collaboration \cite{Pohl1:2016xoo} and the deuteron radius determined from the H-D isotope shift. The empirical value will be compared to our \piEFT/ prediction. A theory compilation for the $\mu$D spectrum, including a review of recent theoretical predictions for the \2PE-exchange effects, can be found in Ref.~\cite{Krauth:2015nja}. At the end of this section, we will present an updated theory prediction of the $\mu$D Lamb shift, based on our \piEFT/ prediction for the \2PE exchange, taking into account all recent theory improvements since the publication of Ref.~\cite{Krauth:2015nja}.

\subsection{Empirical \2PE Exchange}

The theory prediction for the $\mu$D Lamb shift reads \cite[Eq.~(18)]{Krauth:2015nja}:
\beq
E_{2P-2S}=\left[228.77356(75)+0.00310(60)-6.11025(28) \,\left(\frac{r_d}{\mathrm{fm}}\right)^2-E _{2S}^{2\ga}\right]\mathrm{meV}.\eqlab{Krauthteory}
\eeq
Here, the first term is deuteron-radius independent, the next two terms are deuteron-radius dependent, and the last term contains deuteron-structure effects from \2PE exchange.
Note that the prefactor in front of the radius-dependent finite-size term also contains radiative corrections, such as the electronic VP corrections partially discussed in Appendix~\ref{sec:TPE_eVP}, see Ref.~\cite{Krauth:2015nja} for details.\footnote{Note that in the final sum of radius-dependent terms in Ref.~\cite[Table 2]{Krauth:2015nja} the entry \#r8 has been included with a wrong sign.}
The empirical value measured by the CREMA collaboration is:
\beq
E_{2P-2S}=202.8785(31)_\mathrm{stat}(14)_\mathrm{syst}\, \mathrm{meV}.\eqlab{muDexp}
\eeq
With the theory prediction for the Lamb shift in \Eqref{Krauthteory}, the empirical value in \Eqref{muDexp}, and $r_d(\mu\text{H \& iso})$ from \Eqref{Rdisotope},
one obtains an empirical value for the \2PE-exchange effects in the $\mu$D Lamb shift~\cite{Pohl1:2016xoo}:
\begin{equation}
E _{2S}^{2\gamma}(\mathrm{emp.})=-1.7638(68)\,\mathrm{meV}\,.
\end{equation}
In the following, we update this value based on the improved hadronic VP \cite{Karshenboim:2021jsc} and electronic light-by-light scattering contributions \cite{Korzinin:2018tnx}, as well as $r_d(\mu\text{H \& iso})$ from \Eqref{OURRdisotope}. 

For the effect of LO and NLO hadronic VP \cite{Karshenboim:2021jsc}, combined with the mixed electronic and muonic VP, as well as the electronic VP loop in the SE correction \cite{Korzinin:2013uia}, we use ($2P-2S$): $11.64(32)\,\upmu\mathrm{eV}$. This reduces the uncertainty of the old value $11.12(71)\,\upmu\mathrm{eV}$ (sum of \#\#12, 13, 14, 30, 31 in Ref.~\cite[Table 1]{Krauth:2015nja}), thereby improving the uncertainty of the deuteron-radius independent term by a factor $2$. In addition, we include the inelastic $3\ga$-exchange, calculated for the first time in Ref.~\cite{Pachucki:2018yxe}.
Compared to \Eqref{Krauthteory}, the elastic $3\ga$-exchange contribution (\#\#r3, r3$^\prime$ in~\cite[Table 2]{Krauth:2015nja}) has been removed from the radius-dependent term, so that the
sum of elastic and inelastic $3\ga$-exchange ($2P-2S$):  $2.19(88)(27)\,\upmu$eV \cite{Pachucki:2018yxe}, is now listed as an individual term. 
The updated theory prediction for the Lamb shift in $\mu$D then reads \cite{Antognini:2022xoo}:
\beq
E_{2P-2S}   =\left[228.77408(38)-6.10801(28) \,\left(\frac{r_d}{\mathrm{fm}}\right)^2-E _{2S}^{2\ga  }+0.00219(92)\right]\mathrm{meV}. \eqlab{LSnewtheory}\\
\eeq
Inserting the deuteron charge radius determined from the H-D isotope shift, \Eqref{OURRdisotope}, and comparing to the CREMA measurement, \Eqref{muDexp}, we refine the empirical \2PE-exchange effect:
\beq
E _{2S}^{2\ga }(\text{emp.})=-1.7585(56)\,\mathrm{meV}.\eqlab{deuteronstrucNEWemp}
\eeq

\subsection{Comparison of Theoretical Predictions for \2PE Exchange}

{\renewcommand{\arraystretch}{1.5}
\begin{table}[t]
    \centering
      \begin{tabular}{l||l}
       &$E _{2S}^{2\ga}$ [meV] \\
     
      \hline
      \hline
      \multicolumn{2}{c}{Theory prediction}\\
          \hline
      Krauth et al.~'16~\cite{Krauth:2015nja} &$-1.7096(200)$ \\
      Kalinowski '19 \cite[Eq.~(6) + (19)]{Kalinowski:2018rmf}& $-1.740(21)$ \\
           \piEFT/ N3LO & $-1.752(20)$ \\
           \hline
           \hline
          \multicolumn{2}{c}{Empirical ($\mu$H + iso)}\\
          \hline
               Pohl et al. '16 \cite{Pohl1:2016xoo} & $-1.7638(68)$\\     This work & $-1.7585(56)$\\
    \end{tabular}
    \caption{Comparison of prediction for the \2PE-exchange effects in the $\mu$D Lamb shift.
    }
    \label{tab:results_comparison_new}
\end{table}}

 In \secref{SummaryTPEmuD}, we summarized our \piEFT/ results for the deuteron-structure effects in the $\mu$D Lamb shift originating from the forward \2PE exchange, including the accompanied electronic VP contributions, and compared to other theory predictions. Our final result is given in \Eqref{TPE+}.
 For a meaningful comparison to the empirical value for the \2PE-exchange effect, \Eqref{deuteronstrucNEWemp}, we need to add effects from off-forward \2PE exchange (the Coulomb distortions).
Formally of a subleading $O(\alpha^6\ln \alpha)$, they are, however, numerically important. We use the recommended value from the theory compilation in Ref.~\cite{Krauth:2015nja}:
\beq
E _{2S}^\mathrm{Coulomb}=0.2625(15)\,\text{meV},\eqlab{Coulombdist}
\eeq
derived from modern deuteron potentials ($\chi$ET potential and AV18 model~\cite{Wiringa:1994wb}).
This value should be consistent with the \piEFT/ framework, since the deuteron electric dipole polarizability from \piEFT/ \cite{Lensky:2021VVCS} is in agreement with predictions from the applied deuteron potentials \cite{Hernandez:2014pwa}.
Combining Eqs.~\eref{TPE+} and \eref{Coulombdist}, our final result for the \2PE-exchange structure effects on the $2S$-level in $\mu$D reads:
\begin{align}
    E _{2S}^{2\ga} = -1.752(20)\text{ meV},
    \label{eq:deltaE_inel_FF_w_corrections}
\end{align}
which is larger than the value accounted for in Ref.~\cite[Eq.~(17)]{Krauth:2015nja}, but agrees with Ref.~\cite{Kalinowski:2018rmf} within errors, cf.\ Table \ref{tab:results_comparison_new}. It is also in agreement with the empirical value, \Eqref{deuteronstrucNEWemp}, but more than a factor $3$ less precise.  
Our new theory compilation will be used in Section~\ref{sec:radiid} to extract $r_d(\mu\text{D})$ from the experimental value for $E_{2P-2S}$.

\section{Charge Radius Extractions}
\label{sec:radii}

\begin{figure}[bth]
\includegraphics[width=0.45\textwidth]{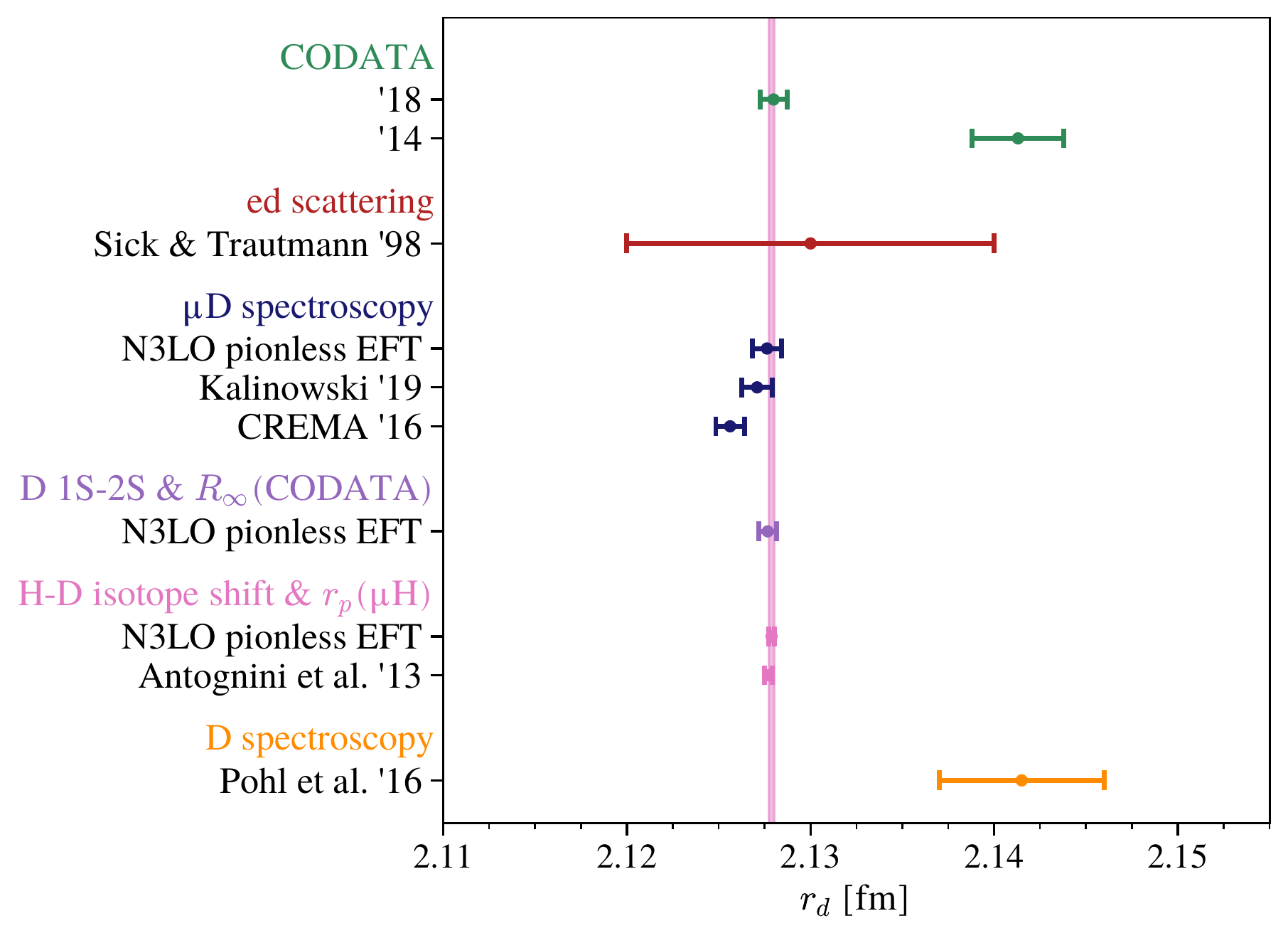}
\caption{Comparison of deuteron charge radius determinations from fits to electron-deuteron scattering data, ordinary and muonic-deuterium spectroscopy, and the $2S-1S$ hydrogen-deuterium isotope shift combined with the proton radius from muonic hydrogen.\figlab{RdSummaryFig}}
\end{figure}

\subsection{Deuteron Charge Radius}
\label{sec:radiid}

This section compares three independent extractions of the deuteron charge radius: from the spectroscopy of the $\mu$D Lamb shift, the $2S-1S$ transition in D and the $2S-1S$ H-D isotope shift, respectively. With the experimental value for the $\mu$D Lamb shift in \Eqref{muDexp}, the theoretical prediction in \Eqref{LSnewtheory}, and our result for the  \2PE-exchange effects, \Eqref{deltaE_inel_FF_w_corrections}, we can extract the deuteron charge radius from $\mu$D spectroscopy:
\begin{equation}
r_d(\mu\text{D}) =2.12763 (13)_\text{exp}(77)_\text{theory}= 2.12763(78)~\text{fm},\eqlab{rdmuDOUR}
\end{equation}
where the uncertainty budget remained the same as in the original extraction from Ref.~\cite{Pohl1:2016xoo}, see \Eqref{rmud}. In addition, we  consider the extraction from the measured $2S-1S$ transition in D \cite{Udem:1997}:
\beq
f^\mathrm{D}_{2S-1S}=2\,466\,732\,407\,522.88(91)\,\mathrm{kHz},
\eeq
and the theory prediction in \Eqref{Dtheory}, which leads to:
\beq
r_d(\text{D}, 2S-1S)=2.12767(49)\,\mathrm{fm}.
\eeq
Note that the entering Rydberg constant, $R_\infty$ in \Eqref{NewRydberg}, is strongly driven by $r_p(\mu\text{H})$. The third extraction from the H-D isotope shift and  $r_p(\mu\text{H})$ has been presented in Section \ref{sec:l1det}:
\beq
r_d(\mu\text{H \& iso})=2.12788(16)\ \text{fm}.\nn\\
\eeq
 All results are shown in  Fig.~\ref{fig:RdSummaryFig}, together will older extractions, results from electron-deuteron scattering and the CODATA recommended values. We can see that the spectroscopy of ordinary and muonic hydrogen isotopes, after the recent theory updates, cf.\ Ref.~\cite{Kalinowski:2018rmf}, gives consistent results for the deuteron charge radius.

\subsection{Proton Charge Radius}

 Analogously to the calculation of $r_d(\mu\text{H \& iso})$, we can use the isotope shift and $r_d(\mu \text{D})$ to extract the proton charge radius: 
\beq
r_p(\mu\text{D \& iso})=0.8404(20)\,\mathrm{fm}.
\eeq
While a previous extraction along these lines, $
r_p(\mu\text{D \& iso})=0.8356(20)\,\mathrm{fm}$ \cite{Pohl1:2016xoo}, had been in tension with $r_p(\mu\text{H})$, the result presented here based on the state-of-the-art theory predictions agrees.
This, again, nicely shows the consistency between the spectroscopic analyses of ordinary and muonic hydrogen isotopes.

\subsection{Proton-Deuteron Squared Charge Radii Difference}

Assuming $m_e \ll M_p \sim M_d$, we can find an approximation for the nuclear-size correction to the H-D isotope shift, \Eqref{fiv}, which is related to the often quoted difference of squared proton and deuteron charge radii. The best such approximation turns out to be
\begin{align}
\Delta f_\mathrm{iv} \approx -\frac{7}{24\pi}\frac{\alpha^4 m_e^3 c^4}{\hbar^3 (1+m_e/M_p)}\left[r_d^2-r_p^2\right]. \eqlab{ivapprox}
\end{align}
We give here the charge radius difference exactly, based on Eqs.~\eref{rpreview} and \eref{OURRdisotope}, and use the relation in \Eqref{ivapprox} only to estimate the uncertainty, which is dominated by the theory of the isotope shift:
\beq
r_d^2-r_p^2=3.820\,61(31)\,\mathrm{fm}^2.\eqlab{radiusdiff}
\eeq
Using \Eqref{ivapprox} instead, the central value would decrease by about $1.5\,\sigma$: $3.820\,13\left({}^{+78}_{-31}\right)\,\mathrm{fm}^2$. These results are in good agreement with the difference between charge radii extracted from the Lamb shift in muonic hydrogen isotopes:
\beq
r_d^2(\mu\mathrm{D})-r_p^2(\mu\mathrm{H})=3.819\,55(337)\,\mathrm{fm}^2.
\eeq
From \Eqref{radiusdiff}, we can see that  the larger CODATA~'14 recommended value for the proton charge radius, $r_p=0.8751(61)\,\mathrm{fm}$ \cite{Mohr:2015ccw}, would impose a larger value for the deuteron radius inconsistent with the $\mu$D Lamb shift.

\section{Conclusion and Outlook}
\seclab{Discussion}

In this work, we calculated the \2PE-exchange corrections to the $S$-levels in ordinary and muonic deuterium in the \piEFT/ framework. The calculation was performed at N3LO, with the only unknown LEC $l_1^{C0_S}$ appearing at this order extracted using the H-D isotope shift, where the correlation between that LEC and the \2PE-exchange correction is negligible. In addition, we evaluated the contribution of the nucleon structure, i.e., the effect of the nucleon polarizability and of the shape of the nucleon FFs, which are the most important single-nucleon effects beyond N3LO. We also included the accompanying electronic vacuum polarization contributions.  

Our predictions for the elastic contribution to the \2PE exchange in $\mu$D from \piEFT/ at N3LO and $\chi$ET at N4LO appear to be several standard deviations larger than the evaluations \cite{Carlson:2013xea,Acharya:2020bxf} based on the deuteron charge FF parametrisation of Ref.~\cite{Abbott:2000ak}, cf.\ Fig.~\ref{fig:TPELitComp} and Table \ref{tab:FF_numbers_TPE}. This suggests that the latter parametrisation does  
not adequately describe the behaviour of the deuteron charge FF at low virtualities. The correlation between the Friar radius $r_{\mathrm{F}d}$ and the deuteron charge radius $r_d$ in \piEFT/, cf.\ Fig.~\ref{fig:RF_correlation}, through the LEC $l_1^{C0_S}$ could serve as a diagnostic criterion for a realistic parametrisation of the deuteron charge FF. We also point out that the \piEFT/ expression for the deuteron charge FF at N3LO~\cite[Sec.~IV]{Lensky:2021VVCS} can be used for an analytic one-parameter fit to the electron-deuteron scattering data in the low-$Q^2$ range relevant to the planned DRad experiment~\cite{Zhou:2020cdt}. 

Supplementing the $\mu$D theory \cite{Krauth:2015nja} with a few
missing electronic VP effects \cite{Kalinowski:2018rmf} and the inelastic $3\gamma$ exchange \cite{Pachucki:2018yxe}, together with the shift of the elastic contribution found in this work, 
the past discrepancy between theory and experiment on the size of \2PE-exchange effects, see Table \ref{tab:results_comparison_new}, is now completely resolved.

The uncertainty of the theoretical result for the $2\gamma$-exchange correction was quantified using Bayesian inference. While our N3LO \piEFT/ prediction was not yet able to improve the theoretical precision, the improved understanding of the elastic contribution is of utmost importance. In addition, by calculating the \2PE-exchange correction to $\mu$D and D, we were able to perform a few consistency checks. In particular, we showed that extractions of the deuteron charge radius from the $\mu$D Lamb shift, the $2S-1S$ transition in D and the $2S-1S$ H-D isotope shift, cf.\ \Figref{RdSummaryFig}, are now in excellent  agreement.

\section*{Acknowledgements}
We thank A.~Hiller Blin for useful communications and the collaboration on the calculation of the deuteron VVCS amplitude, as well as for reading our manuscript and providing extremely valuable comments. We thank V.~Baru and A.~Filin for discussing the details of their work and for sharing with us the results of their $\chi$ET calculation of the deuteron charge form factor. We thank C.~Carlson and M.~Gorchtein for useful communications.
The calculations in this work were performed with the help of \textsc{FORM}~\cite{Vermaseren:2000nd}, and the figures in the article were made with the help of \textsc{SciDraw}~\cite{Caprio:2005dm}. This work was supported by the Deutsche Forschungsgemeinschaft (DFG) through the Emmy Noether Programme under Grant 449369623 and through research unit FOR 5327 under Grant 458854507. FH acknowledges the support of the Swiss National Science Foundation (SNSF) through the Ambizione Grant PZ00P2\_193383.

\appendix

\section{Quantification of Uncertainty}
\label{sec:error}

The uncertainty of an EFT calculation is in many cases, including the present work, dominated by the unknown higher-order terms rather than input parameters. To quantify the uncertainty, we follow the Bayesian approach developed in Refs.~\cite{Furnstahl:2015rha,Perez:2015ufa} and references therein. The results and the details specific to our evaluation are presented in this section.

We start with the EFT expansion of a generic observable $A$ in powers of the expansion parameter $\xi$
(which is $\xi=\gamma/m_\pi$ in \piEFT/):
\begin{align}
    A = A_0\sum\limits_{n=0}^\infty c_n\xi^n,
\label{eq:generic_expansion}
\end{align}
where the parameter $A_0$ sets the scale of $A$, and $c_n$ are the expansion coefficients. The uncertainty of $A$ caused by a truncation at $n=k$ is given by the unknown remainder:
\begin{align}
\Theta A_k = A_0\sum\limits_{n=k+1}^\infty c_n\xi^n.
\label{eq:generic_expansion_truncated}
\end{align}
The typical prior assumption used in EFT calculations is the naturalness of the expansion coefficients, i.e., $c_n$ should be at most $O(1)$. One may refine this assumption using the calculated expansion coefficients as done in, e.g.,
Ref.~\cite{Epelbaum:2014sza} that assigns (written out here for an N3LO calculation, $k=3$):
\begin{align}
    \Theta A_3 & = \max\left\{\xi^4|\Delta A^\mathrm{LO}|, \xi^3|\Delta A^\mathrm{NLO}|,\xi^2|\Delta A^\mathrm{N2LO}|, \xi|\Delta A^\mathrm{N3LO}|\right\}=A_0\,\xi^4 \max\limits_{n\leq 3} |c_n|,
    \label{eq:truncation_error_simplest}
\end{align}
where $\Delta A^\mathrm{LO}$, $\Delta A^\mathrm{NLO}$, etc.\ are the contributions to $A$ at the respective order. This also implicitly assumes that $\Theta A_k$ is dominated by its first term, $A_0\, c_{k+1}\xi^{k+1}$; we will employ this assumption in the following (note that it can be relaxed, in particular, in the Bayesian approach~\cite{Furnstahl:2015rha,Perez:2015ufa}). Looking at the expansion of $E _{2S}$, we note that the expansion of the LSZ factor, $\left[\Sigma'(E_d)\right]^{-1}\propto 1 + (Z-1) + 0 + 0 + \dots$, see Ref.~\cite[Eq.~(49)]{Lensky:2021VVCS}, results in every term appearing at a given order in the four-point function acquiring a factor of $Z$ at the next order, as is also explicitly shown in Table~\ref{tab:TPE_contributionsElInel}. This induces correlations between the coefficients in the expansion. It is therefore natural to slightly modify the expansion for the purpose of quantifying the uncertainty, taking out the known $Z$ factor along with the scale factor $A_0$. This amounts to working with the expansion of the (integrals of) the four-point functions, and the simple estimate of Eq.~\eqref{eq:truncation_error_simplest} becomes:
\begin{align}
    \Theta A_3 & = Z A_0 \,\xi^4 \max\limits_{n\leq 3}|c_n|,
    \label{eq:truncation_error_modified}
\end{align}
where $A_0$ and $c_n$ are now the normalization and the expansion coefficients of the integrals of the four-point function, which can be deduced from Table~\ref{tab:TPE_contributionsElInel}, respectively, for $E _{2S}^\mathrm{elastic}$ and $E _{2S}^{\mathrm{inel},L}$.
{\renewcommand{\arraystretch}{1.5}
\begin{table}[t]
    \centering
    \begin{tabular}{l||r|r|r|r|r}
    &$A_0$ [meV]&$\qquad c_0$&$c_1$&$c_2$&$c_3$\\
    \hline
       \hline
    $E _{2S}^\mathrm{elastic}$ & $-0.204$&$1$&$0.260$&$2.311$&$-1.896$\\
    $E _{2S}^{\mathrm{inel},L}$ & $-0.943$&$1$&$-0.050$&$-0.379$&$0.044$\\
    \hline
    $E _{2S}^\mathrm{sum}$ & $-1.148$&$1$&$0.006$&$0.100$&$-0.301$\\
    \end{tabular}
    \caption{Expansion coefficients and prefactors corresponding to the  modified \piEFT/ expansion.
    }
    \label{tab:expcoeff}
\end{table}}

Moving to the Bayesian quantification, we use the prior probability density functions (PDFs) introduced in Ref.~\cite{Furnstahl:2015rha} as Sets A, B, and C. Each one of the sets of priors consists of two PDFs, $\mathrm{pr}(c)$ and $\mathrm{pr}(c_n|c)$, where the first is associated with the scale $c$ typical of the coefficients $c_n$, while the second describes the probability distribution of $c_n$ given the scale. In the leading-term approximation for the uncertainty, the Bayes' theorem with the given prior PDFs gives the PDF of the uncertainty at order $k$, given the known $c_0,\dots c_k$:
\begin{equation}
    \mathrm{pr}(\Theta A_k|c_0,\dots c_k) = \frac{\int\limits_0^\infty \mathrm{d}c\,  \mathrm{pr}(c_{k+1}|c)\, \mathrm{pr}(c)\prod\limits_{m=0}^k \mathrm{pr}(c_m|c)}{Z\,A_0\,\xi^{k+1}\int\limits_0^\infty \mathrm{d}c\, \mathrm{pr}(c)\prod\limits_{m=0}^k \mathrm{pr}(c_m|c)}
\label{eq:uncertainty_PDF}
\end{equation}
with
\begin{equation}
c_{k+1} = \frac{\Theta A_k}{Z\,A_0\,\xi^{k+1}},
\end{equation}
where we specialize to the case of $E _{2S}$ and the quantities $A_0$ and $c_n$ again pertain to the expansion of the (integrals of) the four-point function (with the $Z$ factored out).

\begin{figure}[t]
    \centering
    \begin{tabular}{ccc}
    \includegraphics[width=0.33\textwidth]{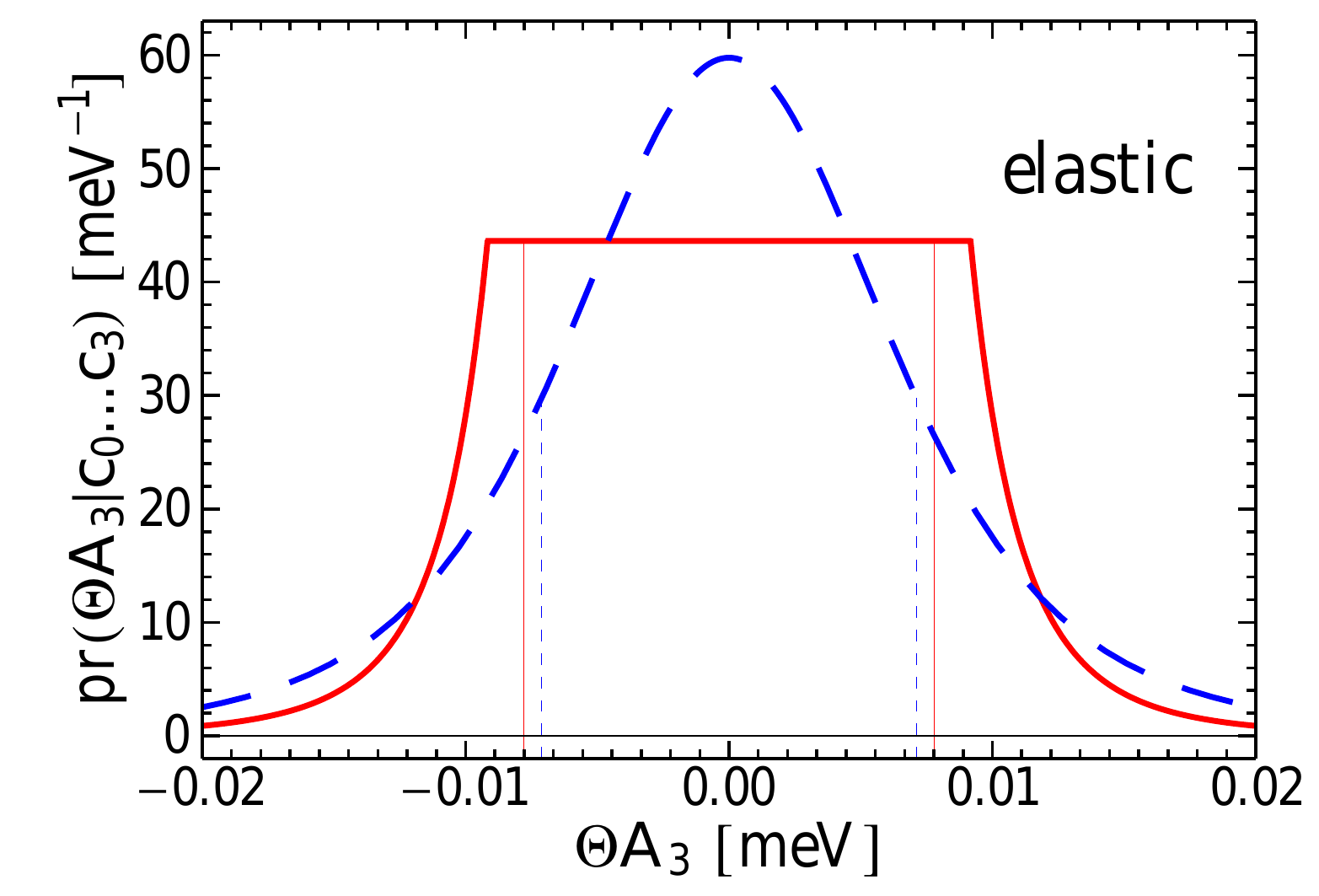}
    &  
    \includegraphics[width=0.33\textwidth]{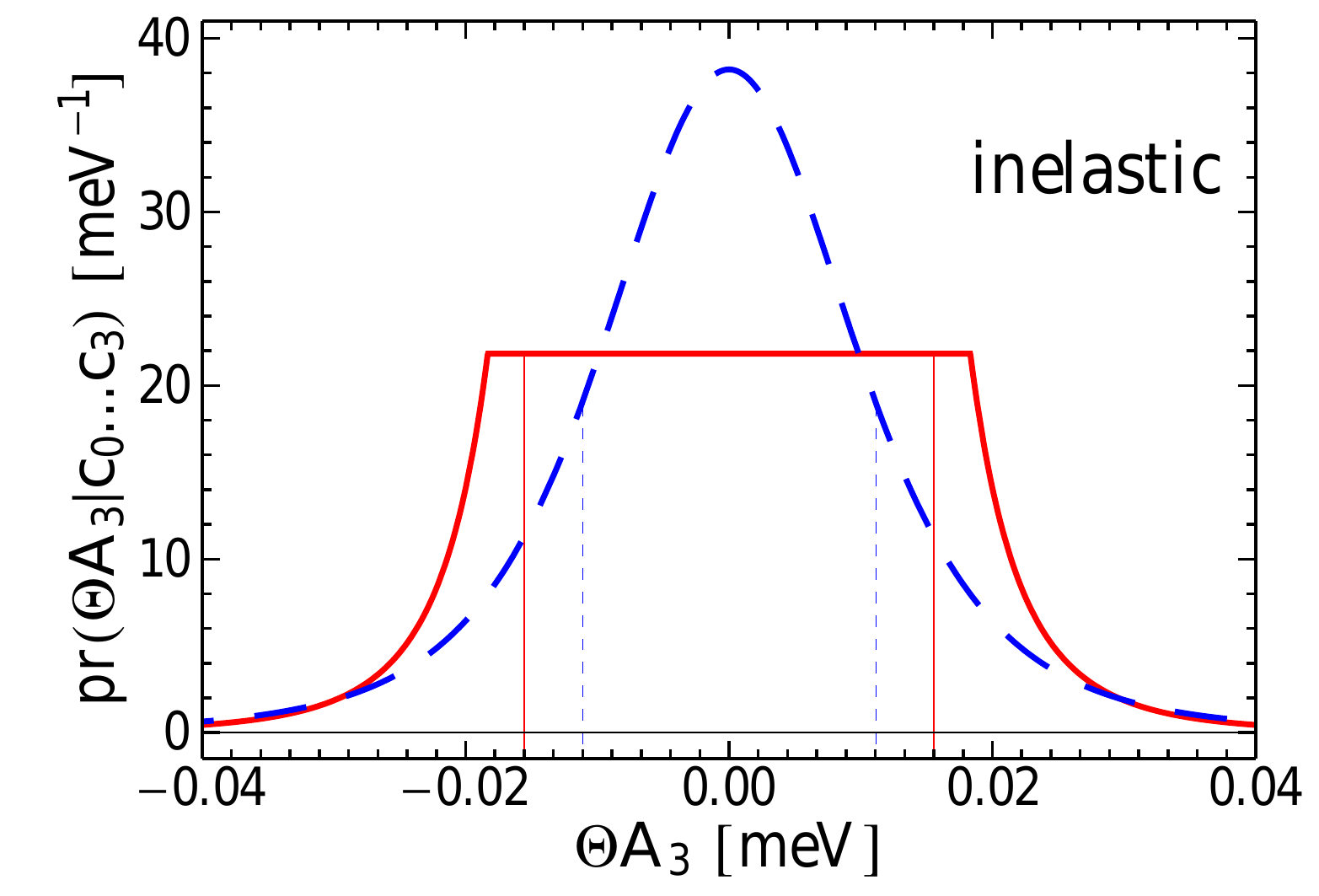}
    &
    \includegraphics[width=0.33\textwidth]{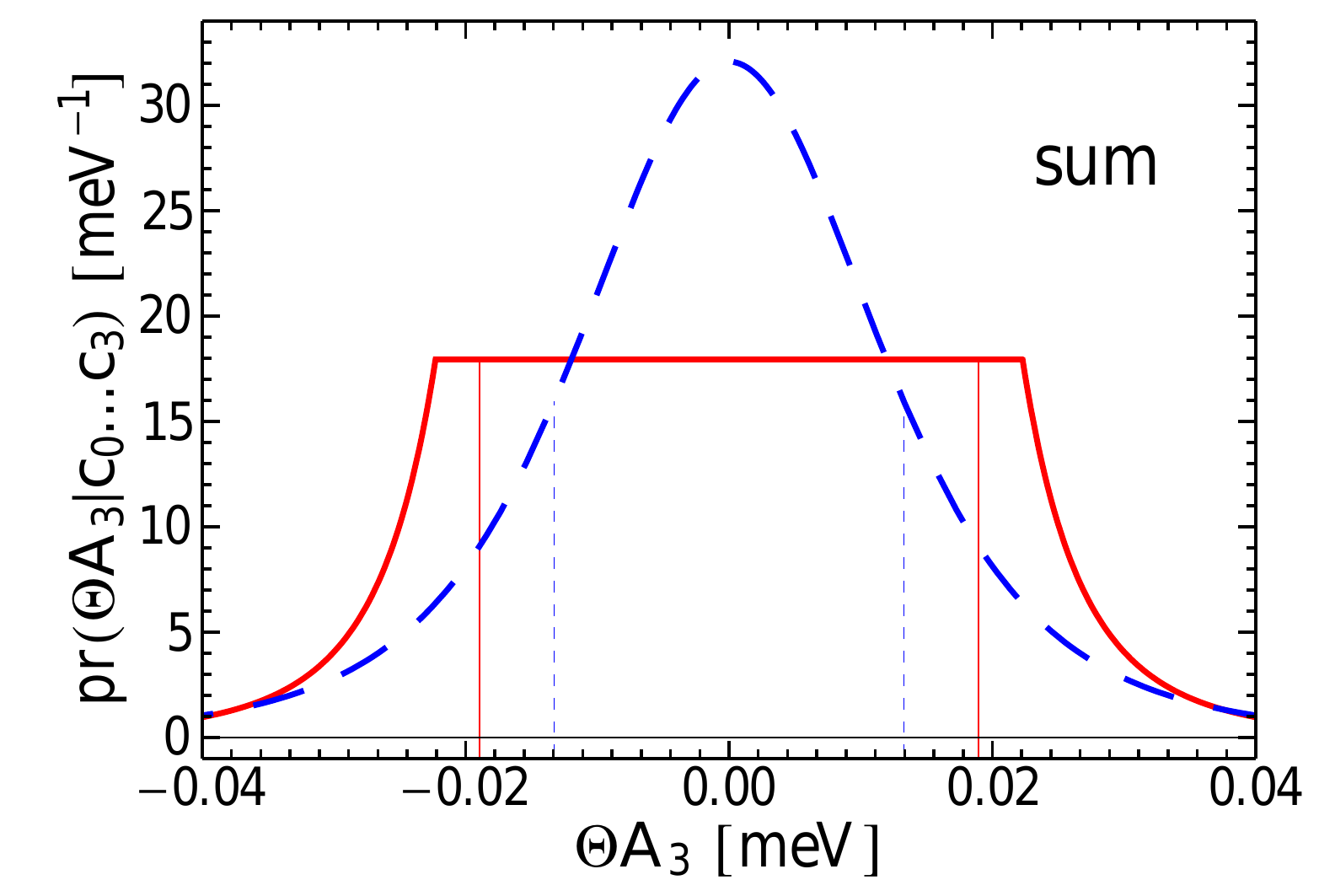}
    \end{tabular}
    \caption{Probability distribution functions for $\Theta(E _{2S}^\mathrm{elastic})$ (left),
    $\Theta(E _{2S}^{\mathrm{inel},L})$ (center), and $\Theta(E _{2S}^\mathrm{sum})$ (right). Red solid (blue dashed) curves correspond to Set A (Set C) priors. Thin vertical lines indicate the corresponding 68\% degree-of-belief intervals.}
    \label{fig:uncertainty_PDFs}
\end{figure}

The expansion coefficients and prefactors that correspond to the modified \piEFT/ expansion of $E _{2S}^\mathrm{elastic}$, $E _{2S}^{\mathrm{inel},L}$, and their sum $E _{2S}^\mathrm{sum} = E _{2S}^\mathrm{elastic}+E _{2S}^{\mathrm{inel},L}$, are given in Table \ref{tab:expcoeff}.
One can see that the expansion coefficients of the elastic contribution are somewhat larger than one, whereas the inelastic part, as well as the total, have smaller expansion coefficients. There are partial cancellations between the higher-order elastic and inelastic contributions, which, together with the LO total term being about five times larger than its elastic counterpart, suppresses the expansion coefficients of the total \2PE-exchange correction. With these coefficients, we start with the estimate of Eq.~\eqref{eq:truncation_error_modified}, which gives:
\begin{equation}
    \Theta \left(\left\{E _{2S}^\mathrm{elastic},E_{2S}^{\mathrm{inel},L},E_{2S}^\mathrm{sum}\right\}\right) = \{0.009,\,0.018,\,0.022 \}\ \text{meV}.
    \label{uncertaintySimple}
\end{equation}
The uncertainties deduced this way are {\it a priori} not Gaussian-distributed quantities
(as one sees explicitly from, e.g., the Bayesian PDFs below, cf.\ Fig.~\ref{fig:uncertainty_PDFs}), therefore the usual addition of uncertainties in quadrature may not be an adequate way to, e.g., calculate the uncertainty of a sum given the uncertainties of its constituents; we therefore estimate the uncertainty of $E _{2S}^\mathrm{sum}$ independently.

Proceeding to the Bayesian estimates, we apply Eq.~\eqref{eq:uncertainty_PDF} with the priors of Ref.~\cite{Furnstahl:2015rha} to the results of our calculation. The specific parameters that we use for the priors are:
\begin{align}
    c_>=100,\ c_<=10^{-3},\ \sigma = 2.0.
\end{align}
The resulting PDFs for $\Theta A_3$ are shown in Fig.~\ref{fig:uncertainty_PDFs}. Set A and Set B result in PDFs that in each case are practically on top of each other, we therefore show only Set A as a representative, along with Set C.

The Bayesian procedure results in the following $68\%$ degree-of-belief (DOB) intervals:
\begin{subequations}
\begin{align}
    \text{Set A}: & \qquad \Theta \left(\left\{E _{2S}^\mathrm{elastic},E_{2S}^{\mathrm{inel},L},E_{2S}^\mathrm{sum}\right\}\right) = \{0.0078,\,0.0156,\,0.0189 \}\ \text{meV},\\
    \text{Set B}: & \qquad \Theta \left(\left\{E _{2S}^\mathrm{elastic},E_{2S}^{\mathrm{inel},L},E_{2S}^\mathrm{sum}\right\}\right) = \{0.0077,\,0.0155,\,0.0188 \}\ \text{meV},\\
    \text{Set C}: & \qquad \Theta \left(\left\{E _{2S}^\mathrm{elastic},E_{2S}^{\mathrm{inel},L},E_{2S}^\mathrm{sum}\right\}\right) = \{0.0071,\,0.0111,\,0.0133 \}\ \text{meV}.
\end{align}
\end{subequations}
One can see that the simple estimate of Eq.~\eqref{uncertaintySimple} gives a more conservative uncertainty than is obtained from the Bayesian framework, which was also the tendency observed in Ref.~\cite{Furnstahl:2015rha}. It is also evident that the priors Set A and B result in somewhat bigger uncertainties than Set C; the difference is rather small for the elastic contribution, but is of the order of 25\% for the inelastic term and the total \2PE-exchange correction. Seeing that the current uncertainty estimate uses the leading omitted term approximation, one can expect an about 30\% change in the uncertainty once higher-order terms are taken into account. The above difference between Sets A and B, on the one hand, and Set C, on the other hand, is thus not unexpectedly big. One can also note that the uncertainty of the elastic term is probably overestimated, since higher-order terms are unlikely to change that term as much as suggested by the projected uncertainty, given that the value of $r_d^2$ is fixed at N3LO. This is also supported by the agreement between the \piEFT/ and $\chi$ET results for the elastic contribution. We take the more conservative results of Sets A and B as our uncertainty estimate for the calculated \2PE-exchange contributions. 

Finally, one may notice that, in particular, the relative smallness of the higher-order coefficients in Table~\ref{tab:expcoeff}, might mean that the assignment $\xi=1/3$ overestimates the size of the expansion parameter (and thus also the truncation uncertainty). As demonstrated in, e.g., Ref.~\cite{Furnstahl:2015rha} with $\chi$ET calculations, Bayesian tools could also be applied to quantify the (assigned) value of the expansion parameter. Such a study would ideally involve additional quantities obtained in the same \piEFT/ framework, and will be presented elsewhere; in this context, see Ref.~\cite{LiMuli:2022kcy} that studies the uncertainties arising from a different expansion employed in calculations of \2PE-exchange corrections in muonic atoms and ions.

\section{Deuteron VVCS Amplitudes with Insertion of Nucleon Form Factors}
\label{app:ff}
In this appendix, we provide the expressions for the longitudinal deuteron VVCS amplitude at LO and NLO resulting from the procedure outlined in Sec.~\ref{sec:nucleon_effects}, where one inserts the full nucleon FFs. The expressions for the four-point function (see Ref.~\cite[Sec.\ II D]{Lensky:2021VVCS} for the definition) with the inserted FFs read:
\begin{align}
    \mathcal{M}_L^\mathrm{(-3),\ FF} & = \frac{e^2 M^3}{\pi}\frac{Q^2}{\bv{q}^2}
    \left\{
    \frac{\left[\bar{G}_E^0(Q^2)\right]^2+\left[\bar{G}_E^1(Q^2)\right]^2}
     {\gamma\left[\bv{q}^2+4(\gamma+\lambda_d)^2\right]}
    +\frac{\left[\bar{G}_E^0(Q^2)\right]^2-\left[\bar{G}_E^1(Q^2)\right]^2}{M|\bv{q}|\nu}\phi(\nu,\bv{q}^2)
    -\frac{4\left[\bar{G}_E^0(Q^2)\right]^2}{\bv{q}^2\left(\gamma-\lambda_d\right)}\phi^2(\nu,\bv{q}^2)
    \right\}\nonumber\\
    &+(\nu\to-\nu),\\
    \mathcal{M}_L^\mathrm{(-2),\ FF} & =\frac{e^2 M^3}{\pi}\frac{Q^2}{\bv{q}^2}
    \frac{2(Z-1)}{\gamma}\frac{\left[\bar{G}_E^0(Q^2)\right]^2\phi(\nu,\bv{q}^2)\left[
    |\bv{q}|-(\gamma+\lambda_d)\phi(\nu,\bv{q}^2)
    \right]}{\bv{q}^2(\gamma-\lambda_d)}
    +(\nu\to-\nu).
\end{align}
Here, the kinematic functions are \cite{Lensky:2021VVCS}:
\begin{equation}
\lambda_d=\sqrt{\gamma^2-M\nu+\frac{\bv{q}^2}{4}},\qquad \phi(\nu,\bv{q}^2)=\arctan\frac{\left|\bv{q}\right|}{2(\gamma+\lambda_d)},
\end{equation}
and the barred nucleon isoscalar and isovector electric FFs are:
\begin{equation}
    \bar{G}_E^{0,1}(Q^2)=\frac{G_E^{0,1}(Q^2)}{\sqrt{1+\frac{Q^2}{4M_p^2}}}.
\end{equation}

\section{Electronic Vacuum Polarization Corrections to Finite-Size and Polarizability Contributions}
\label{sec:TPE_eVP}

In this appendix, we consider the one-loop electronic VP given by:
\begin{align}
    \ol \Pi^{(1)}(Q^2)= \Pi^{(1)}(Q^2)-\Pi^{(1)}(0) & = \frac{\alpha}{3\pi}  \left[2 \left(1-\frac{1}{2\tau_e }\right) \left(\sqrt{1+\frac{1}{\tau_e }}
    \arccoth\sqrt{1+\frac{1}{\tau_e }}
    -1\right)+\frac{1}{3}\right],
\end{align}
with $\tau_e=Q^2/4m_e^2$ and $m_e$ the electron mass, and its well known
corrections to the deuteron structure effects.

\begin{figure}[t]
\includegraphics[width=0.15\textwidth]{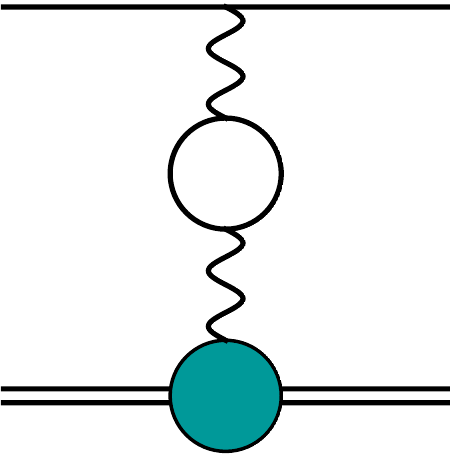}
\caption{One-photon exchange with vacuum-polarization and finite-size correction. \figlab{OPEwVPaFF}}
\end{figure}

We start with the corrections to the $O(\alpha^4)$ deuteron radius term to illustrate the approach, before calculating the corrections to the $O(\alpha^5)$ \2PE-exchange effect, relevant for this paper.
The one-loop electronic VP correction to the deuteron charge radius term, see  Fig.~\ref{fig:OPEwVPaFF}, is described by the following potential:\beq
\Delta V_\text{VP-FF}(r) = - \int \!\frac{\dd \bv{q}}{(2\pi)^3} \, e^{i \bv{q}\cdot \bv{r}}  \, \frac{4\pi \al}{\bv{q}^2} \, \ol \Pi^{(1)}(\bv{q}^2)
\left[G_C(\bv{q}^2) -1\right].
\eeq
Note that this is a contribution to the Breit potential \cite[Ch.\ IX, §83]{LandauLifshitz4}, 
where the retardation effects can be neglected at this order in $\al$, and hence $Q^2$ 
is replaced by $\bv{q}^2$. We shall make use of the dispersion relations (DRs)
for the VP and the FF:
\begin{subequations}
\eqlab{DRs}
\bea
\ol \Pi(Q^2) &=& -\frac{Q^2}{\pi} \fint^\infty_{4m_e^2} \! \!\dd t\, \frac{\im \, \Pi(t)}{t(t+Q^2)},\\
G_C(Q^2) &=& \frac{1}{\pi} \fint^\infty_{t_0} \! \! \dd t'\, \frac{\im \, G_C(t')}{t'+Q^2},\eqlab{GCDRunsub}
\eea
\end{subequations}
where $\fint$ denotes the principal-value integration. The one-loop expression for the absorptive part of electronic VP reads:
\beq
\eqlab{1loopVP}
\im \, \Pi^{(1)}(t) = -\frac{\al}{3}\left(1+\frac{2m_e^2}{t}\right)\sqrt{1-\frac{4m_e^2}{t} }.
\eeq 
The DR for VP
is once-subtracted to ensure the correct normalisation of the electromagnetic
field. Similarly to ensure the correct normalisation of the deuteron charge,
$G_C(0)=1$, we can use the once-subtracted relation for the charge FF:
\beq
G_C(Q^2) -1  =
-\frac{Q^2}{\pi} \fint^\infty_{t_0} \! \! \dd t'\, \frac{\im \, G_C(t')}{t'(t'+Q^2)}.
\eeq
In first-order perturbation theory to the unperturbed Coulomb wave functions, one finds for the Lamb shift:
\beq
\eqlab{geneq2}
E _{2P-2S}^{(1)\langle \text{VP-FF}\rangle}
= -\frac{\al^4 m_r^3}{2\pi} \left( \fint_{4m_e^2}^\infty \! \dd t\, \frac{\im\, \Pi^{(1)}(t)}{(\sqrt{t}+\al m_r)^4}
\, \Big[ G_C(t) -1 \Big]  +
\fint_{t_0}^\infty  \! \dd t'\, \frac{\im\, G_C(t')}{(\sqrt{t'}+\al m_r)^4}
\, \ol\Pi^{(1)}(t') \,\right).
\eeq
It is clear that the dominant effect comes from
the small-$t$ region in the first integral, which starts from the threshold of $e^+ e^-$ production. Unfortunately,
we cannot simply expand $G_C$ around $0$ before integration, since the
integral will eventually diverge. Instead we use again the DR for $G_C$ given in \Eqref{GCDRunsub}. We then change the variable $t\to 4m_e^2 u^2$ and perform the integration 
over $u$. Afterwards, only integrals over $t'$ remain, which start from the threshold of hadron (e.g., $\pi^+ \pi^-$) production $t_0$. Assuming that $2 m_e \ll t_0 \leq t'$, we can expand up to 
 $O(4m_e^2/t' )$. Neglecting terms which are suppressed by additional factors of $m_e^2$, we obtain: 
\begin{subequations}
\bea
E _{2P-2S}^{(1)\langle \text{VP-FF}\rangle}&=&-\frac{1}{6}\, \al^5 m_r^3 A(\kappa)
 r_d^2\\ &=&-0.0135\, \left[\frac{r_d}{\text{fm}}\right]^2 \, \mbox{meV}\simeq -0.06113(1) \, \mbox{meV},
 \eqlab{V1num}
\eea
\end{subequations}
with the auxiliary function: 
\beq
 A(\kappa) =
\frac{1}{12 \pi (1-\kappa^2)^2}  \left[ \kappa^2( 4\kappa^2 - 7) +
   \frac{\kappa( 4 \kappa ^4-10 \kappa ^2+9) }{\sqrt{1-\kappa^2}}
    \arccos \kappa \right]\simeq 0.152309
 \eeq
 at $\kappa=\al m_r/2m_e$. Our formula agrees numerically with Ref.~\cite[Eq.~(28)]{Martynenko:2014bqa}. In \Eqref{V1num}, we used the deuteron radius determined through the isotope shift to illustrate the quantitative size of the effect, where the uncertainity is just propagated from the error of the radius in \Eqref{OURRdisotope}.

\begin{figure}[t]
\includegraphics[width=0.45\textwidth]{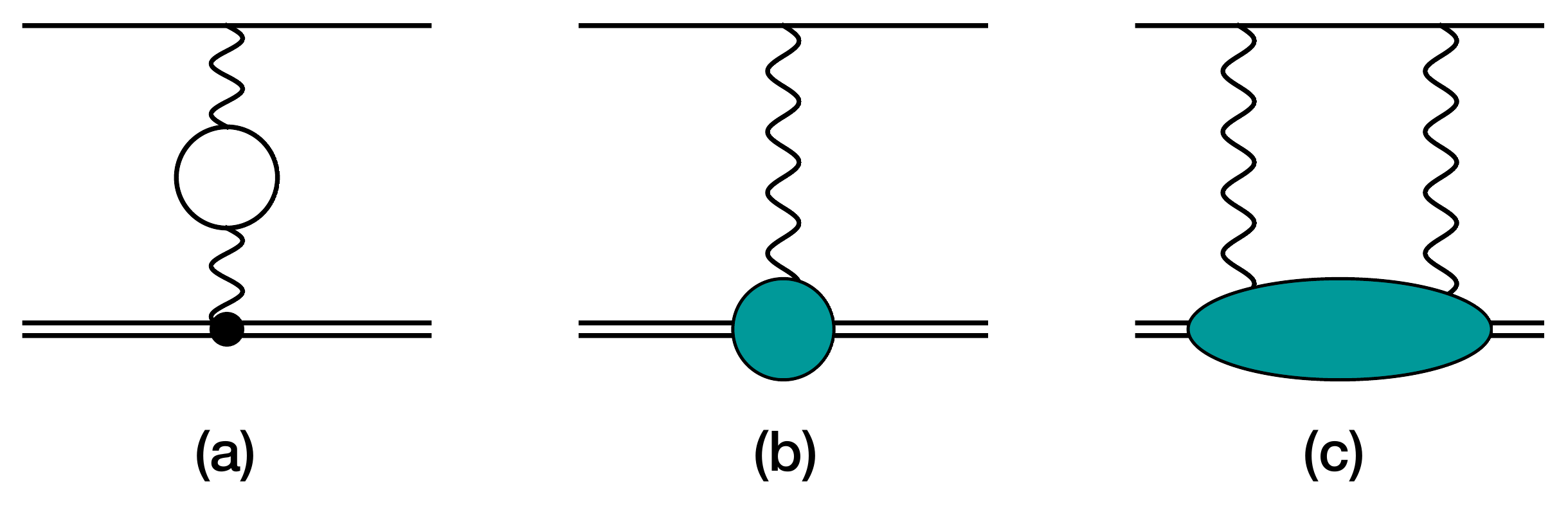}
\caption{(a) One-photon exchange with vacuum polarization; (b) One-photon exchange with finite-size correction; and (c) elastic and inelastic two-photon exchange. \figlab{VPinterference}}
\end{figure}

A similar subleading correction stems from the interference of one-photon exchange potentials with electronic VP, 
\beq
\Delta V_{\mathrm{VP}} (r) = - \int \!\frac{\dd \bv{q}}{(2\pi)^3} \, e^{i \bv{q}\cdot \bv{r}}  \, \frac{4\pi \al}{\bv{q}^{\,\,2}} \, \ol\Pi^{(1)}(\bv{q}^2)=\frac{\al}{\pi} \fint^\infty_{4m_e^2}  \!\dd t\, \frac{\im \, \Pi^{(1)}(t)}{t} \frac{e^{-r \sqrt{t}}}{r},\eqlab{VPpotential}
\eeq
and finite-size corrections, 
 \beq
\Delta V_{\mathrm{FF}} (r) = - \int \!\frac{\dd \bv{q}}{(2\pi)^3} \, e^{i \bv{q}\cdot \bv{r}}  \, \frac{4\pi \al}{\bv{q}^2} \, \left[ G_C(\bv{q}^2)  - 1\right]\simeq\frac{4\pi  \alpha \,r_d^2}{6}  \,\delta(\bv{r}),\eqlab{FFpotential}
\eeq
see Fig.~\ref{fig:VPinterference} (a) and (b), respectively. The latter can be approximated with a delta-function potential proportional to the deuteron radius. To calculate this effect at second order in perturbation theory, we need to know the matrix elements of the delta-function and Yukawa-type potentials between the $\mu$D Coulomb wave functions:
\begin{subequations}
\bea
\langle 2S\vert \delta(\bv{r}) \vert nS \rangle&=& \frac{1}{2\sqrt{2n^3}\pi a^3},\\
\langle 2S\vert \nicefrac{e^{-r\sqrt{t}}}{r} \vert nS \rangle&=&-\frac{4\sqrt{2n}}{a}\frac{4-n^2\left(3+4a^2t\right)}{\left[n^2\left(1+2a\sqrt{t}\right)^2-4\right]^2}\exp\left[-2n \arctanh \frac{2}{n\left(1+2a\sqrt{t}\right)}\right],
\eea
\end{subequations}
and the energy levels of the Coulomb potential:
\beq
E_n=-\frac{\alpha }{2 a n^2}, \qquad E_2=-\frac{\al}{8a},
\eeq
with $n$ the principal quantum number. For the discrete spectrum, we obtain:
\bea
E ^{(2) \mathrm{ disc.}\langle \text{VP} \rangle\langle \text{FF} \rangle}_{2S}&=&2\sum_{n\neq 2}\frac{\langle 2S \vert \Delta V_{\mathrm{VP}}\vert nS\rangle\langle 2S \vert \Delta V_{\mathrm{FF}}\vert nS\rangle}{E_2-E_n}\\
&\simeq&\frac{4\al^2 r_d^2}{3} \sum_{n\neq 2}\frac{1}{E_2-E_n}\langle 2S\vert \delta(\bv{r}) \vert nS \rangle \int_{4m_e^2}^\infty \dd t\, \frac{\im \Pi^{(1)}(t)}{t}\, \langle 2S\vert \nicefrac{e^{-r\sqrt{t}}}{r} \vert nS \rangle\\
&=&-0.008274\, \left[\frac{r_d}{\text{fm}}\right]^2 \, \mbox{meV}\simeq -0.037464(6)\,\mathrm{meV},
\eea
with the deuteron radius in fm units.
For the continuous spectrum, we apply:
\begin{subequations}
\bea
\langle 2S\vert \delta(\bv{r}) \vert kS \rangle&=& \frac{1}{2\sqrt{2}\pi a^3}\sqrt{\frac{k}{1-e^{-2\pi /k}}},\\
\langle 2S\vert \nicefrac{e^{-r\sqrt{t}}}{r} \vert kS \rangle&=&\frac{4\sqrt{2}}{a}\frac{3+4\left(a^2t+k^2\right)}{\left[4k^2+\left(1+2a\sqrt{t}\right)^2\right]^2}\exp\left[-\frac{2}{k} \arctan \frac{2k}{1+2a\sqrt{t}}\right]\sqrt{\frac{k}{1-e^{-2\pi /k}}},
\eea
\end{subequations}
and
\beq
E_k=\frac{\alpha  k^2}{2 a},
\eeq
to get:
\bea
E ^{(2) \mathrm{ cont.}\langle \text{VP} \rangle\langle \text{FF} \rangle}_{2S}&=&2 \int_0^\infty \dd k\, \frac{\langle 2S \vert \Delta V_{\mathrm{VP}}\vert kS\rangle\langle 2S \vert \Delta V_{\mathrm{FF}}\vert kS\rangle}{E_2-E_k}\\
&\simeq&
\frac{\sqrt{2}\al^2 r_d^2}{3\pi a^3} \int_0^\infty \dd k\, \sqrt{\frac{k}{1-e^{-2\pi /k}}} \,\frac{1}{E_2-E_k} \int_{4m_e^2}^\infty \dd t\, \frac{\im \Pi^{(1)}(t)}{t}\, \langle 2S\vert \nicefrac{e^{-r\sqrt{t}}}{r} \vert kS \rangle\\
&=&0.028761\, \left[\frac{r_d}{\text{fm}}\right]^2 \, \mbox{meV}\simeq 0.130226(20)\,\mathrm{meV}.
\eea
In total, the interference of the one-photon-exchange potentials in Fig.~\ref{fig:VPinterference} (a) and (b) amounts to:
\bea
E ^{(2) \langle \text{VP} \rangle\langle \text{FF} \rangle}_{2S}&=&E ^{(2) \mathrm{ disc.}\langle \text{VP} \rangle\langle \text{FF} \rangle}_{2S}+E ^{(2) \mathrm{ cont.}\langle \text{VP} \rangle\langle \text{FF} \rangle}_{2S}\\
&=&0.020487\, \left[\frac{r_d}{\text{fm}}\right]^2 \, \mbox{meV}\simeq0.092763(14)\,\mathrm{meV}.
\eea
This formula agrees numerically with Ref.~\cite[Eq.~(29)]{Martynenko:2014bqa}.

\begin{figure}[t]
\includegraphics[width=0.15\textwidth]{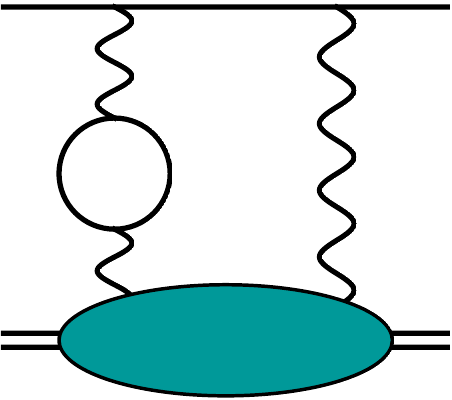}
\caption{Elastic and inelastic two-photon exchange with vacuum-polarization insertion at $O(\al^6)$. \figlab{TPEwithVP}}
\end{figure}

Let us now turn to our main interest: the electronic VP corrections to the \2PE exchange. The simplest correction is due to the insertion of the one-loop electronic VP into the \2PE-exchange diagram, see Fig.~\ref{fig:TPEwithVP}. We multiply the integrand in Eq.~\eqref{eq:TPE_LS} with $\left[1-\ol \Pi^{(1)}(Q^2)\right]^{-2}$ and obtain for the sum of elastic and inelastic contributions:
\begin{align}
    E _{2S}^{(1)\langle2\ga\text{-}\mathrm{VP}\rangle} & =[-0.0071-0.0136]~\mathrm{meV}= -0.0207\text{ meV}.
\end{align}
In addition, there is a correction to the $\mu$D atomic wave function that can be calculated at second order in perturbation theory from the interference of the one-photon exchange potential with VP insertion, \Eqref{VPpotential}, and the forward \2PE-exchange potential:
\beq
\Delta V_{{2\ga}} (r)=\frac{E ^\text{fwd}_{nS}}{\left[\phi_{n}(0)\right]^2}\,\delta(\bv{r}),
\eeq
see Fig.~\ref{fig:VPinterference} (a) and (c), respectively.
Since the latter is a delta-function potential just like our approximated one-photon exchange potential with finite-size correction, \Eqref{FFpotential}, the calculation of 
$E ^{(2) \langle \text{VP} \rangle\langle 2\ga \rangle}_{2S}$ proceeds analogously to the calculation of $E ^{(2) \langle \text{VP} \rangle\langle \text{FF} \rangle}_{2S}$  above. We therefore present here only the results:
\begin{subequations}
\bea
E ^{(2) \mathrm{ disc.}\langle \text{VP} \rangle\langle 2\ga \rangle}_{2S}&=& 0.0013624 \, E ^\text{fwd}_{2S}\simeq-0.00271\,\mathrm{meV},\\
E ^{(2) \mathrm{ cont.}\langle \text{VP} \rangle\langle 2\ga \rangle}_{2S}&=& -0.0047358\, E ^\text{fwd}_{2S}\simeq 0.00941\,\mathrm{meV},\\
E ^{(2) \langle \text{VP} \rangle\langle 2\ga \rangle}_{2S}&=&E ^{(2) \mathrm{ disc.}\langle \text{VP} \rangle\langle 2\ga \rangle}_{2S}+E ^{(2) \mathrm{ cont.}\langle \text{VP} \rangle\langle 2\ga \rangle}_{2S}\\
&=&1.4523\,\frac{\al}{\pi}\, E ^\text{fwd}_{2S}=0.0033734\, E ^\text{fwd}_{2S}=-0.00670(7)\,\mathrm{meV}.\eqlab{TPEwfcorr}
\eea
\end{subequations}
 The formula agrees numerically with the wave function correction in Refs.~\cite[Eqs.~(17) and (18)]{Kalinowski:2018rmf} and \cite[Table II]{Karshenboim:2018iyl}. Here we used the full forward \2PE-exchange result, $E _{2S}^\text{fwd}=-1.987(20)$ meV, collected in \Eqref{TPEforVP}, and propagated its uncertainty into \Eqref{TPEwfcorr}. 

The sum of electronic VP corrections to the \2PE exchange, amounts to:
\beq
    E _{2S}^\mathrm{eVP} =E _{2S}^{(1)\langle2\ga\text{-}\mathrm{VP}\rangle}+E^{(2) \langle \text{VP} \rangle\langle 2\ga \rangle}_{2S}=-0.0274\text{ meV},
    \label{eq:evp_correction_total}
\eeq
which is about a factor one-and-a-half larger than our error estimate for $E _{2S}^\text{fwd}$.
Our result is comparable to the results of Ref.~\cite[Eq.~(19)]{Kalinowski:2018rmf}: $E _{2S}^\mathrm{eVP}=-0.0265(3)$ meV.

\section{Hydrogen-Deuterium Isotope Shift}

\label{app:iso}

In this appendix, we will update the analysis of the H-D isotope shift presented in Ref.~\cite{Jentschura:2011is}, which is based on the framework reviewed in the CODATA 2006 report~\cite{Mohr:2008fa}. More recent summaries of the theory of hydrogen-like atoms can be found, e.g., in Refs.~\cite{Yerokhin2018,Pachucki:2018yxe}. The most relevant physical constants entering the isotope shift calculation are listed below,
with their relative uncertainties given in square brackets:
\begin{align}
m_e&=9.109\,383\,7015(28)\times10^{-31}\;\text{kg}&[3.1\times 10^{-10}],\nn\\
\frac{m_e}{M_p}&=5.446\,170\,214\,87 (33)\times 10^{-4}&[6.0\times 10^{-11}],\nn\\
\frac{m_e}{M_d}&=2.724\,437\,107\,462 (96)\times 10^{-4}&[3.5\times 10^{-11}],\nn\\
\frac{M_d}{M_p}&=1.999\,007\,501\,39 (11)&[5.6\times 10^{-11}],\nn\\
\alpha^{-1}&=137.035\,999\,084 (21)&[1.5\times 10^{-10}],\nn\\
R_\infty c&=3.289\,841\,960\,2508 (64)\times 10^{15}\; \text{Hz}&[1.9\times 10^{-12}].
\eqlab{RydbergCodata18}
\end{align}
The values are taken from the latest CODATA and PDG reports \cite{Zyla:2020zbs,Tiesinga:2021myr}. Compared to the CODATA 2006 report~\cite{Mohr:2008fa}, the relative uncertainties have improved by a factor $3.5$ to $12$. The relative uncertainty of $m_e$ has even shrunk by two orders of magnitude. The Rydberg constant, $R_\infty$, listed above is a result of the CODATA 2018 fit of various measured transitions in hydrogen-like atoms. It is, however, largely driven by the Lamb shift in $\mu$H. A similar result is obtained relying solely on the Lamb shift in $\mu$H and the $2S-1S$ transition in H, see our determination in Appendix \ref{RydbergSection}.

In Ref.~\cite{Jentschura:2011is}, the various contributions to the isotope shift are split into four Sets (i)--(iv).
The bulk of the isotope shift comes from the respective difference of the Dirac eigenvalues once the reduced mass effects are accounted for, identified in Ref.~\cite{Jentschura:2011is} as Set~(i). Based on the newest CODATA 2018 set of physical constants, we evaluate the frequency shift as: 
\begin{align}
\Delta f_\mathrm{i} = & 671\,004\,071.028(85)~\mathrm{kHz}\qquad [671\,004\,071.29(66)~\mathrm{kHz}].\eqlab{fi}
\end{align}
This is $262$ Hz smaller than the old value (given in square brackets) obtained using the CODATA 2006 set of constants, and considerably more precise. As mentioned already, the best values for the Rydberg constant, see Eqs.~\eref{RydbergCodata18} and \eref{NewRydberg}, are determined i.a.\ from the experimental $2S-1S$ transition in H, thus, rely on the same theory input as the $2S-1S$ H-D isotope shift. However, since they are largely driven by $r_p(\mu\text{H})$, it is justified to use these values in our analysis. Both central value and uncertainty estimate, following Ref.~\cite{Jentschura:2011is}, are the same for $\Delta f_\mathrm{i}$ with Eqs.~\eref{RydbergCodata18} and \eref{NewRydberg}, respectively.

We should also mention that since the latest CODATA adjustment, there have been many advances in the experimental determination of the electron-proton mass ratio \cite{Patra2020,Alighanbari2020,Kortunov:2021rfe} and the proton-deuteron mass ratio \cite{PhysRevLett.124.013001,Rau2020}. The presently most precise values (relative uncertainty in square brackets),
\begin{align}
\frac{m_e}{M_p}&=5.446\,170\,214\,805 (98)\times 10^{-4}&[1.8\times 10^{-11}],\nn\\
\frac{M_d}{M_p}&=1.999\,007\,501\,243 (31)&[1.6\times 10^{-11}],\nn
\end{align}
are based on an improved theory of rovibrational spin-averaged transitions in the hydrogen molecular ion $\text{HD}^+$ \cite{PhysRevA.104.032806}.
From this it also follows that: 
\begin{align}
\frac{m_e}{M_d}&=2.724\,437\,107\,624 (65)\times 10^{-4}&[2.4\times 10^{-11}].\nn
\end{align}
Using these values to calculate the effect of the dominant Dirac eigenvalue contribution to the isotope shift, we obtain:
\beq
\Delta f_\mathrm{i} =671\,004\,070.972(29)~\mathrm{kHz},
\eeq
in agreement with \Eqref{fi}, but almost a factor three more precise. Note that the $5\,\sigma$ variance between the latest determination of the fine structure constant $\alpha$ from a rubidium recoil measurement \cite{Morel:2020dww} and the best caesium recoil measurement \cite{Parker:2018vye} has no effect on the isotope shift.

The next by size Set (ii) of Ref.~\cite{Jentschura:2011is} includes ten contributions with frequency shifts $\Delta \nu_j$. We checked that we reproduce the numbers for this set from Ref.~\cite{Jentschura:2011is} individually, using the constants and the formalism given in the CODATA 2006 report~\cite{Mohr:2008fa}, up to $0.01~\mathrm{kHz}$. The new CODATA 2018 constants do not have a significant effect here. However, there have been considerable improvements of the theory, as we show in our updated evaluation below (old values in square brackets):
 \begin{enumerate}
    \item One-loop SE and electronic VP:
    \beq
\Delta \nu_1=-5558.999\,\text{kHz}\qquad [-5558.99\,\text{kHz}].
\eeq
    \item Two-loop SE, electronic VP, and combined effects:
    \beq
\Delta \nu_2=-0.521(1)\,\text{kHz}\qquad [-0.51\,\text{kHz}].\eqlab{nu2}
\eeq
Here, we use updated values for the coefficients \cite{Mohr:2015ccw}:
\begin{align}
B_{50}(nS)&=-21.554 47(13)& [-21.5561(31)],\\
B_{60}(1S)&=-81.3(0.3)(19.7)& [-95.3(0.3)(33.7)],\\
B_{60}(2S)&=-66.2(0.3)(19.7)& [-80.2(0.3)(33.7)].
\end{align}
For the logarithmic coefficient $B_{61}$, we are including previously neglected light-by-light contributions, evaluated in Ref.~\cite{Czarnecki:2016lzl,Szafron:2019tho}, see Ref.~\cite{Karshenboim:2019iuq}:
\begin{align}
B_{61}^\text{LbL}(nS)&=-\frac{43}{36}+\frac{709 \pi^2}{3456}.
\end{align}
We checked that using instead of $B_{60}(1S)$ the all-order in $(\mathcal{Z} \al)$ coefficient \cite{Karshenboim:2019iuq},\footnote{This is in agreement with $G_{60}(1S)=-92(13)$ \cite{Yerokhin2018}.}
\begin{align}
G_{60}(1S)&=-94.5(6.6),
\end{align}
the result changes marginally, to $-0.520$ kHz. This is within the uncertainty estimate in \Eqref{nu2}. For a discussion of the pure SE contribution to $G_{60}$ see Ref.~\cite{Karshenboim:2019dug}.

    \item Three-loop SE, electronic VP, and combined effects:
       \beq
\Delta \nu_3=-0.001\,\text{kHz}\qquad [-0.001\,\text{kHz}].
\eeq
Including $C_{50}$ and $C_{62}$ from Ref.~\cite{Karshenboim:2018mtf,Karshenboim:2019siz}, see Ref.~\cite{Karshenboim:2019iuq}:
\begin{align}
C_{50}(nS)&=-3.3(10.5)&[\pm\, 30],\\
C_{62}(nS)&=-\frac{2}{3}B_{40}(nS)\simeq-0.36,&
\end{align}
in addition to the leading $C_{40}$ term, has no effect at the precision given above.
Note that the four-loop QED contribution has been calculated in Ref.~\cite{Laporta:2019fmy}, but can be neglected at the present level of precision.
    \item Salpeter recoil correction:
    \beq
\Delta \nu_4=1032.65\,\text{kHz}\qquad [1032.65\,\text{kHz}].\eqlab{nu4}
\eeq
    \item Higher-order pure recoil corrections:
 \beq
\Delta \nu_5=(-3.140+0.081)\,\text{kHz}=-3.059(7)\,\text{kHz}\qquad [-3.41(32)\,\text{kHz}].\eqlab{nu5}
\eeq    
In Ref.~\cite{Jentschura:2011is}, pure recoil corrections at first order in the electron-nucleus mass ratio and expanded up to $(\mathcal{Z}\alpha)^7 \log^2 (\mathcal{Z}\alpha)^{-2}$ in the $(\mathcal{Z} \alpha)$ expansion were included. The uncertainty was estimated assuming that the first neglected higher-order term proportional to $(\mathcal{Z}\alpha)^7 \log (\mathcal{Z}\alpha)^{-2}$ is of natural size.
In Ref.~\cite{Yerokhin:2015}, it was shown that the previously neglected coefficient $D_{71}$ multiplying the 
$(\mathcal{Z}\alpha)^7 \log (\mathcal{Z}\alpha)^{-2}$ term is about a factor $16$ larger than the coefficient $D_{72}$ multiplying the supposedly more important
$(\mathcal{Z}\alpha)^7 \log^2 (\mathcal{Z}\alpha)^{-2}$ term. This resolved a discrepancy between numerical all-order and analytical $(\mathcal{Z} \alpha)$-expansion results. Here, we use the all-order result reported in Ref.~\cite{PhysRevA.93.062514}. The two values in \Eqref{nu5} are the recoil corrections for point-like and 
Gaussian distributed nuclear charges, respectively. The uncertainty is due to the latter contribution of the nuclear-finite size. As suggested in Ref.~\cite{Yerokhin:2015}, we estimate that the uncertainty of the dimensionless parameter $\delta_\text{fns}P$ due to the values of the nuclear radii is given by $\nicefrac{2 \delta R}{R}\, \delta_\text{fns}P$, where for $\delta R$ we take the difference between the proton and deuteron radii in the Gauss nuclear model and the experimental values from Eqs.~(\ref{eq:rpreview}) and (\ref{eq:Rdisotope}). Furthermore, we include an uncertainty due to approximations made in the calculation of $\delta_\text{fns}P$ ($19\times 10^{-7}$ for H, $65\times 10^{-7}$ for D) and the difference between the nuclear models in Ref.~\cite{PhysRevA.93.062514,Yerokhin:2015}.

    \item Radiative recoil corrections:\footnote{Note that the reference for this correction given in Ref.~\cite{Jentschura:2011is} is incorrect. It is calculated from Eq.~(56) of Ref.~\cite{Mohr:2008fa}.}
    \beq
\Delta \nu_6=-5.38(32)\,\text{kHz}\qquad [-5.38(11)\,\text{kHz}].
\eeq
As suggested in Ref.~\cite{Pachucki:2018xxx}, the more conservative uncertainty estimate from CODATA is used \cite{Mohr:2015ccw}.

    \item Nuclear SE:
     \beq
\Delta \nu_7=2.98(11)\,\text{kHz}\qquad [2.98(10)\,\text{kHz}].
\eeq
    \item Muonic and hadronic VP:
    \beq
\Delta \nu_8=0.006\,\text{kHz}\qquad [0.006\,\text{kHz}].
\eeq
While the numerical value of the sum of muonic and hadronic VP contributions remains unchanged, it has in fact been improved, including in addition the effect of NLO hadronic VP and the muonic VP correction to the electromagnetic electron vertex \cite{Karshenboim:2021jsc}.
We checked that the scatter between theoretical predictions and experimental extraction of the hadronic VP contribution to the muon anomalous magnetic moment $a_\mu$, used as input in Ref.~\cite{Karshenboim:2021jsc}, agrees with an updated value based on the new experimental result from Fermilab \cite{Muong-2:2021ojo} and the presently recommended Standard Model prediction collected in the Theory Initiative White Paper \cite{Aoyama:2020ynm}. It also covers the recent lattice QCD prediction from the BMW Collaboration \cite{Borsanyi:2020mff}. The further improvement in precision of the scatter is not relevant at the level of precision required for the isotope shift.

    \item Nuclear-polarizability correction:
    
     \beq
\Delta \nu_{9}=\left[18.489 - 0.769\,l_1^{C0_S} \right]~\mathrm{kHz}=18.490(205)\,\text{kHz}\qquad [18.64(2)\,\text{kHz}],\eqlab{isonu9}
\eeq
from combining our results in Eqs.~\eref{nu9D} and \eref{nu9H}. The uncertainty is dominated by $\Delta \nu_{9}^\mathrm{D}$. The decrease of the central value is mainly due to our new \piEFT/ prediction of the inelastic contribution to the \2PE exchange in D. The error estimate in Ref.~\cite{Jentschura:2011is} seems to be too low by about a factor of 4; based on Eqs.~(17) and~(18) from Ref.~\cite{Mohr:2008fa}, one should get $0.08~\mathrm{kHz}$ instead of $0.02$~kHz. In addition, we think that the previously used work \cite{Friar:1997tr} might have underestimated the uncertainty.
The error estimate presented here is a conservative choice, given the smallness of higher-order terms (in particular, the effect of single-nucleon contributions). We also do not take into account the correlation between the proton contributions in H and D, which would lead to a reduction of the uncertainty estimate.

    \item Compensation of the Darwin-Foldy term for the deuteron:
    \beq
\Delta \nu_{10}=11.37\,\text{kHz}\qquad [11.37\,\text{kHz}].
\eeq
\end{enumerate}
 The total contribution to the isotope shift coming from this Set (ii) is given by:
\begin{align}
    \Delta f_\mathrm{ii} =\left[-4502.48 - 0.769\,l_1^{C0_S} \right]~\mathrm{kHz}= -4502.48(40)~\mathrm{kHz}\qquad [-4502.66(60)~\mathrm{kHz}],\eqlab{fii}
\end{align}
where we added the above uncertainties quadratically. 

Finally, the smallest correction comes from Set (iii) in Ref.~\cite{Jentschura:2011is}.  In the following, we will split the set and separate the contributions that we deduce directly  by scaling the leading non-relativistic nuclear-size correction:
\begin{align}
\Delta f_\mathrm{iv} = -\frac{7}{24\pi}\frac{\alpha^4 m_e^3 c^4}{\hbar^3}\left[\frac{r_d^2}{(1+m_e/M_d)^3}-\frac{r_p^2}{(1+m_e/M_p)^3}\right].\eqlab{fiv}
\end{align}
The higher-order nuclear-size corrections of Set (iii) are:
\begin{enumerate}
    \item[(b)] Elastic contribution to the \2PE exchange:
        \begin{align}
    \Delta\nu_{(b)} = \left[0.590 - 1.607\,l_1^{C0_S} \right]~\mathrm{kHz}=0.587(9)\,\text{kHz} \qquad [0.472\,\text{kHz}],\eqlab{isonub}
\end{align}
from combining our results in Eqs.~\eref{nubD} and \eref{nubH}. The previous result only included the dominant Friar-radius correction.
    \item[(c)] Relativistic higher-order corrections: 
    \begin{align}
    \Delta\nu_{(c)} = -2.029(41)\, \mathrm{kHz} \qquad [-2.828\,\text{kHz}].
\end{align}
Here, we use the recent update from Ref.~\cite{Pachucki:2018xxx}, including the $3\gamma$-exchange effects due to finite nuclear size.
\end{enumerate}
\begin{enumerate}
    \item[(d)] SE contribution to the nuclear size correction:
        \begin{align}
    \Delta\nu_{(d)} =\al^2 \left[4 \ln 2 -\frac{23}{4}\right]\,\Delta f_\mathrm{iv}=0.830 \, \mathrm{kHz} \qquad [0.828\,\text{kHz}].
\end{align}
To illustrate the numerical size of the effect, we are using the proton charge radius from the $\mu$H Lamb shift, $r_p(\mu\text{H})$ in \Eqref{rpreview}, and the CODATA 2018 recommended deuteron charge radius, $r_d(\text{CODATA '18})$ in \Eqref{CODATA18rd}.
    \item[(e)] Electronic VP contribution to the nuclear size correction:
        \begin{align}
    \Delta\nu_{(e)} =\frac{3\al^2}{4}\,\Delta f_\mathrm{iv}= -0.209\, \mathrm{kHz} \qquad [-0.209\,\text{kHz}].
\end{align}
\end{enumerate}
In total, Set (iii) amounts to:
\begin{subequations}
\eqlab{fiii}
\bea
    \Delta f_\mathrm{iii}  &=&\Delta\nu_{(b)}+\Delta\nu_{(c)}+\Delta\nu_{(d)}+\Delta\nu_{(e)} = -0.821(42)~\mathrm{kHz}\qquad [ -1.73~\mathrm{kHz}],
\eea
with
\bea
\Delta\nu_{(b)}+\Delta\nu_{(c)}&=&\left[-1.439 + 1.607\,l_1^{C0_S} \right]~\mathrm{kHz} = -1.442(42)~\mathrm{kHz}\qquad [ -2.356~\mathrm{kHz}],\\
 \Delta\nu_{(d)}+\Delta\nu_{(e)} &=&\al^2 \left[4 \ln 2 -5\right]\,\Delta f_\mathrm{iv}=0.621 \, \mathrm{kHz} \qquad [0.619\,\text{kHz}].
\eea
\end{subequations}
Here, we again assumed $r_p(\mu\text{H})$ and $r_d(\text{CODATA '18})$ to illustrate the numerical size of the effect.

Collecting Eqs.~\eref{fi}, \eref{fii}, \eref{fiii} and \eref{fiv}, our final result for the theoretical prediction of the $2S-1S$ H-D isotope shift reads:
\beq
f_\mathrm{iso}=\Delta f_\mathrm{i}+\Delta f_\mathrm{ii}+\Delta f_\mathrm{iii}+\Delta f_\mathrm{iv}=\left[671\,000\,534.811(411)(828)+ 0.838 \,l_1^{C0_S}-1369.346\, \left(\frac{r_d}{\mathrm{fm}}\right)^2\right]~ \mathrm{kHz},\eqlab{fisol1}
\eeq
where the first uncertainty is the combined uncertainty of $\Delta f_\mathrm{i}$-$\Delta f_\mathrm{iii}$, and the second uncertainty is due to $r_p(\mu\text{H})$ in $\Delta f_\mathrm{iv}$.

\section{Rydberg Constant from $2S-1S$ Transition in Hydrogen and Muonic Hydrogen Lamb Shift}\label{RydbergSection}

Analogously to Appendix \ref{app:iso}, we study the $S$-level transition in H:
\beq
E^\text{H} _{2S-1S}=h \, f^\text{H}_{2S-1S}.
\eeq
The individual contributions are listed in Table \ref{tab:Hshift}.  In total, we find:
\bea
f^\mathrm{H}_{2S-1S}&=&\Delta f^\mathrm{H}_\mathrm{i}+\Delta f^\mathrm{H}_\mathrm{ii}+\Delta f^\mathrm{H}_\mathrm{iii}+\Delta f^\mathrm{H}_\mathrm{iv}\nn\\
&=&\left[0.74960091418756\,\frac{R_\infty c}{\mathrm{Hz}}-7\,126\,781\,916(1\,813)-1\,368\,229\left(\frac{r_p}{\mathrm{fm}}\right)^2\right]\,\mathrm{Hz}.\eqlab{Htheory}
\eea
If not specified differently in Appendix \ref{app:iso}, we follow the CODATA procedure for the error estimate \cite{Mohr:2015ccw}. For $\Delta f_i$, we use the exact formula to deduce the uncertainty, see Ref.~\cite[Sec.~III C.]{Jentschura:2011is}. We can use \Eqref{Htheory} and $r_p(\mu\text{H})$ to extract the Rydberg constant from the measured transition \cite{Parthey:2011lfa}:
\beq
f^\mathrm{H}_{2S-1S}=2\,466\,061\,413\,187\,035(10)\,\mathrm{Hz}.
\eeq
Our result:
\beq
R_\infty c=3.289\,841\,960\,2509(27)\times 10^{15}\,\mathrm{Hz}, \eqlab{NewRydberg}
\eeq
is in perfect agreement with Ref.~\cite[Eq.~(22)]{Pohl:2016glp} and \Eqref{RydbergCodata18}. Compared to the latter, the uncertainty is more than a factor $2$ better. We can use this value to extract the deuteron radius from the $2S-1S$ transition in D, since it is largely driven by the measured Lamb shift in $\mu$H.

{\renewcommand{\arraystretch}{1}
\begin{table}[htb]
    \centering
    \begin{tabular}{l|rl}
    Contribution&\multicolumn{2}{|r}{Value in Hz}\\
    \hline
       \hline
 Dirac eigenvalue &$\Delta f^\mathrm{H}_\mathrm{i} =2\,466\,068\,540\,936\,672$&$(2\,026)$\\
 \hline
 \hline
One-loop SE and electronic VP&$\Delta \nu^\mathrm{H}_1=-7\,124\,094\,961$&$(1)$\\
Two-loop SE, electronic VP, and combined effects&$\Delta \nu^\mathrm{H}_2=-636\,881$&$(1\,733)$\\
Three-loop SE, electronic VP, and combined effects& $\Delta \nu^\mathrm{H}_3=-1\,509$&$(370)$\\
Salpeter recoil correction& $\Delta \nu^\mathrm{H}_4=-2\,068\,223$&\\
Higher-order pure recoil corrections& $\Delta \nu^\mathrm{H}_5=6\,354$&$(7)$\\
Radiative recoil corrections& $\Delta \nu^\mathrm{H}_6=10\,781$&$(74)$\\
Nuclear SE& $\Delta \nu^\mathrm{H}_7=-4\,034$&$(141)$\\
Muonic and hadronic VP& $\Delta \nu^\mathrm{H}_8=7\,410$&$(70)$\\
Nuclear polarizability correction& $\Delta \nu^\mathrm{H}_9=41$&$(16)$\\
\hline
Subtotal: Lamb shift contributions & $\Delta f^\mathrm{H}_\mathrm{ii} =-7\,126\,781\,023$&$(1\,780)$\\
\hline
\hline
Elastic \2PE&$ \Delta\nu^\mathrm{H}_{(b)}=35$&$(6)$\\
Relativistic higher-order corrections&$\Delta\nu^\mathrm{H}_{(c)}=-928$&$(344)$\\
SE contribution to the nuclear size correction&$\Delta\nu^\mathrm{H}_{(d)}=217.0\left(\frac{r_p}{\mathrm{fm}}\right)^2 $&\\
Electronic VP contribution to the nuclear size correction&$\Delta\nu^\mathrm{H}_{(e)}=-54.7\left(\frac{r_p}{\mathrm{fm}}\right)^2$&\\
\hline
 \multirow{ 2}{*}{Subtotal: Higher-order nuclear-size correction} & $\Delta\nu^\mathrm{H}_{(b)}+\Delta\nu^\mathrm{H}_{(c)} =-893$&$(344)$\\
 &$\Delta\nu^\mathrm{H}_{(d)}+\Delta\nu^\mathrm{H}_{(e)} =162.3\left(\frac{r_p}{\mathrm{fm}}\right)^2$&\\
\hline
\hline
Leading non-relativistic nuclear-size correction&$\Delta f^\mathrm{H}_\mathrm{iv} =-1\,368\,229 \left(\frac{r_p}{\mathrm{fm}}\right)^2$&
\end{tabular}
    \caption{Individual contributions to the $2S-1S$ transition in H with $R_\infty c$ from \Eqref{NewRydberg}.
    }
    \label{tab:Hshift}
\end{table}}

{\renewcommand{\arraystretch}{1}
\begin{table}[h]
    \centering
    \begin{tabular}{l|rl}
    Contribution&\multicolumn{2}{|r}{Value in Hz}\\
    \hline
       \hline
 Dirac eigenvalue &$\Delta f^\mathrm{D}_\mathrm{i} =2\,466\,739\,545\,007\,700$&$(2\,025)$\\
 \hline
 \hline
One-loop SE and electronic VP&$\Delta \nu^\mathrm{D}_1=-7\,129\,653\,960$&$(1)$\\
Two-loop SE, electronic VP, and combined effects&$\Delta \nu^\mathrm{D}_2=-637\,402$&$(1\,734)$\\
Three-loop SE, electronic VP, and combined effects& $\Delta \nu^\mathrm{D}_3=-1\,510$&$(370)$\\
Salpeter recoil correction& $\Delta \nu^\mathrm{D}_4=-1\,035\,575$&\\
Higher-order pure recoil corrections& $\Delta \nu^\mathrm{D}_5=3\,295$&$(3)$\\
Radiative recoil corrections& $\Delta \nu^\mathrm{D}_6=5\,398$&$(37)$\\
Nuclear SE& $\Delta \nu^\mathrm{D}_7=-1\,059$&$(35)$\\
Muonic and hadronic VP& $\Delta \nu^\mathrm{D}_8=7\,416$&$(70)$\\
Nuclear polarizability correction& $\Delta \nu^\mathrm{D}_9=18\,531$&$(204)$\\
Compensation of the Darwin-Foldy term for the deuteron& $\Delta \nu^\mathrm{D}_{10}=11\,369$&\\
\hline
Subtotal: Lamb shift contributions & $\Delta f^\mathrm{D}_\mathrm{ii} =-7\,131\,283\,498$&$(1\,787)$\\
\hline
\hline
Elastic \2PE&$ \Delta\nu^\mathrm{D}_{(b)}=622$&$(7)$\\
Relativistic higher-order corrections&$\Delta\nu^\mathrm{D}_{(c)}=-2\,961$&$(357)$\\
SE contribution to the nuclear size correction&$\Delta\nu^\mathrm{D}_{(d)}=217.1\left(\frac{r_d}{\mathrm{fm}}\right)^2 $&\\
Electronic VP contribution to the nuclear size correction&$\Delta\nu^\mathrm{D}_{(e)}=-54.7\left(\frac{r_d}{\mathrm{fm}}\right)^2$&\\
\hline
 \multirow{ 2}{*}{Subtotal: Higher-order nuclear-size correction} & $\Delta\nu^\mathrm{D}_{(b)}+\Delta\nu^\mathrm{D}_{(c)} =-2\,340$&$(357)$\\
 &$\Delta\nu^\mathrm{D}_{(d)}+\Delta\nu^\mathrm{D}_{(e)} =162.4\left(\frac{r_d}{\mathrm{fm}}\right)^2$&\\
\hline
\hline
Leading non-relativistic nuclear-size correction&$\Delta f^\mathrm{D}_\mathrm{iv} =-1\,369\,346 \left(\frac{r_d}{\mathrm{fm}}\right)^2$&
\end{tabular}
    \caption{Individual contributions to the $2S-1S$ transition in D with $R_\infty c$ from \Eqref{NewRydberg}.
    }
    \label{tab:Dshift}
\end{table}}

\section{$2S-1S$ Transition in Deuterium}\label{1S2SD}

Analogously to Appendix \ref{RydbergSection}, we study the $S$-level transition in D:
\beq
E^\text{D}_{2S-1S}=h \, f^\text{D}_{2S-1S}.
\eeq
The individual contributions are listed in Table \ref{tab:Dshift}. For the $3\gamma$-exchange contribution, we use Ref.~\cite[Eq.~(104)]{Pachucki:2018xxx} with $r_d(\mu \text{D})$ from \Eqref{rdKalinowksi} and apply $10\,\%$ uncertainty for the inelastic part and $100\,\%$ for the single-nucleon part, where for the latter we also assume a correlation between $1S$ and $2S$ levels. The updated theory prediction for the $2S-1S$ transition in D, including the \2PE exchange from \piEFT/, then reads:
\beq
f^\mathrm{D}_{2S-1S}=\Delta f^\mathrm{D}_\mathrm{i}+\Delta f^\mathrm{D}_\mathrm{ii}+\Delta f^\mathrm{D}_\mathrm{iii}+\Delta f^\mathrm{D}_\mathrm{iv}=\left[2\,466\,732\,413\,721\,862(2\,724)-1\,369\,346\left(\frac{r_d}{\mathrm{fm}}\right)^2\right]\,\mathrm{Hz}.\eqlab{Dtheory}
\eeq
This is used in Section \ref{sec:radiid} to extract the deuteron charge radius.

\section{Neutron Charge Radius and Deuteron Structure Radius}
\label{app:neutroncr}

Another interesting quantity is the deuteron structure radius, defined as:
\beq
r_\mathrm{str}^2=r_d^2-2\,r_0^2.
\eeq
This definition implies that the difference of squared deuteron and proton charge radii is related to the sum of squared deuteron structure radius and neutron charge radius: 
\beq
r_d^2-r_p^2=r_\mathrm{str}^2+\frac{3}{4M_p^2} + r_n^2. \eqlab{neutronradiusdsum}
\eeq
The deuteron structure radius has recently been predicted in $\chi$ET \cite{Filin:2020tcs}:
\beq
r_\mathrm{str}(\chi\mathrm{ET})=1.9729\left({}^{+15}_{-12}\right)\,\mathrm{fm}.\eqlab{LECstr}
\eeq
We can use this value as an alternative reference to fix the unknown LEC from \Eqref{rdl1}:
\begin{align}
    l_1^{C0_S}=-2.41\left({}^{+32}_{-26}\right)(35)\times 10^{-3}=-2.41\left({}^{+48}_{-44}\right)\times 10^{-3},
\end{align}
where the first error is due to $r_\mathrm{str}$ and the second is due to $Z$. 
Using in addition the $2S-1S$ H-D isotope shift in \Eqref{fisol1}, we extract the neutron charge radius as:
\beq
r_n^2=-0.105\left({}^{+5}_{-6}\right)\,\mathrm{fm}^2,
\eeq
in exact agreement with Ref.~\cite{Filin:2020tcs}, but in slight disagreement with the value used by us, see \Eqref{rnOLD}. 
We checked that the effect of the nucleon charge radii entering our \piEFT/ prediction of the \2PE exchange adds up to $0.021$ kHz. 
Since the size of these nucleon-radius contributions is covered by the uncertainty budget in \Eqref{theoryPredictionISO}, we can safely ignore this correlation in the above extraction of $r_n$. Note that even though the value of $l_1^{C0_S}$ in \Eqref{LECstr} agrees well with \Eqref{contact_term_value}, the corresponding values for $r_d$ disagree, since it also depends on $r_n$ through $r_0$, cf.\  \Eqref{rdl1}.


\begin{thebibliography}{91}%
\makeatletter
\providecommand \@ifxundefined [1]{%
 \@ifx{#1\undefined}
}%
\providecommand \@ifnum [1]{%
 \ifnum #1\expandafter \@firstoftwo
 \else \expandafter \@secondoftwo
 \fi
}%
\providecommand \@ifx [1]{%
 \ifx #1\expandafter \@firstoftwo
 \else \expandafter \@secondoftwo
 \fi
}%
\providecommand \natexlab [1]{#1}%
\providecommand \enquote  [1]{``#1''}%
\providecommand \bibnamefont  [1]{#1}%
\providecommand \bibfnamefont [1]{#1}%
\providecommand \citenamefont [1]{#1}%
\providecommand \href@noop [0]{\@secondoftwo}%
\providecommand \href [0]{\begingroup \@sanitize@url \@href}%
\providecommand \@href[1]{\@@startlink{#1}\@@href}%
\providecommand \@@href[1]{\endgroup#1\@@endlink}%
\providecommand \@sanitize@url [0]{\catcode `\\12\catcode `\$12\catcode
  `\&12\catcode `\#12\catcode `\^12\catcode `\_12\catcode `\%12\relax}%
\providecommand \@@startlink[1]{}%
\providecommand \@@endlink[0]{}%
\providecommand \url  [0]{\begingroup\@sanitize@url \@url }%
\providecommand \@url [1]{\endgroup\@href {#1}{\urlprefix }}%
\providecommand \urlprefix  [0]{URL }%
\providecommand \Eprint [0]{\href }%
\providecommand \doibase [0]{https://doi.org/}%
\providecommand \selectlanguage [0]{\@gobble}%
\providecommand \bibinfo  [0]{\@secondoftwo}%
\providecommand \bibfield  [0]{\@secondoftwo}%
\providecommand \translation [1]{[#1]}%
\providecommand \BibitemOpen [0]{}%
\providecommand \bibitemStop [0]{}%
\providecommand \bibitemNoStop [0]{.\EOS\space}%
\providecommand \EOS [0]{\spacefactor3000\relax}%
\providecommand \BibitemShut  [1]{\csname bibitem#1\endcsname}%
\let\auto@bib@innerbib\@empty
\bibitem [{\citenamefont {Pohl}\ \emph {et~al.}(2010)\citenamefont {Pohl} \emph
  {et~al.}}]{Pohl:2010zza}%
  \BibitemOpen
  \bibfield  {author} {\bibinfo {author} {\bibfnamefont {R.}~\bibnamefont
  {Pohl}} \emph {et~al.},\ }\bibfield  {title} {\bibinfo {title} {{The size of
  the proton}},\ }\href {https://doi.org/10.1038/nature09250} {\bibfield
  {journal} {\bibinfo  {journal} {Nature}\ }\textbf {\bibinfo {volume} {466}},\
  \bibinfo {pages} {213} (\bibinfo {year} {2010})}\BibitemShut {NoStop}%
\bibitem [{\citenamefont {Antognini}\ \emph {et~al.}(2013)\citenamefont
  {Antognini} \emph {et~al.}}]{Antognini:1900ns}%
  \BibitemOpen
  \bibfield  {author} {\bibinfo {author} {\bibfnamefont {A.}~\bibnamefont
  {Antognini}} \emph {et~al.},\ }\bibfield  {title} {\bibinfo {title} {{Proton
  Structure from the Measurement of $2S-2P$ Transition Frequencies of Muonic
  Hydrogen}},\ }\href {https://doi.org/10.1126/science.1230016} {\bibfield
  {journal} {\bibinfo  {journal} {Science}\ }\textbf {\bibinfo {volume}
  {339}},\ \bibinfo {pages} {417} (\bibinfo {year} {2013})}\BibitemShut
  {NoStop}%
\bibitem [{\citenamefont {Pohl}\ \emph {et~al.}(2016)\citenamefont {Pohl} \emph
  {et~al.}}]{Pohl1:2016xoo}%
  \BibitemOpen
  \bibfield  {author} {\bibinfo {author} {\bibfnamefont {R.}~\bibnamefont
  {Pohl}} \emph {et~al.} (\bibinfo {collaboration} {CREMA}),\ }\bibfield
  {title} {\bibinfo {title} {{Laser spectroscopy of muonic deuterium}},\ }\href
  {https://doi.org/10.1126/science.aaf2468} {\bibfield  {journal} {\bibinfo
  {journal} {Science}\ }\textbf {\bibinfo {volume} {353}},\ \bibinfo {pages}
  {669} (\bibinfo {year} {2016})}\BibitemShut {NoStop}%
\bibitem [{\citenamefont {Tiesinga}\ \emph {et~al.}(2021)\citenamefont
  {Tiesinga}, \citenamefont {Mohr}, \citenamefont {Newell},\ and\ \citenamefont
  {Taylor}}]{Tiesinga:2021myr}%
  \BibitemOpen
  \bibfield  {author} {\bibinfo {author} {\bibfnamefont {E.}~\bibnamefont
  {Tiesinga}}, \bibinfo {author} {\bibfnamefont {P.~J.}\ \bibnamefont {Mohr}},
  \bibinfo {author} {\bibfnamefont {D.~B.}\ \bibnamefont {Newell}},\ and\
  \bibinfo {author} {\bibfnamefont {B.~N.}\ \bibnamefont {Taylor}},\ }\bibfield
   {title} {\bibinfo {title} {{CODATA recommended values of the fundamental
  physical constants: 2018*}},\ }\href
  {https://doi.org/10.1103/RevModPhys.93.025010} {\bibfield  {journal}
  {\bibinfo  {journal} {Rev. Mod. Phys.}\ }\textbf {\bibinfo {volume} {93}},\
  \bibinfo {pages} {025010} (\bibinfo {year} {2021})}\BibitemShut {NoStop}%
\bibitem [{\citenamefont {Krauth}\ \emph {et~al.}(2016)\citenamefont {Krauth},
  \citenamefont {Diepold}, \citenamefont {Franke}, \citenamefont {Antognini},
  \citenamefont {Kottmann},\ and\ \citenamefont {Pohl}}]{Krauth:2015nja}%
  \BibitemOpen
  \bibfield  {author} {\bibinfo {author} {\bibfnamefont {J.~J.}\ \bibnamefont
  {Krauth}}, \bibinfo {author} {\bibfnamefont {M.}~\bibnamefont {Diepold}},
  \bibinfo {author} {\bibfnamefont {B.}~\bibnamefont {Franke}}, \bibinfo
  {author} {\bibfnamefont {A.}~\bibnamefont {Antognini}}, \bibinfo {author}
  {\bibfnamefont {F.}~\bibnamefont {Kottmann}},\ and\ \bibinfo {author}
  {\bibfnamefont {R.}~\bibnamefont {Pohl}},\ }\bibfield  {title} {\bibinfo
  {title} {{Theory of the n=2 levels in muonic deuterium}},\ }\href
  {https://doi.org/10.1016/j.aop.2015.12.006} {\bibfield  {journal} {\bibinfo
  {journal} {Annals Phys.}\ }\textbf {\bibinfo {volume} {366}},\ \bibinfo
  {pages} {168} (\bibinfo {year} {2016})},\ \Eprint
  {https://arxiv.org/abs/1506.01298} {arXiv:1506.01298 [physics.atom-ph]}
  \BibitemShut {NoStop}%
\bibitem [{\citenamefont {Kalinowski}(2019)}]{Kalinowski:2018rmf}%
  \BibitemOpen
  \bibfield  {author} {\bibinfo {author} {\bibfnamefont {M.}~\bibnamefont
  {Kalinowski}},\ }\bibfield  {title} {\bibinfo {title} {{Deuteron charge
  radius from the Lamb-shift measurement in muonic deuterium}},\ }\href
  {https://doi.org/10.1103/PhysRevA.99.030501} {\bibfield  {journal} {\bibinfo
  {journal} {Phys. Rev. A}\ }\textbf {\bibinfo {volume} {99}},\ \bibinfo
  {pages} {030501} (\bibinfo {year} {2019})},\ \Eprint
  {https://arxiv.org/abs/1812.10993} {arXiv:1812.10993 [physics.atom-ph]}
  \BibitemShut {NoStop}%
\bibitem [{\citenamefont {Pachucki}\ \emph {et~al.}(2018)\citenamefont
  {Pachucki}, \citenamefont {Patk\'o\v{s}},\ and\ \citenamefont
  {Yerokhin}}]{Pachucki:2018yxe}%
  \BibitemOpen
  \bibfield  {author} {\bibinfo {author} {\bibfnamefont {K.}~\bibnamefont
  {Pachucki}}, \bibinfo {author} {\bibfnamefont {V.}~\bibnamefont
  {Patk\'o\v{s}}},\ and\ \bibinfo {author} {\bibfnamefont {V.~A.}\ \bibnamefont
  {Yerokhin}},\ }\bibfield  {title} {\bibinfo {title} {{Three-photon exchange
  nuclear structure correction in hydrogenic systems}},\ }\href
  {https://doi.org/10.1103/PhysRevA.97.062511} {\bibfield  {journal} {\bibinfo
  {journal} {Phys. Rev. A}\ }\textbf {\bibinfo {volume} {97}},\ \bibinfo
  {pages} {062511} (\bibinfo {year} {2018})},\ \Eprint
  {https://arxiv.org/abs/1803.10313} {arXiv:1803.10313 [physics.atom-ph]}
  \BibitemShut {NoStop}%
\bibitem [{\citenamefont {Pohl}\ \emph {et~al.}(2017)\citenamefont {Pohl} \emph
  {et~al.}}]{Pohl:2016glp}%
  \BibitemOpen
  \bibfield  {author} {\bibinfo {author} {\bibfnamefont {R.}~\bibnamefont
  {Pohl}} \emph {et~al.},\ }\bibfield  {title} {\bibinfo {title} {{Deuteron
  charge radius and Rydberg constant from spectroscopy data in atomic
  deuterium}},\ }\href {https://doi.org/10.1088/1681-7575/aa4e59} {\bibfield
  {journal} {\bibinfo  {journal} {Metrologia}\ }\textbf {\bibinfo {volume}
  {54}},\ \bibinfo {pages} {L1} (\bibinfo {year} {2017})},\ \Eprint
  {https://arxiv.org/abs/1607.03165} {arXiv:1607.03165 [physics.atom-ph]}
  \BibitemShut {NoStop}%
\bibitem [{\citenamefont {Sick}\ and\ \citenamefont
  {Trautmann}(1998)}]{Sick:1998cvq}%
  \BibitemOpen
  \bibfield  {author} {\bibinfo {author} {\bibfnamefont {I.}~\bibnamefont
  {Sick}}\ and\ \bibinfo {author} {\bibfnamefont {D.}~\bibnamefont
  {Trautmann}},\ }\bibfield  {title} {\bibinfo {title} {{On the rms radius of
  the deuteron}},\ }\href {https://doi.org/10.1016/S0375-9474(98)00334-0}
  {\bibfield  {journal} {\bibinfo  {journal} {Nucl. Phys. A}\ }\textbf
  {\bibinfo {volume} {637}},\ \bibinfo {pages} {559} (\bibinfo {year}
  {1998})}\BibitemShut {NoStop}%
\bibitem [{\citenamefont {Mohr}\ \emph {et~al.}(2016)\citenamefont {Mohr},
  \citenamefont {Newell},\ and\ \citenamefont {Taylor}}]{Mohr:2015ccw}%
  \BibitemOpen
  \bibfield  {author} {\bibinfo {author} {\bibfnamefont {P.~J.}\ \bibnamefont
  {Mohr}}, \bibinfo {author} {\bibfnamefont {D.~B.}\ \bibnamefont {Newell}},\
  and\ \bibinfo {author} {\bibfnamefont {B.~N.}\ \bibnamefont {Taylor}},\
  }\bibfield  {title} {\bibinfo {title} {{CODATA Recommended Values of the
  Fundamental Physical Constants: 2014}},\ }\href
  {https://doi.org/10.1103/RevModPhys.88.035009} {\bibfield  {journal}
  {\bibinfo  {journal} {Rev. Mod. Phys.}\ }\textbf {\bibinfo {volume} {88}},\
  \bibinfo {pages} {035009} (\bibinfo {year} {2016})},\ \Eprint
  {https://arxiv.org/abs/1507.07956} {arXiv:1507.07956 [physics.atom-ph]}
  \BibitemShut {NoStop}%
\bibitem [{\citenamefont {Kaplan}\ \emph {et~al.}(1996)\citenamefont {Kaplan},
  \citenamefont {Savage},\ and\ \citenamefont {Wise}}]{Kaplan:1996xu}%
  \BibitemOpen
  \bibfield  {author} {\bibinfo {author} {\bibfnamefont {D.~B.}\ \bibnamefont
  {Kaplan}}, \bibinfo {author} {\bibfnamefont {M.~J.}\ \bibnamefont {Savage}},\
  and\ \bibinfo {author} {\bibfnamefont {M.~B.}\ \bibnamefont {Wise}},\
  }\bibfield  {title} {\bibinfo {title} {{Nucleon-nucleon scattering from
  effective field theory}},\ }\href
  {https://doi.org/10.1016/0550-3213(96)00357-4} {\bibfield  {journal}
  {\bibinfo  {journal} {Nucl. Phys. B}\ }\textbf {\bibinfo {volume} {478}},\
  \bibinfo {pages} {629} (\bibinfo {year} {1996})},\ \Eprint
  {https://arxiv.org/abs/nucl-th/9605002} {arXiv:nucl-th/9605002} \BibitemShut
  {NoStop}%
\bibitem [{\citenamefont {Kaplan}\ \emph
  {et~al.}(1998{\natexlab{a}})\citenamefont {Kaplan}, \citenamefont {Savage},\
  and\ \citenamefont {Wise}}]{Kaplan:1998we}%
  \BibitemOpen
  \bibfield  {author} {\bibinfo {author} {\bibfnamefont {D.~B.}\ \bibnamefont
  {Kaplan}}, \bibinfo {author} {\bibfnamefont {M.~J.}\ \bibnamefont {Savage}},\
  and\ \bibinfo {author} {\bibfnamefont {M.~B.}\ \bibnamefont {Wise}},\
  }\bibfield  {title} {\bibinfo {title} {{Two nucleon systems from effective
  field theory}},\ }\href {https://doi.org/10.1016/S0550-3213(98)00440-4}
  {\bibfield  {journal} {\bibinfo  {journal} {Nucl. Phys. B}\ }\textbf
  {\bibinfo {volume} {534}},\ \bibinfo {pages} {329} (\bibinfo {year}
  {1998}{\natexlab{a}})},\ \Eprint {https://arxiv.org/abs/nucl-th/9802075}
  {arXiv:nucl-th/9802075} \BibitemShut {NoStop}%
\bibitem [{\citenamefont {Kaplan}\ \emph
  {et~al.}(1998{\natexlab{b}})\citenamefont {Kaplan}, \citenamefont {Savage},\
  and\ \citenamefont {Wise}}]{Kaplan:1998tg}%
  \BibitemOpen
  \bibfield  {author} {\bibinfo {author} {\bibfnamefont {D.~B.}\ \bibnamefont
  {Kaplan}}, \bibinfo {author} {\bibfnamefont {M.~J.}\ \bibnamefont {Savage}},\
  and\ \bibinfo {author} {\bibfnamefont {M.~B.}\ \bibnamefont {Wise}},\
  }\bibfield  {title} {\bibinfo {title} {{A New expansion for nucleon-nucleon
  interactions}},\ }\href {https://doi.org/10.1016/S0370-2693(98)00210-X}
  {\bibfield  {journal} {\bibinfo  {journal} {Phys. Lett. B}\ }\textbf
  {\bibinfo {volume} {424}},\ \bibinfo {pages} {390} (\bibinfo {year}
  {1998}{\natexlab{b}})},\ \Eprint {https://arxiv.org/abs/nucl-th/9801034}
  {arXiv:nucl-th/9801034} \BibitemShut {NoStop}%
\bibitem [{\citenamefont {Chen}\ and\ \citenamefont
  {Savage}(1999)}]{Chen:1999bg}%
  \BibitemOpen
  \bibfield  {author} {\bibinfo {author} {\bibfnamefont {J.-W.}\ \bibnamefont
  {Chen}}\ and\ \bibinfo {author} {\bibfnamefont {M.~J.}\ \bibnamefont
  {Savage}},\ }\bibfield  {title} {\bibinfo {title} {{$np\to d\gamma$ for big
  bang nucleosynthesis}},\ }\href {https://doi.org/10.1103/PhysRevC.60.065205}
  {\bibfield  {journal} {\bibinfo  {journal} {Phys. Rev. C}\ }\textbf {\bibinfo
  {volume} {60}},\ \bibinfo {pages} {065205} (\bibinfo {year} {1999})},\
  \Eprint {https://arxiv.org/abs/nucl-th/9907042} {arXiv:nucl-th/9907042}
  \BibitemShut {NoStop}%
\bibitem [{\citenamefont {Chen}\ \emph {et~al.}(1999)\citenamefont {Chen},
  \citenamefont {Rupak},\ and\ \citenamefont {Savage}}]{Chen:1999tn}%
  \BibitemOpen
  \bibfield  {author} {\bibinfo {author} {\bibfnamefont {J.-W.}\ \bibnamefont
  {Chen}}, \bibinfo {author} {\bibfnamefont {G.}~\bibnamefont {Rupak}},\ and\
  \bibinfo {author} {\bibfnamefont {M.~J.}\ \bibnamefont {Savage}},\ }\bibfield
   {title} {\bibinfo {title} {{Nucleon-nucleon effective field theory without
  pions}},\ }\href {https://doi.org/10.1016/S0375-9474(99)00298-5} {\bibfield
  {journal} {\bibinfo  {journal} {Nucl. Phys. A}\ }\textbf {\bibinfo {volume}
  {653}},\ \bibinfo {pages} {386} (\bibinfo {year} {1999})},\ \Eprint
  {https://arxiv.org/abs/nucl-th/9902056} {arXiv:nucl-th/9902056} \BibitemShut
  {NoStop}%
\bibitem [{\citenamefont {Rupak}(2000)}]{Rupak:1999rk}%
  \BibitemOpen
  \bibfield  {author} {\bibinfo {author} {\bibfnamefont {G.}~\bibnamefont
  {Rupak}},\ }\bibfield  {title} {\bibinfo {title} {{Precision calculation of
  $np\to d\gamma$ cross-section for big bang nucleosynthesis}},\ }\href
  {https://doi.org/10.1016/S0375-9474(00)00323-7} {\bibfield  {journal}
  {\bibinfo  {journal} {Nucl. Phys. A}\ }\textbf {\bibinfo {volume} {678}},\
  \bibinfo {pages} {405} (\bibinfo {year} {2000})},\ \Eprint
  {https://arxiv.org/abs/nucl-th/9911018} {arXiv:nucl-th/9911018} \BibitemShut
  {NoStop}%
\bibitem [{\citenamefont {Phillips}\ \emph {et~al.}(2000)\citenamefont
  {Phillips}, \citenamefont {Rupak},\ and\ \citenamefont
  {Savage}}]{Phillips:1999hh}%
  \BibitemOpen
  \bibfield  {author} {\bibinfo {author} {\bibfnamefont {D.~R.}\ \bibnamefont
  {Phillips}}, \bibinfo {author} {\bibfnamefont {G.}~\bibnamefont {Rupak}},\
  and\ \bibinfo {author} {\bibfnamefont {M.~J.}\ \bibnamefont {Savage}},\
  }\bibfield  {title} {\bibinfo {title} {{Improving the convergence of N N
  effective field theory}},\ }\href
  {https://doi.org/10.1016/S0370-2693(99)01496-3} {\bibfield  {journal}
  {\bibinfo  {journal} {Phys. Lett. B}\ }\textbf {\bibinfo {volume} {473}},\
  \bibinfo {pages} {209} (\bibinfo {year} {2000})},\ \Eprint
  {https://arxiv.org/abs/nucl-th/9908054} {arXiv:nucl-th/9908054} \BibitemShut
  {NoStop}%
\bibitem [{\citenamefont {Grie{\ss}hammer}\ and\ \citenamefont
  {Rupak}(2002)}]{Griesshammer:2000mi}%
  \BibitemOpen
  \bibfield  {author} {\bibinfo {author} {\bibfnamefont {H.~W.}\ \bibnamefont
  {Grie{\ss}hammer}}\ and\ \bibinfo {author} {\bibfnamefont {G.}~\bibnamefont
  {Rupak}},\ }\bibfield  {title} {\bibinfo {title} {{Nucleon polarizabilities
  from Compton scattering on the deuteron}},\ }\href
  {https://doi.org/10.1016/S0370-2693(02)01238-8} {\bibfield  {journal}
  {\bibinfo  {journal} {Phys. Lett. B}\ }\textbf {\bibinfo {volume} {529}},\
  \bibinfo {pages} {57} (\bibinfo {year} {2002})},\ \Eprint
  {https://arxiv.org/abs/nucl-th/0012096} {arXiv:nucl-th/0012096} \BibitemShut
  {NoStop}%
\bibitem [{\citenamefont {Chen}\ \emph {et~al.}(2005)\citenamefont {Chen},
  \citenamefont {Ji},\ and\ \citenamefont {Li}}]{Chen:2004wv}%
  \BibitemOpen
  \bibfield  {author} {\bibinfo {author} {\bibfnamefont {J.-W.}\ \bibnamefont
  {Chen}}, \bibinfo {author} {\bibfnamefont {X.-d.}\ \bibnamefont {Ji}},\ and\
  \bibinfo {author} {\bibfnamefont {Y.-c.}\ \bibnamefont {Li}},\ }\bibfield
  {title} {\bibinfo {title} {{Deuteron Compton scattering in effective field
  theory and spin-independent nucleon polarizabilities}},\ }\href
  {https://doi.org/10.1016/j.physletb.2005.06.001} {\bibfield  {journal}
  {\bibinfo  {journal} {Phys. Lett. B}\ }\textbf {\bibinfo {volume} {620}},\
  \bibinfo {pages} {33} (\bibinfo {year} {2005})},\ \Eprint
  {https://arxiv.org/abs/nucl-th/0408003} {arXiv:nucl-th/0408003} \BibitemShut
  {NoStop}%
\bibitem [{\citenamefont {Furnstahl}\ \emph {et~al.}(2015)\citenamefont
  {Furnstahl}, \citenamefont {Klco}, \citenamefont {Phillips},\ and\
  \citenamefont {Wesolowski}}]{Furnstahl:2015rha}%
  \BibitemOpen
  \bibfield  {author} {\bibinfo {author} {\bibfnamefont {R.~J.}\ \bibnamefont
  {Furnstahl}}, \bibinfo {author} {\bibfnamefont {N.}~\bibnamefont {Klco}},
  \bibinfo {author} {\bibfnamefont {D.~R.}\ \bibnamefont {Phillips}},\ and\
  \bibinfo {author} {\bibfnamefont {S.}~\bibnamefont {Wesolowski}},\ }\bibfield
   {title} {\bibinfo {title} {{Quantifying truncation errors in effective field
  theory}},\ }\href {https://doi.org/10.1103/PhysRevC.92.024005} {\bibfield
  {journal} {\bibinfo  {journal} {Phys. Rev. C}\ }\textbf {\bibinfo {volume}
  {92}},\ \bibinfo {pages} {024005} (\bibinfo {year} {2015})},\ \Eprint
  {https://arxiv.org/abs/1506.01343} {arXiv:1506.01343 [nucl-th]} \BibitemShut
  {NoStop}%
\bibitem [{\citenamefont {Lensky}\ \emph {et~al.}(2021)\citenamefont {Lensky},
  \citenamefont {Hiller~Blin},\ and\ \citenamefont
  {Pascalutsa}}]{Lensky:2021VVCS}%
  \BibitemOpen
  \bibfield  {author} {\bibinfo {author} {\bibfnamefont {V.}~\bibnamefont
  {Lensky}}, \bibinfo {author} {\bibfnamefont {A.}~\bibnamefont
  {Hiller~Blin}},\ and\ \bibinfo {author} {\bibfnamefont {V.}~\bibnamefont
  {Pascalutsa}},\ }\bibfield  {title} {\bibinfo {title} {{Forward
  doubly-virtual Compton scattering off an unpolarized deuteron in pionless
  effective field theory}},\ }\href
  {https://doi.org/10.1103/PhysRevC.104.054003} {\bibfield  {journal} {\bibinfo
   {journal} {Phys. Rev. C}\ }\textbf {\bibinfo {volume} {104}},\ \bibinfo
  {pages} {054003} (\bibinfo {year} {2021})},\ \Eprint
  {https://arxiv.org/abs/2109.08223} {arXiv:2109.08223 [nucl-th]} \BibitemShut
  {NoStop}%
\bibitem [{\citenamefont {Lensky}\ \emph
  {et~al.}(2022{\natexlab{a}})\citenamefont {Lensky}, \citenamefont
  {Hagelstein},\ and\ \citenamefont {Pascalutsa}}]{Lensky:2022fif}%
  \BibitemOpen
  \bibfield  {author} {\bibinfo {author} {\bibfnamefont {V.}~\bibnamefont
  {Lensky}}, \bibinfo {author} {\bibfnamefont {F.}~\bibnamefont {Hagelstein}},\
  and\ \bibinfo {author} {\bibfnamefont {V.}~\bibnamefont {Pascalutsa}},\
  }\bibfield  {title} {\bibinfo {title} {{A reassessment of nuclear effects in
  muonic deuterium using pionless effective field theory at N3LO}},\ }\href
  {https://doi.org/10.1016/j.physletb.2022.137500} {\bibfield  {journal}
  {\bibinfo  {journal} {Phys. Lett. B}\ }\textbf {\bibinfo {volume} {835}},\
  \bibinfo {pages} {137500} (\bibinfo {year} {2022}{\natexlab{a}})},\ \Eprint
  {https://arxiv.org/abs/2206.14066} {arXiv:2206.14066 [nucl-th]} \BibitemShut
  {NoStop}%
\bibitem [{\citenamefont {Carlson}\ and\ \citenamefont
  {Vanderhaeghen}(2011)}]{Carlson:2011zd}%
  \BibitemOpen
  \bibfield  {author} {\bibinfo {author} {\bibfnamefont {C.~E.}\ \bibnamefont
  {Carlson}}\ and\ \bibinfo {author} {\bibfnamefont {M.}~\bibnamefont
  {Vanderhaeghen}},\ }\bibfield  {title} {\bibinfo {title} {{Higher order
  proton structure corrections to the Lamb shift in muonic hydrogen}},\ }\href
  {https://doi.org/10.1103/PhysRevA.84.020102} {\bibfield  {journal} {\bibinfo
  {journal} {Phys. Rev. A}\ }\textbf {\bibinfo {volume} {84}},\ \bibinfo
  {pages} {020102} (\bibinfo {year} {2011})},\ \Eprint
  {https://arxiv.org/abs/1101.5965} {arXiv:1101.5965 [hep-ph]} \BibitemShut
  {NoStop}%
\bibitem [{\citenamefont {Carlson}\ \emph {et~al.}(2014)\citenamefont
  {Carlson}, \citenamefont {Gorchtein},\ and\ \citenamefont
  {Vanderhaeghen}}]{Carlson:2013xea}%
  \BibitemOpen
  \bibfield  {author} {\bibinfo {author} {\bibfnamefont {C.~E.}\ \bibnamefont
  {Carlson}}, \bibinfo {author} {\bibfnamefont {M.}~\bibnamefont {Gorchtein}},\
  and\ \bibinfo {author} {\bibfnamefont {M.}~\bibnamefont {Vanderhaeghen}},\
  }\bibfield  {title} {\bibinfo {title} {{Nuclear structure contribution to the
  Lamb shift in muonic deuterium}},\ }\href
  {https://doi.org/10.1103/PhysRevA.89.022504} {\bibfield  {journal} {\bibinfo
  {journal} {Phys. Rev. A}\ }\textbf {\bibinfo {volume} {89}},\ \bibinfo
  {pages} {022504} (\bibinfo {year} {2014})},\ \Eprint
  {https://arxiv.org/abs/1311.6512} {arXiv:1311.6512 [nucl-th]} \BibitemShut
  {NoStop}%
\bibitem [{\citenamefont {Lensky}\ \emph
  {et~al.}(2022{\natexlab{b}})\citenamefont {Lensky}, \citenamefont
  {Hagelstein}, \citenamefont {Blin},\ and\ \citenamefont
  {Pascalutsa}}]{Lensky:2022tue}%
  \BibitemOpen
  \bibfield  {author} {\bibinfo {author} {\bibfnamefont {V.}~\bibnamefont
  {Lensky}}, \bibinfo {author} {\bibfnamefont {F.}~\bibnamefont {Hagelstein}},
  \bibinfo {author} {\bibfnamefont {A.~H.}\ \bibnamefont {Blin}},\ and\
  \bibinfo {author} {\bibfnamefont {V.}~\bibnamefont {Pascalutsa}},\ }\bibfield
   {title} {\bibinfo {title} {{Deuteron VVCS and nuclear structure effects in
  muonic deuterium at N3LO in pionless EFT}},\ }in\ \href@noop {} {\emph
  {\bibinfo {booktitle} {{10th International workshop on Chiral Dynamics}}}}\
  (\bibinfo {year} {2022})\ \Eprint {https://arxiv.org/abs/2203.13030}
  {arXiv:2203.13030 [nucl-th]} \BibitemShut {NoStop}%
\bibitem [{\citenamefont {Filin}\ \emph {et~al.}(2020)\citenamefont {Filin},
  \citenamefont {Baru}, \citenamefont {Epelbaum}, \citenamefont {Krebs},
  \citenamefont {M\"oller},\ and\ \citenamefont {Reinert}}]{Filin:2019eoe}%
  \BibitemOpen
  \bibfield  {author} {\bibinfo {author} {\bibfnamefont {A.~A.}\ \bibnamefont
  {Filin}}, \bibinfo {author} {\bibfnamefont {V.}~\bibnamefont {Baru}},
  \bibinfo {author} {\bibfnamefont {E.}~\bibnamefont {Epelbaum}}, \bibinfo
  {author} {\bibfnamefont {H.}~\bibnamefont {Krebs}}, \bibinfo {author}
  {\bibfnamefont {D.}~\bibnamefont {M\"oller}},\ and\ \bibinfo {author}
  {\bibfnamefont {P.}~\bibnamefont {Reinert}},\ }\bibfield  {title} {\bibinfo
  {title} {{Extraction of the neutron charge radius from a precision
  calculation of the deuteron structure radius}},\ }\href
  {https://doi.org/10.1103/PhysRevLett.124.082501} {\bibfield  {journal}
  {\bibinfo  {journal} {Phys. Rev. Lett.}\ }\textbf {\bibinfo {volume} {124}},\
  \bibinfo {pages} {082501} (\bibinfo {year} {2020})},\ \Eprint
  {https://arxiv.org/abs/1911.04877} {arXiv:1911.04877 [nucl-th]} \BibitemShut
  {NoStop}%
\bibitem [{\citenamefont {Filin}\ \emph {et~al.}(2021)\citenamefont {Filin},
  \citenamefont {M\"oller}, \citenamefont {Baru}, \citenamefont {Epelbaum},
  \citenamefont {Krebs},\ and\ \citenamefont {Reinert}}]{Filin:2020tcs}%
  \BibitemOpen
  \bibfield  {author} {\bibinfo {author} {\bibfnamefont {A.~A.}\ \bibnamefont
  {Filin}}, \bibinfo {author} {\bibfnamefont {D.}~\bibnamefont {M\"oller}},
  \bibinfo {author} {\bibfnamefont {V.}~\bibnamefont {Baru}}, \bibinfo {author}
  {\bibfnamefont {E.}~\bibnamefont {Epelbaum}}, \bibinfo {author}
  {\bibfnamefont {H.}~\bibnamefont {Krebs}},\ and\ \bibinfo {author}
  {\bibfnamefont {P.}~\bibnamefont {Reinert}},\ }\bibfield  {title} {\bibinfo
  {title} {{High-accuracy calculation of the deuteron charge and quadrupole
  form factors in chiral effective field theory}},\ }\href
  {https://doi.org/10.1103/PhysRevC.103.024313} {\bibfield  {journal} {\bibinfo
   {journal} {Phys. Rev. C}\ }\textbf {\bibinfo {volume} {103}},\ \bibinfo
  {pages} {024313} (\bibinfo {year} {2021})},\ \Eprint
  {https://arxiv.org/abs/2009.08911} {arXiv:2009.08911 [nucl-th]} \BibitemShut
  {NoStop}%
\bibitem [{\citenamefont {Abbott}\ \emph {et~al.}(2000)\citenamefont {Abbott}
  \emph {et~al.}}]{Abbott:2000ak}%
  \BibitemOpen
  \bibfield  {author} {\bibinfo {author} {\bibfnamefont {D.}~\bibnamefont
  {Abbott}} \emph {et~al.} (\bibinfo {collaboration} {JLAB t20}),\ }\bibfield
  {title} {\bibinfo {title} {{Phenomenology of the deuteron electromagnetic
  form-factors}},\ }\href {https://doi.org/10.1007/PL00013629} {\bibfield
  {journal} {\bibinfo  {journal} {Eur. Phys. J.}\ }\textbf {\bibinfo {volume}
  {A7}},\ \bibinfo {pages} {421} (\bibinfo {year} {2000})},\ \Eprint
  {https://arxiv.org/abs/nucl-ex/0002003} {arXiv:nucl-ex/0002003 [nucl-ex]}
  \BibitemShut {NoStop}%
\bibitem [{\citenamefont {Acharya}\ \emph {et~al.}(2021)\citenamefont
  {Acharya}, \citenamefont {Lensky}, \citenamefont {Bacca}, \citenamefont
  {Gorchtein},\ and\ \citenamefont {Vanderhaeghen}}]{Acharya:2020bxf}%
  \BibitemOpen
  \bibfield  {author} {\bibinfo {author} {\bibfnamefont {B.}~\bibnamefont
  {Acharya}}, \bibinfo {author} {\bibfnamefont {V.}~\bibnamefont {Lensky}},
  \bibinfo {author} {\bibfnamefont {S.}~\bibnamefont {Bacca}}, \bibinfo
  {author} {\bibfnamefont {M.}~\bibnamefont {Gorchtein}},\ and\ \bibinfo
  {author} {\bibfnamefont {M.}~\bibnamefont {Vanderhaeghen}},\ }\bibfield
  {title} {\bibinfo {title} {{Dispersive evaluation of the Lamb shift in muonic
  deuterium from chiral effective field theory}},\ }\href
  {https://doi.org/10.1103/PhysRevC.103.024001} {\bibfield  {journal} {\bibinfo
   {journal} {Phys. Rev. C}\ }\textbf {\bibinfo {volume} {103}},\ \bibinfo
  {pages} {024001} (\bibinfo {year} {2021})},\ \Eprint
  {https://arxiv.org/abs/2010.11155} {arXiv:2010.11155 [nucl-th]} \BibitemShut
  {NoStop}%
\bibitem [{\citenamefont {Hagelstein}\ \emph {et~al.}(2016)\citenamefont
  {Hagelstein}, \citenamefont {Miskimen},\ and\ \citenamefont
  {Pascalutsa}}]{Hagelstein:2015egb}%
  \BibitemOpen
  \bibfield  {author} {\bibinfo {author} {\bibfnamefont {F.}~\bibnamefont
  {Hagelstein}}, \bibinfo {author} {\bibfnamefont {R.}~\bibnamefont
  {Miskimen}},\ and\ \bibinfo {author} {\bibfnamefont {V.}~\bibnamefont
  {Pascalutsa}},\ }\bibfield  {title} {\bibinfo {title} {{Nucleon
  Polarizabilities: from Compton Scattering to Hydrogen Atom}},\ }\href
  {https://doi.org/10.1016/j.ppnp.2015.12.001} {\bibfield  {journal} {\bibinfo
  {journal} {Prog. Part. Nucl. Phys.}\ }\textbf {\bibinfo {volume} {88}},\
  \bibinfo {pages} {29} (\bibinfo {year} {2016})},\ \Eprint
  {https://arxiv.org/abs/1512.03765} {arXiv:1512.03765 [nucl-th]} \BibitemShut
  {NoStop}%
\bibitem [{\citenamefont {Zhou}\ \emph {et~al.}(2021)\citenamefont {Zhou} \emph
  {et~al.}}]{Zhou:2020cdt}%
  \BibitemOpen
  \bibfield  {author} {\bibinfo {author} {\bibfnamefont {J.}~\bibnamefont
  {Zhou}} \emph {et~al.},\ }\bibfield  {title} {\bibinfo {title} {{Advanced
  extraction of the deuteron charge radius from electron-deuteron scattering
  data}},\ }\href {https://doi.org/10.1103/PhysRevC.103.024002} {\bibfield
  {journal} {\bibinfo  {journal} {Phys. Rev. C}\ }\textbf {\bibinfo {volume}
  {103}},\ \bibinfo {pages} {024002} (\bibinfo {year} {2021})},\ \Eprint
  {https://arxiv.org/abs/2010.09003} {arXiv:2010.09003 [nucl-ex]} \BibitemShut
  {NoStop}%
\bibitem [{\citenamefont {Hernandez}\ \emph {et~al.}(2019)\citenamefont
  {Hernandez}, \citenamefont {Ji}, \citenamefont {Bacca},\ and\ \citenamefont
  {Barnea}}]{Hernandez:2019zcm}%
  \BibitemOpen
  \bibfield  {author} {\bibinfo {author} {\bibfnamefont {O.~J.}\ \bibnamefont
  {Hernandez}}, \bibinfo {author} {\bibfnamefont {C.}~\bibnamefont {Ji}},
  \bibinfo {author} {\bibfnamefont {S.}~\bibnamefont {Bacca}},\ and\ \bibinfo
  {author} {\bibfnamefont {N.}~\bibnamefont {Barnea}},\ }\bibfield  {title}
  {\bibinfo {title} {{Probing uncertainties of nuclear structure corrections in
  light muonic atoms}},\ }\href {https://doi.org/10.1103/PhysRevC.100.064315}
  {\bibfield  {journal} {\bibinfo  {journal} {Phys. Rev. C}\ }\textbf {\bibinfo
  {volume} {100}},\ \bibinfo {pages} {064315} (\bibinfo {year} {2019})},\
  \Eprint {https://arxiv.org/abs/1909.05717} {arXiv:1909.05717 [nucl-th]}
  \BibitemShut {NoStop}%
\bibitem [{\citenamefont {Emmons}\ \emph {et~al.}(2021)\citenamefont {Emmons},
  \citenamefont {Ji},\ and\ \citenamefont {Platter}}]{Emmons:2020aov}%
  \BibitemOpen
  \bibfield  {author} {\bibinfo {author} {\bibfnamefont {S.~B.}\ \bibnamefont
  {Emmons}}, \bibinfo {author} {\bibfnamefont {C.}~\bibnamefont {Ji}},\ and\
  \bibinfo {author} {\bibfnamefont {L.}~\bibnamefont {Platter}},\ }\bibfield
  {title} {\bibinfo {title} {{Pionless Effective Field Theory Evaluation of
  Nuclear Polarizability in Muonic Deuterium}},\ }\href
  {https://doi.org/10.1088/1361-6471/abcb58} {\bibfield  {journal} {\bibinfo
  {journal} {J. Phys. G}\ }\textbf {\bibinfo {volume} {48}},\ \bibinfo {pages}
  {035101} (\bibinfo {year} {2021})},\ \Eprint
  {https://arxiv.org/abs/2009.08347} {arXiv:2009.08347 [nucl-th]} \BibitemShut
  {NoStop}%
\bibitem [{\citenamefont {Ji}\ \emph {et~al.}(2018)\citenamefont {Ji},
  \citenamefont {Bacca}, \citenamefont {Barnea}, \citenamefont {Hernandez},\
  and\ \citenamefont {Nevo-Dinur}}]{Ji:2018ozm}%
  \BibitemOpen
  \bibfield  {author} {\bibinfo {author} {\bibfnamefont {C.}~\bibnamefont
  {Ji}}, \bibinfo {author} {\bibfnamefont {S.}~\bibnamefont {Bacca}}, \bibinfo
  {author} {\bibfnamefont {N.}~\bibnamefont {Barnea}}, \bibinfo {author}
  {\bibfnamefont {O.~J.}\ \bibnamefont {Hernandez}},\ and\ \bibinfo {author}
  {\bibfnamefont {N.}~\bibnamefont {Nevo-Dinur}},\ }\bibfield  {title}
  {\bibinfo {title} {{\textit{Ab initio} calculation of nuclear structure
  corrections in muonic atoms}},\ }\href
  {https://doi.org/10.1088/1361-6471/aad3eb} {\bibfield  {journal} {\bibinfo
  {journal} {J. Phys. G}\ }\textbf {\bibinfo {volume} {45}},\ \bibinfo {pages}
  {093002} (\bibinfo {year} {2018})},\ \Eprint
  {https://arxiv.org/abs/1806.03101} {arXiv:1806.03101 [nucl-th]} \BibitemShut
  {NoStop}%
\bibitem [{\citenamefont {Borah}\ \emph {et~al.}(2020)\citenamefont {Borah},
  \citenamefont {Hill}, \citenamefont {Lee},\ and\ \citenamefont
  {Tomalak}}]{Borah:2020gte}%
  \BibitemOpen
  \bibfield  {author} {\bibinfo {author} {\bibfnamefont {K.}~\bibnamefont
  {Borah}}, \bibinfo {author} {\bibfnamefont {R.~J.}\ \bibnamefont {Hill}},
  \bibinfo {author} {\bibfnamefont {G.}~\bibnamefont {Lee}},\ and\ \bibinfo
  {author} {\bibfnamefont {O.}~\bibnamefont {Tomalak}},\ }\bibfield  {title}
  {\bibinfo {title} {{Parameterization and applications of the low-$Q^2$
  nucleon vector form factors}},\ }\href
  {https://doi.org/10.1103/PhysRevD.102.074012} {\bibfield  {journal} {\bibinfo
   {journal} {Phys. Rev. D}\ }\textbf {\bibinfo {volume} {102}},\ \bibinfo
  {pages} {074012} (\bibinfo {year} {2020})},\ \Eprint
  {https://arxiv.org/abs/2003.13640} {arXiv:2003.13640 [hep-ph]} \BibitemShut
  {NoStop}%
\bibitem [{\citenamefont {Kopecky}\ \emph {et~al.}(1995)\citenamefont
  {Kopecky}, \citenamefont {Riehs}, \citenamefont {Harvey},\ and\ \citenamefont
  {Hill}}]{Kopecky:1995zz}%
  \BibitemOpen
  \bibfield  {author} {\bibinfo {author} {\bibfnamefont {S.}~\bibnamefont
  {Kopecky}}, \bibinfo {author} {\bibfnamefont {P.}~\bibnamefont {Riehs}},
  \bibinfo {author} {\bibfnamefont {J.~A.}\ \bibnamefont {Harvey}},\ and\
  \bibinfo {author} {\bibfnamefont {N.~W.}\ \bibnamefont {Hill}},\ }\bibfield
  {title} {\bibinfo {title} {{New Measurement of the Charge Radius of the
  Neutron}},\ }\href {https://doi.org/10.1103/PhysRevLett.74.2427} {\bibfield
  {journal} {\bibinfo  {journal} {Phys. Rev. Lett.}\ }\textbf {\bibinfo
  {volume} {74}},\ \bibinfo {pages} {2427} (\bibinfo {year}
  {1995})}\BibitemShut {NoStop}%
\bibitem [{\citenamefont {Kopecky}\ \emph {et~al.}(1997)\citenamefont
  {Kopecky}, \citenamefont {Krenn}, \citenamefont {Riehs}, \citenamefont
  {Steiner}, \citenamefont {Harvey}, \citenamefont {Hill},\ and\ \citenamefont
  {Pernicka}}]{Kopecky:1997rw}%
  \BibitemOpen
  \bibfield  {author} {\bibinfo {author} {\bibfnamefont {S.}~\bibnamefont
  {Kopecky}}, \bibinfo {author} {\bibfnamefont {M.}~\bibnamefont {Krenn}},
  \bibinfo {author} {\bibfnamefont {P.}~\bibnamefont {Riehs}}, \bibinfo
  {author} {\bibfnamefont {S.}~\bibnamefont {Steiner}}, \bibinfo {author}
  {\bibfnamefont {J.~A.}\ \bibnamefont {Harvey}}, \bibinfo {author}
  {\bibfnamefont {N.~W.}\ \bibnamefont {Hill}},\ and\ \bibinfo {author}
  {\bibfnamefont {M.}~\bibnamefont {Pernicka}},\ }\bibfield  {title} {\bibinfo
  {title} {{Neutron charge radius determined from the energy dependence of the
  neutron transmission of liquid Pb-208 and Bi-209}},\ }\href
  {https://doi.org/10.1103/PhysRevC.56.2229} {\bibfield  {journal} {\bibinfo
  {journal} {Phys. Rev. C}\ }\textbf {\bibinfo {volume} {56}},\ \bibinfo
  {pages} {2229} (\bibinfo {year} {1997})}\BibitemShut {NoStop}%
\bibitem [{\citenamefont {Bradford}\ \emph {et~al.}(2006)\citenamefont
  {Bradford}, \citenamefont {Bodek}, \citenamefont {Budd},\ and\ \citenamefont
  {Arrington}}]{Bradford:2006yz}%
  \BibitemOpen
  \bibfield  {author} {\bibinfo {author} {\bibfnamefont {R.}~\bibnamefont
  {Bradford}}, \bibinfo {author} {\bibfnamefont {A.}~\bibnamefont {Bodek}},
  \bibinfo {author} {\bibfnamefont {H.~S.}\ \bibnamefont {Budd}},\ and\
  \bibinfo {author} {\bibfnamefont {J.}~\bibnamefont {Arrington}},\ }\bibfield
  {title} {\bibinfo {title} {{A New parameterization of the nucleon elastic
  form-factors}},\ }\href {https://doi.org/10.1016/j.nuclphysbps.2006.08.028}
  {\bibfield  {journal} {\bibinfo  {journal} {Nucl. Phys. B Proc. Suppl.}\
  }\textbf {\bibinfo {volume} {159}},\ \bibinfo {pages} {127} (\bibinfo {year}
  {2006})},\ \Eprint {https://arxiv.org/abs/hep-ex/0602017}
  {arXiv:hep-ex/0602017} \BibitemShut {NoStop}%
\bibitem [{\citenamefont {Alarcon}\ \emph {et~al.}(2014)\citenamefont
  {Alarcon}, \citenamefont {Lensky},\ and\ \citenamefont
  {Pascalutsa}}]{Alarcon:2013cba}%
  \BibitemOpen
  \bibfield  {author} {\bibinfo {author} {\bibfnamefont {J.~M.}\ \bibnamefont
  {Alarcon}}, \bibinfo {author} {\bibfnamefont {V.}~\bibnamefont {Lensky}},\
  and\ \bibinfo {author} {\bibfnamefont {V.}~\bibnamefont {Pascalutsa}},\
  }\bibfield  {title} {\bibinfo {title} {{Chiral perturbation theory of muonic
  hydrogen Lamb shift: polarizability contribution}},\ }\href
  {https://doi.org/10.1140/epjc/s10052-014-2852-0} {\bibfield  {journal}
  {\bibinfo  {journal} {Eur. Phys. J. C}\ }\textbf {\bibinfo {volume} {74}},\
  \bibinfo {pages} {2852} (\bibinfo {year} {2014})},\ \Eprint
  {https://arxiv.org/abs/1312.1219} {arXiv:1312.1219 [hep-ph]} \BibitemShut
  {NoStop}%
\bibitem [{\citenamefont {Lensky}\ \emph {et~al.}(2018)\citenamefont {Lensky},
  \citenamefont {Hagelstein}, \citenamefont {Pascalutsa},\ and\ \citenamefont
  {Vanderhaeghen}}]{Lensky:2017bwi}%
  \BibitemOpen
  \bibfield  {author} {\bibinfo {author} {\bibfnamefont {V.}~\bibnamefont
  {Lensky}}, \bibinfo {author} {\bibfnamefont {F.}~\bibnamefont {Hagelstein}},
  \bibinfo {author} {\bibfnamefont {V.}~\bibnamefont {Pascalutsa}},\ and\
  \bibinfo {author} {\bibfnamefont {M.}~\bibnamefont {Vanderhaeghen}},\
  }\bibfield  {title} {\bibinfo {title} {{Sum rules across the unpolarized
  Compton processes involving generalized polarizabilities and moments of
  nucleon structure functions}},\ }\href
  {https://doi.org/10.1103/PhysRevD.97.074012} {\bibfield  {journal} {\bibinfo
  {journal} {Phys. Rev. D}\ }\textbf {\bibinfo {volume} {97}},\ \bibinfo
  {pages} {074012} (\bibinfo {year} {2018})},\ \Eprint
  {https://arxiv.org/abs/1712.03886} {arXiv:1712.03886 [hep-ph]} \BibitemShut
  {NoStop}%
\bibitem [{\citenamefont {Tomalak}(2019)}]{Tomalak:2018uhr}%
  \BibitemOpen
  \bibfield  {author} {\bibinfo {author} {\bibfnamefont {O.}~\bibnamefont
  {Tomalak}},\ }\bibfield  {title} {\bibinfo {title} {{Two-Photon Exchange
  Correction to the Lamb Shift and Hyperfine Splitting of S Levels}},\ }\href
  {https://doi.org/10.1140/epja/i2019-12743-1} {\bibfield  {journal} {\bibinfo
  {journal} {Eur. Phys. J. A}\ }\textbf {\bibinfo {volume} {55}},\ \bibinfo
  {pages} {64} (\bibinfo {year} {2019})},\ \Eprint
  {https://arxiv.org/abs/1808.09204} {arXiv:1808.09204 [hep-ph]} \BibitemShut
  {NoStop}%
\bibitem [{\citenamefont {Eskin}\ \emph {et~al.}(2016)\citenamefont {Eskin},
  \citenamefont {Faustov}, \citenamefont {Martynenko},\ and\ \citenamefont
  {Martynenko}}]{Eskin:2015brf}%
  \BibitemOpen
  \bibfield  {author} {\bibinfo {author} {\bibfnamefont {A.~V.}\ \bibnamefont
  {Eskin}}, \bibinfo {author} {\bibfnamefont {R.~N.}\ \bibnamefont {Faustov}},
  \bibinfo {author} {\bibfnamefont {A.~P.}\ \bibnamefont {Martynenko}},\ and\
  \bibinfo {author} {\bibfnamefont {F.~A.}\ \bibnamefont {Martynenko}},\
  }\bibfield  {title} {\bibinfo {title} {{Hadronic deuteron polarizability
  contribution to the Lamb shift in muonic deuterium}},\ }\href
  {https://doi.org/10.1142/S0217732316501042} {\bibfield  {journal} {\bibinfo
  {journal} {Mod. Phys. Lett. A}\ }\textbf {\bibinfo {volume} {31}},\ \bibinfo
  {pages} {1650104} (\bibinfo {year} {2016})},\ \Eprint
  {https://arxiv.org/abs/1511.02477} {arXiv:1511.02477 [hep-ph]} \BibitemShut
  {NoStop}%
\bibitem [{\citenamefont {Birse}\ and\ \citenamefont
  {McGovern}(2012)}]{Birse:2012eb}%
  \BibitemOpen
  \bibfield  {author} {\bibinfo {author} {\bibfnamefont {M.~C.}\ \bibnamefont
  {Birse}}\ and\ \bibinfo {author} {\bibfnamefont {J.~A.}\ \bibnamefont
  {McGovern}},\ }\bibfield  {title} {\bibinfo {title} {{Proton polarisability
  contribution to the Lamb shift in muonic hydrogen at fourth order in chiral
  perturbation theory}},\ }\href {https://doi.org/10.1140/epja/i2012-12120-8}
  {\bibfield  {journal} {\bibinfo  {journal} {Eur. Phys. J. A}\ }\textbf
  {\bibinfo {volume} {48}},\ \bibinfo {pages} {120} (\bibinfo {year} {2012})},\
  \Eprint {https://arxiv.org/abs/1206.3030} {arXiv:1206.3030 [hep-ph]}
  \BibitemShut {NoStop}%
\bibitem [{\citenamefont {Peset}\ and\ \citenamefont
  {Pineda}(2014)}]{Peset:2014jxa}%
  \BibitemOpen
  \bibfield  {author} {\bibinfo {author} {\bibfnamefont {C.}~\bibnamefont
  {Peset}}\ and\ \bibinfo {author} {\bibfnamefont {A.}~\bibnamefont {Pineda}},\
  }\bibfield  {title} {\bibinfo {title} {{The two-photon exchange contribution
  to muonic hydrogen from chiral perturbation theory}},\ }\href
  {https://doi.org/10.1016/j.nuclphysb.2014.07.027} {\bibfield  {journal}
  {\bibinfo  {journal} {Nucl. Phys. B}\ }\textbf {\bibinfo {volume} {887}},\
  \bibinfo {pages} {69} (\bibinfo {year} {2014})},\ \Eprint
  {https://arxiv.org/abs/1406.4524} {arXiv:1406.4524 [hep-ph]} \BibitemShut
  {NoStop}%
\bibitem [{\citenamefont {Jentschura}\ \emph {et~al.}(2011)\citenamefont
  {Jentschura}, \citenamefont {Matveev}, \citenamefont {Parthey}, \citenamefont
  {Alnis}, \citenamefont {Pohl}, \citenamefont {Udem}, \citenamefont
  {Kolachevsky},\ and\ \citenamefont {H\"ansch}}]{Jentschura:2011is}%
  \BibitemOpen
  \bibfield  {author} {\bibinfo {author} {\bibfnamefont {U.}~\bibnamefont
  {Jentschura}}, \bibinfo {author} {\bibfnamefont {A.}~\bibnamefont {Matveev}},
  \bibinfo {author} {\bibfnamefont {C.}~\bibnamefont {Parthey}}, \bibinfo
  {author} {\bibfnamefont {J.}~\bibnamefont {Alnis}}, \bibinfo {author}
  {\bibfnamefont {R.}~\bibnamefont {Pohl}}, \bibinfo {author} {\bibfnamefont
  {T.}~\bibnamefont {Udem}}, \bibinfo {author} {\bibfnamefont {N.}~\bibnamefont
  {Kolachevsky}},\ and\ \bibinfo {author} {\bibfnamefont {T.}~\bibnamefont
  {H\"ansch}},\ }\bibfield  {title} {\bibinfo {title} {Hydrogen-deuterium
  isotope shift: From the {$1S\ensuremath{-}2S$}-transition frequency to the
  proton-deuteron charge-radius difference},\ }\href
  {https://doi.org/10.1103/PhysRevA.83.042505} {\bibfield  {journal} {\bibinfo
  {journal} {Phys. Rev. A}\ }\textbf {\bibinfo {volume} {83}},\ \bibinfo
  {pages} {042505} (\bibinfo {year} {2011})}\BibitemShut {NoStop}%
\bibitem [{\citenamefont {Gorchtein}(2021)}]{Gorchtein:2020prv}%
  \BibitemOpen
  \bibfield  {author} {\bibinfo {author} {\bibfnamefont {M.}~\bibnamefont
  {Gorchtein}},\ }\href@noop {} {\bibinfo {title} {private communication}}
  (\bibinfo {year} {2021})\BibitemShut {NoStop}%
\bibitem [{\citenamefont {Pachucki}(2011)}]{Pachucki:2011xr}%
  \BibitemOpen
  \bibfield  {author} {\bibinfo {author} {\bibfnamefont {K.}~\bibnamefont
  {Pachucki}},\ }\bibfield  {title} {\bibinfo {title} {{Nuclear structure
  corrections in muonic deuterium}},\ }\href
  {https://doi.org/10.1103/PhysRevLett.106.193007} {\bibfield  {journal}
  {\bibinfo  {journal} {Phys. Rev. Lett.}\ }\textbf {\bibinfo {volume} {106}},\
  \bibinfo {pages} {193007} (\bibinfo {year} {2011})},\ \Eprint
  {https://arxiv.org/abs/1102.3296} {arXiv:1102.3296 [hep-ph]} \BibitemShut
  {NoStop}%
\bibitem [{\citenamefont {Mohr}\ \emph {et~al.}(2008)\citenamefont {Mohr},
  \citenamefont {Taylor},\ and\ \citenamefont {Newell}}]{Mohr:2008fa}%
  \BibitemOpen
  \bibfield  {author} {\bibinfo {author} {\bibfnamefont {P.~J.}\ \bibnamefont
  {Mohr}}, \bibinfo {author} {\bibfnamefont {B.~N.}\ \bibnamefont {Taylor}},\
  and\ \bibinfo {author} {\bibfnamefont {D.~B.}\ \bibnamefont {Newell}},\
  }\bibfield  {title} {\bibinfo {title} {{CODATA Recommended Values of the
  Fundamental Physical Constants: 2006}},\ }\href
  {https://doi.org/10.1103/RevModPhys.80.633} {\bibfield  {journal} {\bibinfo
  {journal} {Rev. Mod. Phys.}\ }\textbf {\bibinfo {volume} {80}},\ \bibinfo
  {pages} {633} (\bibinfo {year} {2008})},\ \Eprint
  {https://arxiv.org/abs/0801.0028} {arXiv:0801.0028 [physics.atom-ph]}
  \BibitemShut {NoStop}%
\bibitem [{\citenamefont {Friar}\ and\ \citenamefont
  {Payne}(1997)}]{Friar:1997tr}%
  \BibitemOpen
  \bibfield  {author} {\bibinfo {author} {\bibfnamefont {J.~L.}\ \bibnamefont
  {Friar}}\ and\ \bibinfo {author} {\bibfnamefont {G.~L.}\ \bibnamefont
  {Payne}},\ }\bibfield  {title} {\bibinfo {title} {{Higher order nuclear
  polarizability corrections in atomic hydrogen}},\ }\href
  {https://doi.org/10.1103/PhysRevC.56.619} {\bibfield  {journal} {\bibinfo
  {journal} {Phys. Rev. C}\ }\textbf {\bibinfo {volume} {56}},\ \bibinfo
  {pages} {619} (\bibinfo {year} {1997})},\ \Eprint
  {https://arxiv.org/abs/nucl-th/9704032} {arXiv:nucl-th/9704032} \BibitemShut
  {NoStop}%
\bibitem [{\citenamefont {Khriplovich}\ and\ \citenamefont
  {Sen'kov}(2000)}]{Khriplovich:1999ak}%
  \BibitemOpen
  \bibfield  {author} {\bibinfo {author} {\bibfnamefont {I.~B.}\ \bibnamefont
  {Khriplovich}}\ and\ \bibinfo {author} {\bibfnamefont {R.~A.}\ \bibnamefont
  {Sen'kov}},\ }\bibfield  {title} {\bibinfo {title} {{Comment on `Proton
  polarization shifts in electronic and muonic hydrogen'}},\ }\href
  {https://doi.org/10.1016/S0370-2693(00)00497-4} {\bibfield  {journal}
  {\bibinfo  {journal} {Phys. Lett. B}\ }\textbf {\bibinfo {volume} {481}},\
  \bibinfo {pages} {447} (\bibinfo {year} {2000})},\ \Eprint
  {https://arxiv.org/abs/nucl-th/9903077} {arXiv:nucl-th/9903077} \BibitemShut
  {NoStop}%
\bibitem [{\citenamefont {Antognini}\ \emph {et~al.}(2022)\citenamefont
  {Antognini}, \citenamefont {Hagelstein},\ and\ \citenamefont
  {Pascalutsa}}]{Antognini:2022xoo}%
  \BibitemOpen
  \bibfield  {author} {\bibinfo {author} {\bibfnamefont {A.}~\bibnamefont
  {Antognini}}, \bibinfo {author} {\bibfnamefont {F.}~\bibnamefont
  {Hagelstein}},\ and\ \bibinfo {author} {\bibfnamefont {V.}~\bibnamefont
  {Pascalutsa}},\ }\bibfield  {title} {\bibinfo {title} {{The proton structure
  in and out of muonic hydrogen}},\ }\href
  {https://doi.org/10.1146/annurev-nucl-101920-024709} {\bibfield  {journal}
  {\bibinfo  {journal} {Ann.\ Rev.\ Nucl.\ Part.\ Sci.}\ }\textbf {\bibinfo
  {volume} {72}},\ \bibinfo {pages} {389} (\bibinfo {year} {2022})},\ \Eprint
  {https://arxiv.org/abs/2205.10076} {arXiv:2205.10076 [nucl-th]} \BibitemShut
  {NoStop}%
\bibitem [{\citenamefont {Korzinin}\ \emph {et~al.}(2013)\citenamefont
  {Korzinin}, \citenamefont {Ivanov},\ and\ \citenamefont
  {Karshenboim}}]{Korzinin:2013uia}%
  \BibitemOpen
  \bibfield  {author} {\bibinfo {author} {\bibfnamefont {E.~Y.}\ \bibnamefont
  {Korzinin}}, \bibinfo {author} {\bibfnamefont {V.~G.}\ \bibnamefont
  {Ivanov}},\ and\ \bibinfo {author} {\bibfnamefont {S.~G.}\ \bibnamefont
  {Karshenboim}},\ }\bibfield  {title} {\bibinfo {title}
  {{$\alpha^2(Z\alpha)^4m$ contributions to the Lamb shift and the fine
  structure in light muonic atoms}},\ }\href
  {https://doi.org/10.1103/PhysRevD.88.125019} {\bibfield  {journal} {\bibinfo
  {journal} {Phys. Rev. D}\ }\textbf {\bibinfo {volume} {88}},\ \bibinfo
  {pages} {125019} (\bibinfo {year} {2013})},\ \Eprint
  {https://arxiv.org/abs/1311.5784} {arXiv:1311.5784 [physics.atom-ph]}
  \BibitemShut {NoStop}%
\bibitem [{\citenamefont {Karshenboim}\ \emph {et~al.}(2015)\citenamefont
  {Karshenboim}, \citenamefont {Korzinin} \emph {et~al.}}]{Karshenboim:2015}%
  \BibitemOpen
  \bibfield  {author} {\bibinfo {author} {\bibfnamefont {S.~G.}\ \bibnamefont
  {Karshenboim}}, \bibinfo {author} {\bibfnamefont {E.~Y.}\ \bibnamefont
  {Korzinin}}, \emph {et~al.},\ }\bibfield  {title} {\bibinfo {title} {Theory
  of lamb shift in muonic hydrogen},\ }\href
  {https://doi.org/http://dx.doi.org/10.1063/1.4921197} {\bibfield  {journal}
  {\bibinfo  {journal} {J. Phys. Chem. Ref. Data}\ }\textbf {\bibinfo {volume}
  {44}},\ \bibinfo {eid} {031202} (\bibinfo {year} {2015})}\BibitemShut
  {NoStop}%
\bibitem [{\citenamefont {Karshenboim}\ \emph {et~al.}(2018)\citenamefont
  {Karshenboim}, \citenamefont {Korzinin}, \citenamefont {Shelyuto},\ and\
  \citenamefont {Ivanov}}]{Karshenboim:2018iyl}%
  \BibitemOpen
  \bibfield  {author} {\bibinfo {author} {\bibfnamefont {S.~G.}\ \bibnamefont
  {Karshenboim}}, \bibinfo {author} {\bibfnamefont {E.~Y.}\ \bibnamefont
  {Korzinin}}, \bibinfo {author} {\bibfnamefont {V.~A.}\ \bibnamefont
  {Shelyuto}},\ and\ \bibinfo {author} {\bibfnamefont {V.~G.}\ \bibnamefont
  {Ivanov}},\ }\bibfield  {title} {\bibinfo {title} {{$\alpha(Z \alpha)^5m$
  finite-nuclear-size contribution to the energy levels in light muonic
  atoms}},\ }\href {https://doi.org/10.1103/PhysRevA.98.062512} {\bibfield
  {journal} {\bibinfo  {journal} {Phys. Rev. A}\ }\textbf {\bibinfo {volume}
  {98}},\ \bibinfo {pages} {062512} (\bibinfo {year} {2018})}\BibitemShut
  {NoStop}%
\bibitem [{\citenamefont {Karshenboim}\ and\ \citenamefont
  {Shelyuto}(2021)}]{Karshenboim:2021jsc}%
  \BibitemOpen
  \bibfield  {author} {\bibinfo {author} {\bibfnamefont {S.~G.}\ \bibnamefont
  {Karshenboim}}\ and\ \bibinfo {author} {\bibfnamefont {V.~A.}\ \bibnamefont
  {Shelyuto}},\ }\bibfield  {title} {\bibinfo {title} {{Hadronic
  vacuum-polarization contribution to various QED observables}},\ }\href
  {https://doi.org/10.1140/epjd/s10053-021-00052-4} {\bibfield  {journal}
  {\bibinfo  {journal} {Eur. Phys. J. D}\ }\textbf {\bibinfo {volume} {75}},\
  \bibinfo {pages} {49} (\bibinfo {year} {2021})}\BibitemShut {NoStop}%
\bibitem [{\citenamefont {Epelbaum}\ \emph {et~al.}(2020)\citenamefont
  {Epelbaum}, \citenamefont {Krebs},\ and\ \citenamefont
  {Reinert}}]{Epelbaum:2019kcf}%
  \BibitemOpen
  \bibfield  {author} {\bibinfo {author} {\bibfnamefont {E.}~\bibnamefont
  {Epelbaum}}, \bibinfo {author} {\bibfnamefont {H.}~\bibnamefont {Krebs}},\
  and\ \bibinfo {author} {\bibfnamefont {P.}~\bibnamefont {Reinert}},\
  }\bibfield  {title} {\bibinfo {title} {{High-precision nuclear forces from
  chiral EFT: State-of-the-art, challenges and outlook}},\ }\href
  {https://doi.org/10.3389/fphy.2020.00098} {\bibfield  {journal} {\bibinfo
  {journal} {Front. in Phys.}\ }\textbf {\bibinfo {volume} {8}},\ \bibinfo
  {pages} {98} (\bibinfo {year} {2020})},\ \Eprint
  {https://arxiv.org/abs/1911.11875} {arXiv:1911.11875 [nucl-th]} \BibitemShut
  {NoStop}%
\bibitem [{\citenamefont {Korzinin}\ \emph {et~al.}(2018)\citenamefont
  {Korzinin}, \citenamefont {Shelyuto}, \citenamefont {Ivanov}, \citenamefont
  {Szafron},\ and\ \citenamefont {Karshenboim}}]{Korzinin:2018tnx}%
  \BibitemOpen
  \bibfield  {author} {\bibinfo {author} {\bibfnamefont {E.~Y.}\ \bibnamefont
  {Korzinin}}, \bibinfo {author} {\bibfnamefont {V.~A.}\ \bibnamefont
  {Shelyuto}}, \bibinfo {author} {\bibfnamefont {V.~G.}\ \bibnamefont
  {Ivanov}}, \bibinfo {author} {\bibfnamefont {R.}~\bibnamefont {Szafron}},\
  and\ \bibinfo {author} {\bibfnamefont {S.~G.}\ \bibnamefont {Karshenboim}},\
  }\bibfield  {title} {\bibinfo {title} {{Light-by-light-scattering
  contributions to the Lamb shift in light muonic atoms}},\ }\href
  {https://doi.org/10.1103/PhysRevA.98.062519} {\bibfield  {journal} {\bibinfo
  {journal} {Phys. Rev. A}\ }\textbf {\bibinfo {volume} {98}},\ \bibinfo
  {pages} {062519} (\bibinfo {year} {2018})},\ \Eprint
  {https://arxiv.org/abs/1812.09159} {arXiv:1812.09159 [physics.atom-ph]}
  \BibitemShut {NoStop}%
\bibitem [{\citenamefont {Wiringa}\ \emph {et~al.}(1995)\citenamefont
  {Wiringa}, \citenamefont {Stoks},\ and\ \citenamefont
  {Schiavilla}}]{Wiringa:1994wb}%
  \BibitemOpen
  \bibfield  {author} {\bibinfo {author} {\bibfnamefont {R.~B.}\ \bibnamefont
  {Wiringa}}, \bibinfo {author} {\bibfnamefont {V.~G.~J.}\ \bibnamefont
  {Stoks}},\ and\ \bibinfo {author} {\bibfnamefont {R.}~\bibnamefont
  {Schiavilla}},\ }\bibfield  {title} {\bibinfo {title} {{An Accurate
  nucleon-nucleon potential with charge independence breaking}},\ }\href
  {https://doi.org/10.1103/PhysRevC.51.38} {\bibfield  {journal} {\bibinfo
  {journal} {Phys. Rev. C}\ }\textbf {\bibinfo {volume} {51}},\ \bibinfo
  {pages} {38} (\bibinfo {year} {1995})},\ \Eprint
  {https://arxiv.org/abs/nucl-th/9408016} {arXiv:nucl-th/9408016} \BibitemShut
  {NoStop}%
\bibitem [{\citenamefont {Hernandez}\ \emph {et~al.}(2014)\citenamefont
  {Hernandez}, \citenamefont {Ji}, \citenamefont {Bacca}, \citenamefont
  {Nevo~Dinur},\ and\ \citenamefont {Barnea}}]{Hernandez:2014pwa}%
  \BibitemOpen
  \bibfield  {author} {\bibinfo {author} {\bibfnamefont {O.~J.}\ \bibnamefont
  {Hernandez}}, \bibinfo {author} {\bibfnamefont {C.}~\bibnamefont {Ji}},
  \bibinfo {author} {\bibfnamefont {S.}~\bibnamefont {Bacca}}, \bibinfo
  {author} {\bibfnamefont {N.}~\bibnamefont {Nevo~Dinur}},\ and\ \bibinfo
  {author} {\bibfnamefont {N.}~\bibnamefont {Barnea}},\ }\bibfield  {title}
  {\bibinfo {title} {{Improved estimates of the nuclear structure corrections
  in $\mu$D}},\ }\href {https://doi.org/10.1016/j.physletb.2014.07.039}
  {\bibfield  {journal} {\bibinfo  {journal} {Phys. Lett.}\ }\textbf {\bibinfo
  {volume} {B736}},\ \bibinfo {pages} {344} (\bibinfo {year} {2014})},\ \Eprint
  {https://arxiv.org/abs/1406.5230} {arXiv:1406.5230 [nucl-th]} \BibitemShut
  {NoStop}%
\bibitem [{\citenamefont {Udem}(1997)}]{Udem:1997}%
  \BibitemOpen
  \bibfield  {author} {\bibinfo {author} {\bibfnamefont {T.}~\bibnamefont
  {Udem}},\ }\emph {\bibinfo {title} {{Phasenkoh{\"a}rente optische
  Frequenzmessungen am Wasserstoffatom. Bestimmung der Rydberg–Konstanten und
  der $1S$ Lamb–Verschiebung}}},\ \href
  {http://www2.mpq.mpg.de/~thu/home/udem_doktor.pdf} {Ph.D. thesis},\ \bibinfo
  {school} {{Ludwig-Maximilians-Universität, Munich, Germany}} (\bibinfo
  {year} {1997})\BibitemShut {NoStop}%
\bibitem [{\citenamefont {Vermaseren}(2000)}]{Vermaseren:2000nd}%
  \BibitemOpen
  \bibfield  {author} {\bibinfo {author} {\bibfnamefont {J.~A.~M.}\
  \bibnamefont {Vermaseren}},\ }\href@noop {} {\bibinfo {title} {{New features
  of \textsc{FORM}}}} (\bibinfo {year} {2000}),\ \Eprint
  {https://arxiv.org/abs/math-ph/0010025} {arXiv:math-ph/0010025} \BibitemShut
  {NoStop}%
\bibitem [{\citenamefont {Caprio}(2005)}]{Caprio:2005dm}%
  \BibitemOpen
  \bibfield  {author} {\bibinfo {author} {\bibfnamefont {M.~A.}\ \bibnamefont
  {Caprio}},\ }\bibfield  {title} {\bibinfo {title} {{\textsc{LevelScheme}: A
  level scheme drawing and scientific figure preparation system for
  \textsc{Mathematica}}},\ }\href {https://doi.org/10.1016/j.cpc.2005.04.010}
  {\bibfield  {journal} {\bibinfo  {journal} {Comput. Phys. Commun.}\ }\textbf
  {\bibinfo {volume} {171}},\ \bibinfo {pages} {107} (\bibinfo {year}
  {2005})},\ \Eprint {https://arxiv.org/abs/physics/0505065}
  {arXiv:physics/0505065} \BibitemShut {NoStop}%
\bibitem [{\citenamefont {Coello~P\'erez}\ and\ \citenamefont
  {Papenbrock}(2015)}]{Perez:2015ufa}%
  \BibitemOpen
  \bibfield  {author} {\bibinfo {author} {\bibfnamefont {E.~A.}\ \bibnamefont
  {Coello~P\'erez}}\ and\ \bibinfo {author} {\bibfnamefont {T.}~\bibnamefont
  {Papenbrock}},\ }\bibfield  {title} {\bibinfo {title} {{Effective field
  theory for nuclear vibrations with quantified uncertainties}},\ }\href
  {https://doi.org/10.1103/PhysRevC.92.064309} {\bibfield  {journal} {\bibinfo
  {journal} {Phys. Rev. C}\ }\textbf {\bibinfo {volume} {92}},\ \bibinfo
  {pages} {064309} (\bibinfo {year} {2015})},\ \Eprint
  {https://arxiv.org/abs/1510.02401} {arXiv:1510.02401 [nucl-th]} \BibitemShut
  {NoStop}%
\bibitem [{\citenamefont {Epelbaum}\ \emph {et~al.}(2015)\citenamefont
  {Epelbaum}, \citenamefont {Krebs},\ and\ \citenamefont
  {Mei{\ss}ner}}]{Epelbaum:2014sza}%
  \BibitemOpen
  \bibfield  {author} {\bibinfo {author} {\bibfnamefont {E.}~\bibnamefont
  {Epelbaum}}, \bibinfo {author} {\bibfnamefont {H.}~\bibnamefont {Krebs}},\
  and\ \bibinfo {author} {\bibfnamefont {U.-G.}\ \bibnamefont {Mei{\ss}ner}},\
  }\bibfield  {title} {\bibinfo {title} {{Precision nucleon-nucleon potential
  at fifth order in the chiral expansion}},\ }\href
  {https://doi.org/10.1103/PhysRevLett.115.122301} {\bibfield  {journal}
  {\bibinfo  {journal} {Phys. Rev. Lett.}\ }\textbf {\bibinfo {volume} {115}},\
  \bibinfo {pages} {122301} (\bibinfo {year} {2015})},\ \Eprint
  {https://arxiv.org/abs/1412.4623} {arXiv:1412.4623 [nucl-th]} \BibitemShut
  {NoStop}%
\bibitem [{\citenamefont {Muli}\ \emph {et~al.}(2022)\citenamefont {Muli},
  \citenamefont {Acharya}, \citenamefont {Hernandez},\ and\ \citenamefont
  {Bacca}}]{LiMuli:2022kcy}%
  \BibitemOpen
  \bibfield  {author} {\bibinfo {author} {\bibfnamefont {S.~S.~L.}\
  \bibnamefont {Muli}}, \bibinfo {author} {\bibfnamefont {B.}~\bibnamefont
  {Acharya}}, \bibinfo {author} {\bibfnamefont {O.~J.}\ \bibnamefont
  {Hernandez}},\ and\ \bibinfo {author} {\bibfnamefont {S.}~\bibnamefont
  {Bacca}},\ }\bibfield  {title} {\bibinfo {title} {{Bayesian analysis of
  nuclear polarizability corrections to the Lamb shift of muonic H-atoms and
  He-ions}},\ }\href {https://doi.org/10.1088/1361-6471/ac81e0} {\bibfield
  {journal} {\bibinfo  {journal} {J. Phys. G}\ }\textbf {\bibinfo {volume}
  {49}},\ \bibinfo {pages} {105101} (\bibinfo {year} {2022})},\ \Eprint
  {https://arxiv.org/abs/2203.10792} {arXiv:2203.10792 [nucl-th]} \BibitemShut
  {NoStop}%
\bibitem [{\citenamefont {Berestetskii}\ \emph {et~al.}(1982)\citenamefont
  {Berestetskii}, \citenamefont {Lifshitz},\ and\ \citenamefont
  {Pitaevskii}}]{LandauLifshitz4}%
  \BibitemOpen
  \bibfield  {author} {\bibinfo {author} {\bibfnamefont {V.~B.}\ \bibnamefont
  {Berestetskii}}, \bibinfo {author} {\bibfnamefont {E.~M.}\ \bibnamefont
  {Lifshitz}},\ and\ \bibinfo {author} {\bibfnamefont {L.~P.}\ \bibnamefont
  {Pitaevskii}},\ }\href@noop {} {\emph {\bibinfo {title} {Course of
  Theoretical Physics, Vol. 4. Quantum Electrodynamics}}}\ (\bibinfo
  {publisher} {Pergamon Press},\ \bibinfo {year} {1982})\BibitemShut {NoStop}%
\bibitem [{\citenamefont {Martynenko}\ \emph {et~al.}(2014)\citenamefont
  {Martynenko}, \citenamefont {Krutov},\ and\ \citenamefont
  {Shamsutdinov}}]{Martynenko:2014bqa}%
  \BibitemOpen
  \bibfield  {author} {\bibinfo {author} {\bibfnamefont {A.~P.}\ \bibnamefont
  {Martynenko}}, \bibinfo {author} {\bibfnamefont {A.~A.}\ \bibnamefont
  {Krutov}},\ and\ \bibinfo {author} {\bibfnamefont {R.~N.}\ \bibnamefont
  {Shamsutdinov}},\ }\bibfield  {title} {\bibinfo {title} {{Theory of the lamb
  shift in muonic deuterium}},\ }\href
  {https://doi.org/10.1134/S1063778814060106} {\bibfield  {journal} {\bibinfo
  {journal} {Phys. Atom. Nucl.}\ }\textbf {\bibinfo {volume} {77}},\ \bibinfo
  {pages} {786} (\bibinfo {year} {2014})}\BibitemShut {NoStop}%
\bibitem [{\citenamefont {Yerokhin}\ \emph {et~al.}(2019)\citenamefont
  {Yerokhin}, \citenamefont {Pachucki},\ and\ \citenamefont
  {Patkóš}}]{Yerokhin2018}%
  \BibitemOpen
  \bibfield  {author} {\bibinfo {author} {\bibfnamefont {V.~A.}\ \bibnamefont
  {Yerokhin}}, \bibinfo {author} {\bibfnamefont {K.}~\bibnamefont {Pachucki}},\
  and\ \bibinfo {author} {\bibfnamefont {V.}~\bibnamefont {Patkóš}},\
  }\bibfield  {title} {\bibinfo {title} {Theory of the lamb shift in hydrogen
  and light hydrogen-like ions},\ }\href
  {https://doi.org/https://doi.org/10.1002/andp.201800324} {\bibfield
  {journal} {\bibinfo  {journal} {Annalen der Physik}\ }\textbf {\bibinfo
  {volume} {531}},\ \bibinfo {pages} {1800324} (\bibinfo {year}
  {2019})}\BibitemShut {NoStop}%
\bibitem [{\citenamefont {Zyla}\ \emph {et~al.}(2020)\citenamefont {Zyla} \emph
  {et~al.}}]{Zyla:2020zbs}%
  \BibitemOpen
  \bibfield  {author} {\bibinfo {author} {\bibfnamefont {P.}~\bibnamefont
  {Zyla}} \emph {et~al.} (\bibinfo {collaboration} {Particle Data Group}),\
  }\bibfield  {title} {\bibinfo {title} {{Review of Particle Physics}},\ }\href
  {https://doi.org/10.1093/ptep/ptaa104} {\bibfield  {journal} {\bibinfo
  {journal} {PTEP}\ }\textbf {\bibinfo {volume} {2020}},\ \bibinfo {pages}
  {083C01} (\bibinfo {year} {2020})}\BibitemShut {NoStop}%
\bibitem [{\citenamefont {Patra}\ \emph {et~al.}(2020)\citenamefont {Patra},
  \citenamefont {Germann}, \citenamefont {Karr}, \citenamefont {Haidar},
  \citenamefont {Hilico}, \citenamefont {Korobov}, \citenamefont {Cozijn},
  \citenamefont {Eikema}, \citenamefont {Ubachs},\ and\ \citenamefont
  {Koelemeij}}]{Patra2020}%
  \BibitemOpen
  \bibfield  {author} {\bibinfo {author} {\bibfnamefont {S.}~\bibnamefont
  {Patra}}, \bibinfo {author} {\bibfnamefont {M.}~\bibnamefont {Germann}},
  \bibinfo {author} {\bibfnamefont {J.-P.}\ \bibnamefont {Karr}}, \bibinfo
  {author} {\bibfnamefont {M.}~\bibnamefont {Haidar}}, \bibinfo {author}
  {\bibfnamefont {L.}~\bibnamefont {Hilico}}, \bibinfo {author} {\bibfnamefont
  {V.~I.}\ \bibnamefont {Korobov}}, \bibinfo {author} {\bibfnamefont
  {F.~M.~J.}\ \bibnamefont {Cozijn}}, \bibinfo {author} {\bibfnamefont
  {K.~S.~E.}\ \bibnamefont {Eikema}}, \bibinfo {author} {\bibfnamefont
  {W.}~\bibnamefont {Ubachs}},\ and\ \bibinfo {author} {\bibfnamefont
  {J.~C.~J.}\ \bibnamefont {Koelemeij}},\ }\bibfield  {title} {\bibinfo {title}
  {Proton-electron mass ratio from laser spectroscopy of $\mathrm{HD}^+$ at the
  part-per-trillion level},\ }\href {https://doi.org/10.1126/science.aba0453}
  {\bibfield  {journal} {\bibinfo  {journal} {Science}\ }\textbf {\bibinfo
  {volume} {369}},\ \bibinfo {pages} {1238} (\bibinfo {year}
  {2020})}\BibitemShut {NoStop}%
\bibitem [{\citenamefont {Alighanbari}\ \emph {et~al.}(2020)\citenamefont
  {Alighanbari}, \citenamefont {Giri}, \citenamefont {Constantin},
  \citenamefont {Korobov},\ and\ \citenamefont {Schiller}}]{Alighanbari2020}%
  \BibitemOpen
  \bibfield  {author} {\bibinfo {author} {\bibfnamefont {S.}~\bibnamefont
  {Alighanbari}}, \bibinfo {author} {\bibfnamefont {G.~S.}\ \bibnamefont
  {Giri}}, \bibinfo {author} {\bibfnamefont {F.~L.}\ \bibnamefont
  {Constantin}}, \bibinfo {author} {\bibfnamefont {V.~I.}\ \bibnamefont
  {Korobov}},\ and\ \bibinfo {author} {\bibfnamefont {S.}~\bibnamefont
  {Schiller}},\ }\bibfield  {title} {\bibinfo {title} {Precise test of quantum
  electrodynamics and determination of fundamental constants with
  $\mathrm{HD}^+$ ions},\ }\href {https://doi.org/10.1038/s41586-020-2261-5}
  {\bibfield  {journal} {\bibinfo  {journal} {Nature}\ }\textbf {\bibinfo
  {volume} {581}},\ \bibinfo {pages} {152} (\bibinfo {year}
  {2020})}\BibitemShut {NoStop}%
\bibitem [{\citenamefont {Kortunov}\ \emph {et~al.}(2021)\citenamefont
  {Kortunov}, \citenamefont {Alighanbari}, \citenamefont {Hansen},
  \citenamefont {Giri}, \citenamefont {Korobov},\ and\ \citenamefont
  {Schiller}}]{Kortunov:2021rfe}%
  \BibitemOpen
  \bibfield  {author} {\bibinfo {author} {\bibfnamefont {I.~V.}\ \bibnamefont
  {Kortunov}}, \bibinfo {author} {\bibfnamefont {S.}~\bibnamefont
  {Alighanbari}}, \bibinfo {author} {\bibfnamefont {M.~G.}\ \bibnamefont
  {Hansen}}, \bibinfo {author} {\bibfnamefont {G.~S.}\ \bibnamefont {Giri}},
  \bibinfo {author} {\bibfnamefont {V.~I.}\ \bibnamefont {Korobov}},\ and\
  \bibinfo {author} {\bibfnamefont {S.}~\bibnamefont {Schiller}},\ }\bibfield
  {title} {\bibinfo {title} {{Proton-electron mass ratio by high-resolution
  optical spectroscopy of ion ensembles in the resolved-carrier regime}},\
  }\href {https://doi.org/10.1038/s41567-020-01150-7} {\bibfield  {journal}
  {\bibinfo  {journal} {Nature Phys.}\ }\textbf {\bibinfo {volume} {17}},\
  \bibinfo {pages} {569} (\bibinfo {year} {2021})},\ \Eprint
  {https://arxiv.org/abs/2103.11741} {arXiv:2103.11741 [quant-ph]} \BibitemShut
  {NoStop}%
\bibitem [{\citenamefont {Fink}\ and\ \citenamefont
  {Myers}(2020)}]{PhysRevLett.124.013001}%
  \BibitemOpen
  \bibfield  {author} {\bibinfo {author} {\bibfnamefont {D.~J.}\ \bibnamefont
  {Fink}}\ and\ \bibinfo {author} {\bibfnamefont {E.~G.}\ \bibnamefont
  {Myers}},\ }\bibfield  {title} {\bibinfo {title} {Deuteron-to-proton mass
  ratio from the cyclotron frequency ratio of $\mathrm{H}_{2}^{+}$ to
  $\mathrm{D}^{+}$ with $\mathrm{H}_{2}^{+}$ in a resolved vibrational state},\
  }\href {https://doi.org/10.1103/PhysRevLett.124.013001} {\bibfield  {journal}
  {\bibinfo  {journal} {Phys. Rev. Lett.}\ }\textbf {\bibinfo {volume} {124}},\
  \bibinfo {pages} {013001} (\bibinfo {year} {2020})}\BibitemShut {NoStop}%
\bibitem [{\citenamefont {Rau}\ \emph {et~al.}(2020)\citenamefont {Rau},
  \citenamefont {Hei{\ss}e}, \citenamefont {K{\"o}hler-Langes}, \citenamefont
  {Sasidharan}, \citenamefont {Haas}, \citenamefont {Renisch}, \citenamefont
  {D{\"u}llmann}, \citenamefont {Quint}, \citenamefont {Sturm},\ and\
  \citenamefont {Blaum}}]{Rau2020}%
  \BibitemOpen
  \bibfield  {author} {\bibinfo {author} {\bibfnamefont {S.}~\bibnamefont
  {Rau}}, \bibinfo {author} {\bibfnamefont {F.}~\bibnamefont {Hei{\ss}e}},
  \bibinfo {author} {\bibfnamefont {F.}~\bibnamefont {K{\"o}hler-Langes}},
  \bibinfo {author} {\bibfnamefont {S.}~\bibnamefont {Sasidharan}}, \bibinfo
  {author} {\bibfnamefont {R.}~\bibnamefont {Haas}}, \bibinfo {author}
  {\bibfnamefont {D.}~\bibnamefont {Renisch}}, \bibinfo {author} {\bibfnamefont
  {C.~E.}\ \bibnamefont {D{\"u}llmann}}, \bibinfo {author} {\bibfnamefont
  {W.}~\bibnamefont {Quint}}, \bibinfo {author} {\bibfnamefont
  {S.}~\bibnamefont {Sturm}},\ and\ \bibinfo {author} {\bibfnamefont
  {K.}~\bibnamefont {Blaum}},\ }\bibfield  {title} {\bibinfo {title} {Penning
  trap mass measurements of the deuteron and the $\mathrm{HD}^+$ molecular
  ion},\ }\href {https://doi.org/10.1038/s41586-020-2628-7} {\bibfield
  {journal} {\bibinfo  {journal} {Nature}\ }\textbf {\bibinfo {volume} {585}},\
  \bibinfo {pages} {43} (\bibinfo {year} {2020})}\BibitemShut {NoStop}%
\bibitem [{\citenamefont {Korobov}\ and\ \citenamefont
  {Karr}(2021)}]{PhysRevA.104.032806}%
  \BibitemOpen
  \bibfield  {author} {\bibinfo {author} {\bibfnamefont {V.~I.}\ \bibnamefont
  {Korobov}}\ and\ \bibinfo {author} {\bibfnamefont {J.-P.}\ \bibnamefont
  {Karr}},\ }\bibfield  {title} {\bibinfo {title} {Rovibrational spin-averaged
  transitions in the hydrogen molecular ions},\ }\href
  {https://doi.org/10.1103/PhysRevA.104.032806} {\bibfield  {journal} {\bibinfo
   {journal} {Phys. Rev. A}\ }\textbf {\bibinfo {volume} {104}},\ \bibinfo
  {pages} {032806} (\bibinfo {year} {2021})}\BibitemShut {NoStop}%
\bibitem [{\citenamefont {Morel}\ \emph {et~al.}(2020)\citenamefont {Morel},
  \citenamefont {Yao}, \citenamefont {Clad\'e},\ and\ \citenamefont
  {Guellati-Kh\'elifa}}]{Morel:2020dww}%
  \BibitemOpen
  \bibfield  {author} {\bibinfo {author} {\bibfnamefont {L.}~\bibnamefont
  {Morel}}, \bibinfo {author} {\bibfnamefont {Z.}~\bibnamefont {Yao}}, \bibinfo
  {author} {\bibfnamefont {P.}~\bibnamefont {Clad\'e}},\ and\ \bibinfo {author}
  {\bibfnamefont {S.}~\bibnamefont {Guellati-Kh\'elifa}},\ }\bibfield  {title}
  {\bibinfo {title} {{Determination of the fine-structure constant with an
  accuracy of 81 parts per trillion}},\ }\href
  {https://doi.org/10.1038/s41586-020-2964-7} {\bibfield  {journal} {\bibinfo
  {journal} {Nature}\ }\textbf {\bibinfo {volume} {588}},\ \bibinfo {pages}
  {61} (\bibinfo {year} {2020})}\BibitemShut {NoStop}%
\bibitem [{\citenamefont {Parker}\ \emph {et~al.}(2018)\citenamefont {Parker},
  \citenamefont {Yu}, \citenamefont {Zhong}, \citenamefont {Estey},\ and\
  \citenamefont {M\"uller}}]{Parker:2018vye}%
  \BibitemOpen
  \bibfield  {author} {\bibinfo {author} {\bibfnamefont {R.~H.}\ \bibnamefont
  {Parker}}, \bibinfo {author} {\bibfnamefont {C.}~\bibnamefont {Yu}}, \bibinfo
  {author} {\bibfnamefont {W.}~\bibnamefont {Zhong}}, \bibinfo {author}
  {\bibfnamefont {B.}~\bibnamefont {Estey}},\ and\ \bibinfo {author}
  {\bibfnamefont {H.}~\bibnamefont {M\"uller}},\ }\bibfield  {title} {\bibinfo
  {title} {{Measurement of the fine-structure constant as a test of the
  Standard Model}},\ }\href {https://doi.org/10.1126/science.aap7706}
  {\bibfield  {journal} {\bibinfo  {journal} {Science}\ }\textbf {\bibinfo
  {volume} {360}},\ \bibinfo {pages} {191} (\bibinfo {year} {2018})},\ \Eprint
  {https://arxiv.org/abs/1812.04130} {arXiv:1812.04130 [physics.atom-ph]}
  \BibitemShut {NoStop}%
\bibitem [{\citenamefont {Czarnecki}\ and\ \citenamefont
  {Szafron}(2016)}]{Czarnecki:2016lzl}%
  \BibitemOpen
  \bibfield  {author} {\bibinfo {author} {\bibfnamefont {A.}~\bibnamefont
  {Czarnecki}}\ and\ \bibinfo {author} {\bibfnamefont {R.}~\bibnamefont
  {Szafron}},\ }\bibfield  {title} {\bibinfo {title} {{Light-by-light
  scattering in the Lamb shift and the bound electron g factor}},\ }\href
  {https://doi.org/10.1103/PhysRevA.94.060501} {\bibfield  {journal} {\bibinfo
  {journal} {Phys. Rev. A}\ }\textbf {\bibinfo {volume} {94}},\ \bibinfo
  {pages} {060501} (\bibinfo {year} {2016})},\ \Eprint
  {https://arxiv.org/abs/1611.04875} {arXiv:1611.04875 [physics.atom-ph]}
  \BibitemShut {NoStop}%
\bibitem [{\citenamefont {Szafron}\ \emph {et~al.}(2019)\citenamefont
  {Szafron}, \citenamefont {Korzinin}, \citenamefont {Shelyuto}, \citenamefont
  {Ivanov},\ and\ \citenamefont {Karshenboim}}]{Szafron:2019tho}%
  \BibitemOpen
  \bibfield  {author} {\bibinfo {author} {\bibfnamefont {R.}~\bibnamefont
  {Szafron}}, \bibinfo {author} {\bibfnamefont {E.~Y.}\ \bibnamefont
  {Korzinin}}, \bibinfo {author} {\bibfnamefont {V.~A.}\ \bibnamefont
  {Shelyuto}}, \bibinfo {author} {\bibfnamefont {V.~G.}\ \bibnamefont
  {Ivanov}},\ and\ \bibinfo {author} {\bibfnamefont {S.~G.}\ \bibnamefont
  {Karshenboim}},\ }\bibfield  {title} {\bibinfo {title} {{Virtual Delbr\"uck
  scattering and the Lamb shift in light hydrogenlike atoms}},\ }\href
  {https://doi.org/10.1103/PhysRevA.100.032507} {\bibfield  {journal} {\bibinfo
   {journal} {Phys. Rev. A}\ }\textbf {\bibinfo {volume} {100}},\ \bibinfo
  {pages} {032507} (\bibinfo {year} {2019})},\ \Eprint
  {https://arxiv.org/abs/1909.04116} {arXiv:1909.04116 [physics.atom-ph]}
  \BibitemShut {NoStop}%
\bibitem [{\citenamefont {Karshenboim}\ \emph
  {et~al.}(2019{\natexlab{a}})\citenamefont {Karshenboim}, \citenamefont
  {Ozawa}, \citenamefont {Shelyuto}, \citenamefont {Szafron},\ and\
  \citenamefont {Ivanov}}]{Karshenboim:2019iuq}%
  \BibitemOpen
  \bibfield  {author} {\bibinfo {author} {\bibfnamefont {S.~G.}\ \bibnamefont
  {Karshenboim}}, \bibinfo {author} {\bibfnamefont {A.}~\bibnamefont {Ozawa}},
  \bibinfo {author} {\bibfnamefont {V.~A.}\ \bibnamefont {Shelyuto}}, \bibinfo
  {author} {\bibfnamefont {R.}~\bibnamefont {Szafron}},\ and\ \bibinfo {author}
  {\bibfnamefont {V.~G.}\ \bibnamefont {Ivanov}},\ }\bibfield  {title}
  {\bibinfo {title} {{The Lamb shift of the {$1S$} state in hydrogen: Two-loop
  and three-loop contributions}},\ }\href
  {https://doi.org/10.1016/j.physletb.2019.06.023} {\bibfield  {journal}
  {\bibinfo  {journal} {Phys. Lett. B}\ }\textbf {\bibinfo {volume} {795}},\
  \bibinfo {pages} {432} (\bibinfo {year} {2019}{\natexlab{a}})},\ \Eprint
  {https://arxiv.org/abs/1906.11105} {arXiv:1906.11105 [physics.atom-ph]}
  \BibitemShut {NoStop}%
\bibitem [{\citenamefont {Karshenboim}\ \emph
  {et~al.}(2019{\natexlab{b}})\citenamefont {Karshenboim}, \citenamefont
  {Ozawa},\ and\ \citenamefont {Ivanov}}]{Karshenboim:2019dug}%
  \BibitemOpen
  \bibfield  {author} {\bibinfo {author} {\bibfnamefont {S.~G.}\ \bibnamefont
  {Karshenboim}}, \bibinfo {author} {\bibfnamefont {A.}~\bibnamefont {Ozawa}},\
  and\ \bibinfo {author} {\bibfnamefont {V.~G.}\ \bibnamefont {Ivanov}},\
  }\bibfield  {title} {\bibinfo {title} {{Higher-order logarithmic corrections
  and the two-loop self-energy of a {$1S$} electron in hydrogen}},\ }\href
  {https://doi.org/10.1103/PhysRevA.100.032515} {\bibfield  {journal} {\bibinfo
   {journal} {Phys. Rev. A}\ }\textbf {\bibinfo {volume} {100}},\ \bibinfo
  {pages} {032515} (\bibinfo {year} {2019}{\natexlab{b}})}\BibitemShut
  {NoStop}%
\bibitem [{\citenamefont {Karshenboim}\ and\ \citenamefont
  {Ivanov}(2018)}]{Karshenboim:2018mtf}%
  \BibitemOpen
  \bibfield  {author} {\bibinfo {author} {\bibfnamefont {S.~G.}\ \bibnamefont
  {Karshenboim}}\ and\ \bibinfo {author} {\bibfnamefont {V.~G.}\ \bibnamefont
  {Ivanov}},\ }\bibfield  {title} {\bibinfo {title} {{Higher-order logarithmic
  contributions to the Lamb shift in hydrogen, deuterium, and {He$^+$}}},\
  }\href {https://doi.org/10.1103/PhysRevA.98.022522} {\bibfield  {journal}
  {\bibinfo  {journal} {Phys. Rev. A}\ }\textbf {\bibinfo {volume} {98}},\
  \bibinfo {pages} {022522} (\bibinfo {year} {2018})}\BibitemShut {NoStop}%
\bibitem [{\citenamefont {Karshenboim}\ and\ \citenamefont
  {Shelyuto}(2019)}]{Karshenboim:2019siz}%
  \BibitemOpen
  \bibfield  {author} {\bibinfo {author} {\bibfnamefont {S.~G.}\ \bibnamefont
  {Karshenboim}}\ and\ \bibinfo {author} {\bibfnamefont {V.~A.}\ \bibnamefont
  {Shelyuto}},\ }\bibfield  {title} {\bibinfo {title} {{Three-loop radiative
  corrections to the {$1S$} Lamb shift in hydrogen}},\ }\href
  {https://doi.org/10.1103/PhysRevA.100.032513} {\bibfield  {journal} {\bibinfo
   {journal} {Phys. Rev. A}\ }\textbf {\bibinfo {volume} {100}},\ \bibinfo
  {pages} {032513} (\bibinfo {year} {2019})}\BibitemShut {NoStop}%
\bibitem [{\citenamefont {Laporta}(2020)}]{Laporta:2019fmy}%
  \BibitemOpen
  \bibfield  {author} {\bibinfo {author} {\bibfnamefont {S.}~\bibnamefont
  {Laporta}},\ }\bibfield  {title} {\bibinfo {title} {{High-precision
  calculation of the 4-loop QED contribution to the slope of the Dirac form
  factor}},\ }\href {https://doi.org/10.1016/j.physletb.2019.135137} {\bibfield
   {journal} {\bibinfo  {journal} {Phys. Lett. B}\ }\textbf {\bibinfo {volume}
  {800}},\ \bibinfo {pages} {135137} (\bibinfo {year} {2020})},\ \Eprint
  {https://arxiv.org/abs/1910.01248} {arXiv:1910.01248 [hep-ph]} \BibitemShut
  {NoStop}%
\bibitem [{\citenamefont {Yerokhin}\ and\ \citenamefont
  {Shabaev}(2015)}]{Yerokhin:2015}%
  \BibitemOpen
  \bibfield  {author} {\bibinfo {author} {\bibfnamefont {V.~A.}\ \bibnamefont
  {Yerokhin}}\ and\ \bibinfo {author} {\bibfnamefont {V.~M.}\ \bibnamefont
  {Shabaev}},\ }\bibfield  {title} {\bibinfo {title} {{Nuclear Recoil Effect in
  the Lamb Shift of Light Hydrogenlike Atoms}},\ }\href
  {https://doi.org/10.1103/PhysRevLett.115.233002} {\bibfield  {journal}
  {\bibinfo  {journal} {Phys. Rev. Lett.}\ }\textbf {\bibinfo {volume} {115}},\
  \bibinfo {pages} {233002} (\bibinfo {year} {2015})}\BibitemShut {NoStop}%
\bibitem [{\citenamefont {Yerokhin}\ and\ \citenamefont
  {Shabaev}(2016)}]{PhysRevA.93.062514}%
  \BibitemOpen
  \bibfield  {author} {\bibinfo {author} {\bibfnamefont {V.~A.}\ \bibnamefont
  {Yerokhin}}\ and\ \bibinfo {author} {\bibfnamefont {V.~M.}\ \bibnamefont
  {Shabaev}},\ }\bibfield  {title} {\bibinfo {title} {Nuclear recoil
  corrections to the lamb shift of hydrogen and light hydrogenlike ions},\
  }\href {https://doi.org/10.1103/PhysRevA.93.062514} {\bibfield  {journal}
  {\bibinfo  {journal} {Phys. Rev. A}\ }\textbf {\bibinfo {volume} {93}},\
  \bibinfo {pages} {062514} (\bibinfo {year} {2016})}\BibitemShut {NoStop}%
\bibitem [{\citenamefont {{Pachucki}}\ \emph {et~al.}(2018)\citenamefont
  {{Pachucki}}, \citenamefont {{Patk{\'o}{\v{s}}}},\ and\ \citenamefont
  {{Yerokhin}}}]{Pachucki:2018xxx}%
  \BibitemOpen
  \bibfield  {author} {\bibinfo {author} {\bibfnamefont {K.}~\bibnamefont
  {{Pachucki}}}, \bibinfo {author} {\bibfnamefont {V.}~\bibnamefont
  {{Patk{\'o}{\v{s}}}}},\ and\ \bibinfo {author} {\bibfnamefont {V.~A.}\
  \bibnamefont {{Yerokhin}}},\ }\bibfield  {title} {\bibinfo {title}
  {{Three-photon-exchange nuclear structure correction in hydrogenic
  systems}},\ }\href {https://doi.org/10.1103/PhysRevA.97.062511} {\bibfield
  {journal} {\bibinfo  {journal} {\pra}\ }\textbf {\bibinfo {volume} {97}},\
  \bibinfo {eid} {062511} (\bibinfo {year} {2018})},\ \Eprint
  {https://arxiv.org/abs/1803.10313} {arXiv:1803.10313 [physics.atom-ph]}
  \BibitemShut {NoStop}%
\bibitem [{\citenamefont {Abi}\ \emph {et~al.}(2021)\citenamefont {Abi} \emph
  {et~al.}}]{Muong-2:2021ojo}%
  \BibitemOpen
  \bibfield  {author} {\bibinfo {author} {\bibfnamefont {B.}~\bibnamefont
  {Abi}} \emph {et~al.} (\bibinfo {collaboration} {Muon $g-2$}),\ }\bibfield
  {title} {\bibinfo {title} {{Measurement of the Positive Muon Anomalous
  Magnetic Moment to 0.46 ppm}},\ }\href
  {https://doi.org/10.1103/PhysRevLett.126.141801} {\bibfield  {journal}
  {\bibinfo  {journal} {Phys. Rev. Lett.}\ }\textbf {\bibinfo {volume} {126}},\
  \bibinfo {pages} {141801} (\bibinfo {year} {2021})},\ \Eprint
  {https://arxiv.org/abs/2104.03281} {arXiv:2104.03281 [hep-ex]} \BibitemShut
  {NoStop}%
\bibitem [{\citenamefont {Aoyama}\ \emph {et~al.}(2020)\citenamefont {Aoyama}
  \emph {et~al.}}]{Aoyama:2020ynm}%
  \BibitemOpen
  \bibfield  {author} {\bibinfo {author} {\bibfnamefont {T.}~\bibnamefont
  {Aoyama}} \emph {et~al.},\ }\bibfield  {title} {\bibinfo {title} {{The
  anomalous magnetic moment of the muon in the Standard Model}},\ }\href
  {https://doi.org/10.1016/j.physrep.2020.07.006} {\bibfield  {journal}
  {\bibinfo  {journal} {Phys. Rept.}\ }\textbf {\bibinfo {volume} {887}},\
  \bibinfo {pages} {1} (\bibinfo {year} {2020})},\ \Eprint
  {https://arxiv.org/abs/2006.04822} {arXiv:2006.04822 [hep-ph]} \BibitemShut
  {NoStop}%
\bibitem [{\citenamefont {Borsanyi}\ \emph {et~al.}(2021)\citenamefont
  {Borsanyi} \emph {et~al.}}]{Borsanyi:2020mff}%
  \BibitemOpen
  \bibfield  {author} {\bibinfo {author} {\bibfnamefont {S.}~\bibnamefont
  {Borsanyi}} \emph {et~al.},\ }\bibfield  {title} {\bibinfo {title} {{Leading
  hadronic contribution to the muon magnetic moment from lattice QCD}},\ }\href
  {https://doi.org/10.1038/s41586-021-03418-1} {\bibfield  {journal} {\bibinfo
  {journal} {Nature}\ }\textbf {\bibinfo {volume} {593}},\ \bibinfo {pages}
  {51} (\bibinfo {year} {2021})},\ \Eprint {https://arxiv.org/abs/2002.12347}
  {arXiv:2002.12347 [hep-lat]} \BibitemShut {NoStop}%
\bibitem [{\citenamefont {Parthey}\ \emph {et~al.}(2011)\citenamefont {Parthey}
  \emph {et~al.}}]{Parthey:2011lfa}%
  \BibitemOpen
  \bibfield  {author} {\bibinfo {author} {\bibfnamefont {C.~G.}\ \bibnamefont
  {Parthey}} \emph {et~al.},\ }\bibfield  {title} {\bibinfo {title} {{Improved
  Measurement of the Hydrogen {$1S - 2S$} Transition Frequency}},\ }\href
  {https://doi.org/10.1103/PhysRevLett.107.203001} {\bibfield  {journal}
  {\bibinfo  {journal} {Phys. Rev. Lett.}\ }\textbf {\bibinfo {volume} {107}},\
  \bibinfo {pages} {203001} (\bibinfo {year} {2011})},\ \Eprint
  {https://arxiv.org/abs/1107.3101} {arXiv:1107.3101 [physics.atom-ph]}
  \BibitemShut {NoStop}%
\end{thebibliography}
\end{document}